\documentclass[eqsecnum,twocolumn,aps,tightenlines,floats,prd,showpacs]{revtex4-1}
\def\bea{\begin{eqnarray}}
\def\eea{\end{eqnarray}}
\def\bal{\begin{align}}
\def\eal{\end{align}}
\def\etal{{\it et al.\/}}

\def\eqq2{3.31}

\usepackage{graphics}
\usepackage{graphicx}
\usepackage{epsf} 
\usepackage{amsmath}
\usepackage{amssymb}
\usepackage{bbold}
\usepackage{slashed}
\usepackage{appendix}
\usepackage{bm}
\usepackage{dcolumn}
\usepackage{color,soul}
\usepackage{rotating}

\usepackage{footnote} 

\begin{document}  


\phantom{0}
\hspace{5.5in}\parbox{1.5in}{ \leftline{JLAB-THY-19-3022}
                \leftline{}\leftline{}\leftline{}\leftline{}
}
\title
{\bf  Covariant Spectator Theory of $np$ scattering: Deuteron form factors}

\author{Franz Gross$^{1,2}$ }
\email{email address: gross@jlab.org}
\affiliation{
$^1$Thomas Jefferson National Accelerator Facility, Newport News, VA 23606 \vspace{-0.15in}}
\affiliation{
$^2$College of William and Mary, Williamsburg, Virginia 23185}

\date{\today}

\begin{abstract} 

The deuteron form factors are calculated using two model wave functions obtained from the 2007 CST high precision fits to $np$ scattering data. Included in the calculation are a new class of isoscalar $np$ interaction currents which are  automatically generated by the nuclear force model used in these fits. If the nuclear model WJC2 is used, a precision fit ($\chi^2$/datum $\eqsim1$) to the Sick Global Analysis (GA) of all $ed$ elastic scattering data can be obtained by adjusting the unknown off-shell nucleon form factors $F_3(Q^2)$ (discussed before) and $F_4(Q^2)$ (introduced in this paper), and predicting the high $Q^2$ behavior of the neutron charge form factor $G_{En}(Q^2)$ well beyond the region where it has been measured directly.  Relativistic corrections, isoscalar interaction currents, and off-shell effects are defined, discussed, and their size displayed.  A rationale for extending $ed$ elastic scattering measurements  to higher $Q^2$ is presented.


\end{abstract}
 
\phantom{0}
\vspace{0.76in}
\vspace*{-0.1in}  

\maketitle


\section{Introduction} 
\label{sec:intro}

\subsection{Background}

This work is the last in a series of  papers \cite{Gross:2014zra, Gross:2014wqa,Gross:2014tya}  (referred to as Refs.~I, II, and III)  that present the fourth generation calculation of the deuteron form factors using what is sometimes called the Covariant Spectator Theory (CST) \cite{Gross:1969rv,Gross:1972ye,Gross:1982nz}.  The third generation, done by Van Orden and collaborators in 1995 \cite{VanOrden:1995eg}, calculated the form factors from a variation of model IIB   (originally obtained from a 1991 fit to the $np$ database  \cite{Gross:1991pm},  with an improved fit giving $\chi^2$/datum $\simeq 2.5$ \cite{VanOrden:1995eg})  already provided an excellent description of the deuteron form factors.  The current calculation  is needed only because a better CST fit to the $np$ database was found in 2007. This fit, with a  $\chi^2$/datum $\sim 1$, included momentum dependent terms in the kernel and requires a completely new treatment.  For a brief review of the previous CST history of calculations of the form factors, see the introduction to 
Ref.~I.  For a more comprehensive survey of the field see several recent reviews \cite{Garcon:2001sz,Gilman:2001yh,Marcucci:2015rca}.

The CST, in common with other  treatments based on the assumption that $NN$ scattering can be explained by the ladders and crossed ladders arising from the exchange of mesons between nucleons \cite{Gross:1982nz,Pena:1996tf}, treats nucleons and mesons as the elementary degrees of freedom, with the internal structure of the nucleons and mesons treated phenomenologically.  This means that, in particular, the electromagnetic form factors of the nucleons that are bound into a deuteron are  not calculated, but must be obtained from direct measurements of electron-nucleon scattering.  If the form factors cannot be measured directly, they can be treated as undetermined functions that can be fixed by fitting the theory to electron-deuteron ($ed$) scattering data.

In common with the 1991 fit that lead to model IIB, the  new fit to the 2007 $np$ data base \cite{Gross:2008ps,Gross:2010qm}  was obtained using the CST two body equation (sometimes called the Gross equation) with a one boson exchange (OBE) kernel.  However, we found that a high precision fit (with $\chi^2$/datum $\sim1$) could be obtained only if  the $NN\sigma_0$ vertices associated with the exchange of a scalar-isoscalar meson, denoted $\sigma_0$,  included momentum dependent  terms in the form 
\bea
\Lambda^{\sigma_0}(p,p')= 
g_{\sigma_0}{ \bf 1}-\nu_{\sigma_0}\big[\Theta(p)+\Theta(p')\big]\qquad
\label{eq:1.2}
\eea
where $\nu_{\sigma_0}$ is a parameter  fixed by fitting the $NN$ scattering data, $p$ and $p'$ are the four-momenta of the outgoing and incoming nucleons, respectively, and the $\Theta$ are projection operators 
\bea
\Theta(p)=\frac{m-\slashed{p}}{2m} \, ,\label{eq:theta}
\eea
which are non-zero for off-shell particles, and hence are a feature of both the Bethe-Salpeter and CST equations.  

\begin{figure*}
\centerline{
\mbox{
\includegraphics[width=5.6in]{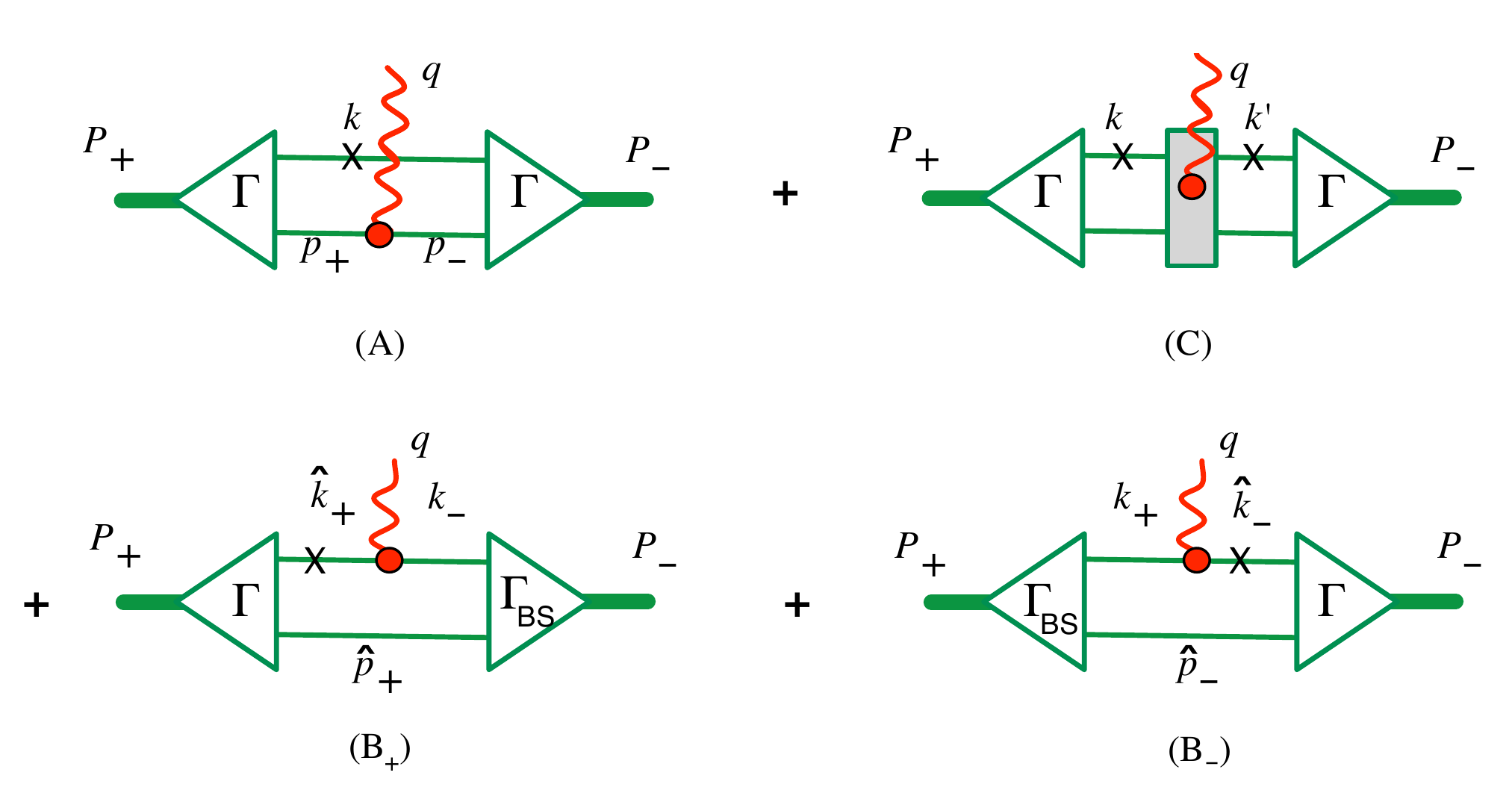}
}
}
\caption{\footnotesize\baselineskip=10pt  Diagrammatic representation of the current operator of the Covariant Spectator Theory with particle 1 on-shell (the on-shell particle is labeled with a $\times$).  Diagrams (A), (B$_+$), and (B$_-$) are the complete impulse approximation (CIA), while (C) is the interaction current term.  
Note that both particles are off-shell in the initial state in diagram (B$_+$) and in the final state in diagram (B$_-$).  
}
\label{Fig1}
\end{figure*} 

\begin{figure*}
\centerline{
\mbox{
\includegraphics[width=7in]{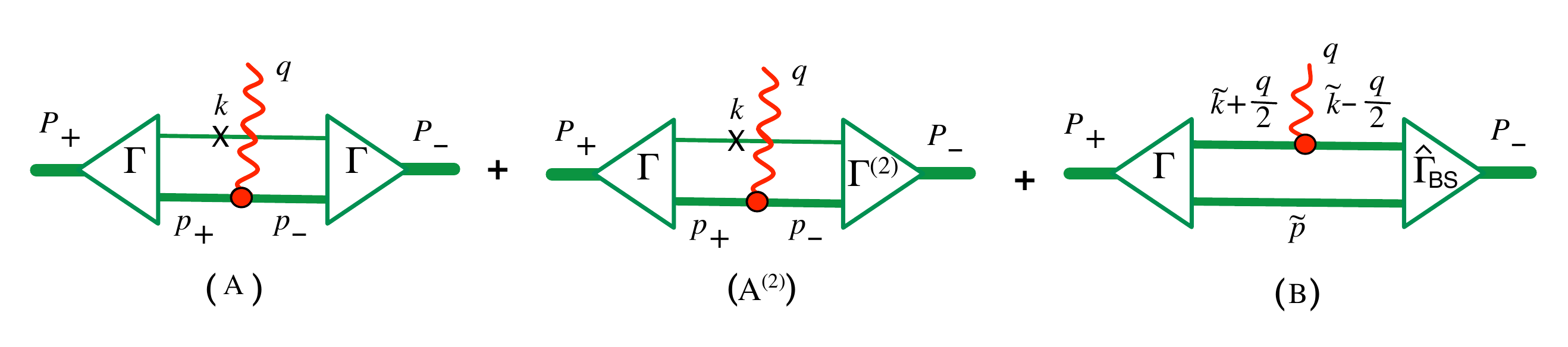}
}
}
\caption{\footnotesize\baselineskip=10pt Using the consequences of current conservation the diagrams in Fig.~\ref{Fig1} can be transformed into the three diagrams shown above.  Diagram (A) is unchanged, the two diagrams diagrams (B$_\pm$) of Fig.~\ref{Fig1} are combined into the single diagram (B), 
and the interaction current contributions of Fig.~\ref{Fig1}C are distributed to diagrams (A$^{(2)}$) and parts of the (B) diagram, as discussed in the text. Off-shell nucleon lines are thicker than on-shell lines.   Diagram (B) represents the sum of two diagrams, one with particle 1 off-shell in the initial state and one with particle 1 off-shell in the final state (in this case $\widehat\Gamma_{\rm BS}$  and $\Gamma$ would be interchanged in the diagram), collectively represented by the single diagram with a thick line for particle 1.  
}
\label{Fig2}
\end{figure*} 

Two high precision models were found with somewhat different properties. Model WJC1 (originaly designated WJC-1), designed to give the best fit possible, has 27 parameters, $\chi^2/{\rm datum}\simeq 1.06$, and a large $\nu_{\sigma_0}=-15.2$.  Model WJC2 (originaly designated WJC-2), designed to give a excellent fit with as few parameters as possible,  has only 15 parameters, $\chi^2/{\rm datum}\simeq 1.12$, and a smaller $\nu_{\sigma_0}=-2.6$.  Both models also predict the correct triton binding energy (see Figs.~12 and 13 of Ref.~\cite{Gross:2008ps} and Ref.~\cite{Stadler:1996ut}).  The deuteron wave functions predicted by both of these models \cite{Gross:2010qm} have small P-state components of relativistic origin, and the normalization of the wave functions includes a term coming from the energy dependence of the kernel, which contributes $-5.5\%$ for WJC1 and $-2.3\%$ for WJC2.

This momentum dependence of the kernel implies the existence of a  new class of $np$ isoscalar interaction currents that will contribute to the electromagnetic interaction of the deuteron, leading to the need for this fouth generation calculation.  These currents were fixed in 
Ref.\ I, and used to predict the deuteron magnetic moment 
(Ref.~II) and the quadrupole moment 
(Ref.~III).  
This paper completes this series of papers by calculating the dependence of the form factors on the momentum transfer of the scattered electron, $Q^2$.  

In the process of fitting the $ed$ data, we are able to determine two off-shell nucleon form factors and predict the high momentum behavior of the neutron electric form factor, $G_{En}(Q^2)$, beyond the region where it has been measured.  These and other major conclusions of this paper are discussed in detail in Sec.~\ref{sec:5} below.

\subsection{Organization of the paper}

This paper is organized into six sections, with most of the theoretical details moved to the  Appendices.  The rest of this section describes the ingredients of the calculation as simply as possible, with emphasis on the important off-shell nucleon current.  A more complete discussion can be found in the Appendices and in Refs.~I--III.  
The major results are described in Sec.~\ref{sec:cal}, 
which gives predictions that are extracted from the $ed$ measurements. 
Sec.~\ref{sec:Rel} shows how the individual deuteron form factors are built up from the different theoretical contributions, and Sec.~\ref{sec:Rel2} discusses the size and importance of relativistic effects. 
The results for the deuteron static moments are reviewed in Sec.~\ref{sec:qis0}, and finally I draw major conclusions in Sec.~\ref{sec:5}.  The reader eager to get to the conclusions may jump to Sec.~\ref{sec:5}, and backtrack as needed to fill in the many  missing details.

Seven appendices summarize many details needed for a precise understanding of this paper.  Appendix \ref{app:theory}
 reviews the theoretical definitions of the deuteron form factors, deuteron current and deuteron wave and vertex functions, and examines how the arguments of the amplitudes are shifted by the relativistic boosts that enter into the calculation of the form factors.  Appendix  \ref{app:NRff} derives the form of the nonrelativistic deuteron charge form factor, $G_C$, in momentum  space.  Appendices \ref{app:F3F4nextract} and \ref{app:GEnextract} discuss some details of the extraction of the nucleon form factors from the theory, and Appendices \ref{app:Ctraces} and \ref{app:x10corr} describe some theoretical transformations that facilitate the calculations.  Finally,
Appendix \ref{sec:errorsinII} discusses some errors that were found in Ref.~II.

\subsection{Diagrammatic form of the deuteron current}

In CST when a OBE kernel is used to describe the $NN$ interaction, the two-body current is given initially  by the four diagrams shown in Fig.~\ref{Fig1}.  Here, by convention, it is assumed that particle 1 is on-shell (I could have chosen particle 2 to be on-shell with corresponding changes in the diagrams), and the necessary (anti)symmetry between the two particles is contained in the kernel, which is explicitly (anti)symmetrized.   In these diagrams the deuteron structure is represented by a vertex function, $\Gamma$, in which particle 1 is on shell and particle 2 off-shell, and a vertex function $\Gamma_{\rm BS}$ in which {\it both\/} particles are off-shell.  The vertex function $\Gamma$ is calculated directly from the deuteron bound state equation and $\Gamma_{\rm BS}$  can be calculated from $\Gamma$.   

I showed in Ref.\ I  how the current operator shown in Fig.~\ref{Fig1}(C) can be re-expressed using an effective vertex function $\Gamma^{(2)}$ and a subtracted vertex function $\widehat \Gamma_{\rm BS}$.   When this is done, diagram  \ref{Fig2}(A$^{(2)}$) and parts of  \ref{Fig2}(B) include the interaction current contributions originally included in diagram \ref{Fig1}(C).  The new diagram  \ref{Fig2}(B) not only includes part of the interaction current, but is also a generic way of combining the two diagrams \ref{Fig1}(B$_\pm$).   
The three diagrams of Fig.~\ref{Fig2} 
are completely equivalent to the five diagrams shown in Fig.~2 of Ref.~I and Fig.~1 of Ref.~II (but the labeling of the momenta in the B$_\pm$ diagrams differs from the choice here).

Diagrams \ref{Fig2}(A) and \ref{Fig2}(A$^{(2)}$) describe the interaction of the photon with particle 2, allowing particle 1 to be on-shell in both the initial and final state.  The internal momenta are 
\bea
&&k=\{E_k,{\bf k}\}\equiv \widehat k
\nonumber\\
&&p_\pm=P_\pm-k\, ,
\eea
where $P_+$ ($P_-$) are the four-momenta of the outgoing (incoming) deuterons, and the hat symbol over a four-vector means that the four-vector is on-shell. 
Diagram \ref{Fig2}(B)  describes all the interactions of the photon with particle 1, so that  both particles must off-shell in either the initial  or in the final state.  Here the internal momenta are
\bea
\widetilde k&=&\{k_0,{\bf k}\}
\nonumber\\
\widetilde p&=&P_\pm- (\widetilde k\pm \frac12 q)\, .   
\eea
The final (initial) nucleon is on shell when $k_0=E_+$ ($E_-$),  with 
\bea
E_\pm=\sqrt{m^2+\left({\bf k}\pm \frac{{\bf q}}{2}\right)^2}\, . \label{eq:epm}
\eea

\subsection{Strong form factor $h$ and the bound nucleon current} 

In this section I describe two central features of the CST calculation of the deuteron observables and form factors from the diagrams in Fig.~\ref{Fig2}.  These are (i)  the presence of a strong nucleon form factor, $h$, and (ii) the structure of the bound nucleon current, which depends on {\it four} form factors: not only the usual Dirac and Pauli form factors $F_1$ and $F_2$, but also two off-shell form factors $F_3$ and $F_4$.  The first of these, $F_3$, has been discussed extensively in previous work, but $F_4$ has never been introduced before and is a major new feature of this paper.  


\subsubsection{The strong nucleon form factor $h(p)$}

	In all strong, nonperturbative theories of hadronic structure there is a need to include form factors that cut off  high momentum contributions and provide convergent results.  
In the CST-OBE models studied so far, the form factors at the meson-$NN$ vertices are assumed to be   {\it products\/} of strong form factors for each particle entering or leaving the vertex.  This means that for each nucleon of momentum $p$ entering or leaving a vertex, there is a universal {\it strong\/} nucleon form factor $h(p)$ (a function of $p^2$ only) present at that vertex.  This form factor is normalized so that when $p=\hat p$ (so that $p$ is on shell), $h(\hat p)=1$. 

Because it is universal, the strong form factor  associated with each external nucleon line can be factored out from the $NN$ scattering kernel giving
\bea
&&\overline{V}(k,k';P)=h(k)h(p)\widetilde V(k,k';P)h(k')h(p')\, ,
\label{eq:Vwithh}
\eea
where $k$ ($k'$) are the four-momentum of the outgoing (incoming) particle 1,  $\widetilde V$ is the {\it reduced\/} kernel  
and, for both primed and unprimed variables, $p=P-k$. 
Note that the expression for the kernel is written allowing for the possibility that any (or all four) of the particles could be off-shell.  
Similarly, removing the strong form factors from the vertex function gives a {\it reduced\/} vertex function $\widetilde \Gamma_{BS}$, where
\bea
\widehat\Gamma_{BS}(k,P)=h(k)h(p)\widetilde\Gamma_{BS}(k,P)\, .
\eea

Since $h(p)$ is included in the kernel, and $h(\hat k)=1$ when particle 1 is on-shell, the dependence of the results on variations of $h(p)$ when particle 1 is on-shell has already been studied in the fits to the $NN$ scattering and presents nothing new.   However, when electromagnetic current conservation is imposed, the presence of  $h(p)$ leads to a modification of the nucleon current.  This $h$ dependence is a new feature of the relativistic theory that is interesting to study.  
In addition, when particle 1 is off-shell, so that $k\ne \hat k$, the dependence of the calculation on $h(p)$ for $k_0-E\ne0$ is another feature of the relativistic theory that is new.  

I will report on some of these effects later in Sec.~\ref{sec:Rel}; for now I only want to highlight existence of the strong form factor $h$, because its presence drives the discussion of the bound nucleon current.

\subsubsection{Structure of the bound nucleon current} \label{Sec:A2}


Using interactions that depend only on $\Delta$, the momentum transfer by the interacting particles, Feynman showed a long time ago that current conservation could be proved if the off-shell bound nucleon current satisfied the  Ward-Takahashi (WT) identity
\bea
q_\mu \, j_0^\mu(p',p)&&=e_0\Big[S^{-1}(p)-S^{-1}(p')\Big]
\, , \label{eq:WT0}
\eea
where $S(p)$ is the propagator of a bare nucleon, which in my notation (with the $i$'s removed) is
\bea
S(p)=\frac1{m-\slashed{p}-i\epsilon}\, .
\eea

When a strong nucleon form factor is present, the interactions in a one boson exchange (OBE) model will be of the form  $h(p)V(\Delta)h(p')$, and can be made to depend only on $\Delta$ if the strong nucleon form factors coming from the initial and final interactions that connect each propagator are moved from the interactions to the propagators connecting them. 
Since each  propagator connects two interactions, the new (dressed) nucleon propagator then has the form
\bea
S_d(p)=h^2(p) S(p)\, .  \label{eq:Sd}
\eea
Now a similar proof of current conservation is possible \cite{Gross:1987bu} provided a {\it reduced\/}  current $j_R^\mu(p',p)$, is constructed 
%
\bea
j^\mu&&(p',p)=h(p')h(p)j_R^\mu(p',p) \, ,
\eea
and required to satisfy a generalized WT identity
\bea
q_\mu \, j_R^\mu(p',p)&&=e_0\Big[S^{-1}_{d}(p)-S^{-1}_{d}(p')\Big]
\, . \label{eq:210ab}
\eea
%

There are many solutions to (\ref{eq:210ab}).  The one I use in this paper is
\bea
j^\mu&&(p',p)=
e_0\,f_0(p',p)\bigg[{\cal F}_1^\mu +F_2(Q^2)\frac{i\sigma^{\mu\nu}q_\nu}{2m}\bigg]
\nonumber\\
&&+e_0\,g_0(p',p)\Theta(p')\bigg[{\cal F}_3^\mu +F_4(Q^2)\frac{i\sigma^{\mu\nu}q_\nu}{2m}\bigg]\Theta(p)\qquad \label{3.1}
\eea
where $f_0, g_0$ are (uniquely determined) off-shell functions discussed below,  $e_0=\frac12$ is the isoscalar charge, the off-shell projection operator $\Theta$ was defined in (\ref{eq:theta}),
\bea
{\cal F}_i^\mu&=&[F_i(Q^2)-1]\widetilde\gamma^\mu +\gamma^\mu
\nonumber\\&=&
F_i(Q^2)\widetilde\gamma^\mu+\frac{\slashed{q}q^\mu}{q^2}\, ,\qquad
\label{eq:Fimua}
\eea
and the transverse gamma matrix is
\begin{equation}
\widetilde\gamma^\mu=\gamma^\mu-\frac{\slashed{q}q^\mu}{q^2}\, , \label{eq:gammatilde}
\end{equation}
with $q=p'-p$.  Except for the addition of the new form factor $F_4$, this is precisely the current that has been used in all previous work.  
 
\subsubsection{Uniqueness of the bound nucleon current, and the principle of balance}

The longitudinal parts of the current (\ref{3.1}) are largely determined by the generalized WT identity (\ref{eq:210ab}).  Still, as (\ref{eq:Fimua}) displays, the  
important physics contained in the form factors $F_1$ and $F_3$ is purely {\it transverse\/}, and the longitudinal part that is constrained by the WT identities will {\it not contribute to any observable\/} since it is proportional to $q^\mu$ which vanishes when contracted into any conserved current or any of the three polarization vectors of an off-shell photon.  
The form factors themselves are completely unconstrained by current conservation, except for the requirement that $F_1(0)=F_3(0)=1$ (with the real normalization set by $e_0$).  This is as it should be; the structure of the nucleon should not be fixed by the general requirement of current conservation.

Similarly, the transverse Pauli-like terms $F_2$ and $F_4$ are completely unconstrained, and there are may other off-shell terms that we could add to the current.  What principal is to constrain these?

In Ref.\ I, I introduced the principles of {\it simplicity\/} and {\it picture independence\/} in an attempt to limit possible contributions.  I found that, using current conservation and these principles,  all contributions from the structure of the meson-nucleon vertices could be expressed in terms of the nucleon structure $F_1$ alone; no new interactions, such as the famous $\rho\pi\gamma$ interaction current, needed to  be added.  
However, these arguments placed no constraint on the $F_2$ term.  Clearly it must be included because the free nucleon cannot be described without it, but the choice of whether or not to multiply the $F_2$ term by $f_0$ is not dictated by these principles. 
Similarly, I emphasize that the introduction of $F_4$ is  {\it not required by the principles of simplicity or picture independence\/}.   
To justify the introduction of $F_4$ and to explain the use of the same $f_0$ for both $F_1$ and $F_2$, and the same $g_0$ terms for $F_3$ and $F_4$, a new principle is needed.

The new principle will be referred to as the principle of {\it balance\/} between Dirac and Pauli interactions.  The principle states that whenever a Dirac-like charge term ($F_1$ and $F_3$ in this case) is required, a similar Pauli-like term ($F_2$ and $F_4$) will be included.  This ensures that the off shell current, expressed in terms of the $F_3$ and $F_4$ from factors, could also be expressed in terms of off-shell charge ($G_{E}^{\rm off}$) and magnetic ($G_{M}^{\rm off}$) form factors, without a constraint on the structure of either (except for the previously discussed constraint $F_3(0)=1$).

\subsubsection{Properties of $f_0$ and $g_0$}

The simplest solution to (\ref {eq:210ab})  is
\bea
f_0(p',p)&=&\frac{h'}{h} \left[\frac{m^2-p^2}{p'^2-p^2}\right]+\frac{h}{h'} \left[\frac{m^2-p'^2}{p^2-p'^2}\right]
\nonumber\\
g_0(p',p)&=&\frac{4m^2}{p'^2-p^2}\left[\frac{h}{h'}-\frac{h'}{h}\right] \label{eq:f0g0}
\, .
\eea
where I use the the shorthand notation $h=h(p)$ and $h'=h(p')$.  Both $f_0$ and $g_0$ are symmetric in $p',p$, and important limits are
\bea
f_{00}(p^2)&=&\lim_{p'^2\to p^2} f_0(p',p)=1 + 2a(p^2)(m^2-p^2)\qquad
\nonumber\\
f_{01}(p^2)&=&\lim_{p'^2\to m^2} f_0(p',p)= \lim_{p'^2\to m^2} f_0(p,p')=\frac1{h}\qquad
\nonumber\\
g_{00}(p^2)&=&\lim_{p'^2\to p^2} g_0(p',p)=-8m^2 a(p^2)
\eea
where
\bea
a(p^2)=\frac1{h}\frac{dh}{dp^2}\, .
\eea

\subsection{Definitions of deuteron observables}

Precise definitions of the deuteron form factors will be reviewed in Appendix \ref{app:theory}.  For an understanding of the results to be presented in Sec.~\ref{sec:cal}, it is only important to review that electron-deuteron scattering is described by three independent deuteron form factors \cite{Garcon:2001sz,Gilman:2001yh}: $G_C$ (charge), $G_M$ (magnetic), and $G_Q$ (quadrupole).  Denoted generically by $G_X$ (with $X =\{C,M,Q\}$).  These form factors are a sum of products of {\it isoscalar\/} nucleon form factors, $F_{i}(Q^2)$ (where the subscript $s$ labeling these as isoscalar will be omitted throughout this paper for simplicity), and {\it body\/} form factors, $D_{X,i}(Q^2)$, so that 
\bea
G_X(Q^2)&=&\sum_{i=1}^4 F_{i}(Q^2) D_{X,i}(Q^2) \label{eq:expbody}
\eea
It is important to realize that the theory presented in this paper calculates the body form factors only; the nucleon form factors must be obtained from another source.

The deuteron form factors can be measured by the analysis of three independent experiments.  Two of these can be obtained from  the unpolarized elastic scattering of electrons from the  deuteron.  In one photon exchange approximation, this elastic scattering is given by
\begin{equation}
{ {d\sigma}\over{d\Omega} } = {{d\sigma}\over{d\Omega}} \Bigg|_{NS} 
\Bigl[ A(Q^2) + B(Q^2) \tan^2(\theta/2) 
\Bigr] 
\label{edxsect}
\end{equation}
where 
\begin{equation}
{{d\sigma}\over{d\Omega}} \Bigg|_{NS} = { {\alpha^2 E^{\prime}
\cos^2(\theta/2)} \over
 {4 E^3 \sin^4(\theta/2)} }=\sigma_M {E'\over E}=\frac{\sigma_M} 
{1+{2E\over m_d}\sin^2{\textstyle{1\over2}}\theta}\, ,
\label{mott}
\end{equation}
is the  cross section for scattering from a particle without internal
structure ($\sigma_M$ is the Mott cross section), and $\theta$, $E, E'$, 
and $d\Omega$ are the electron scattering angle, the  incident and final
electron  energies, and the solid angle of the scattered electron, all in
the lab system.
The structure functions $A$ and
$B$, which can be separated by comparing unpolarized measurements in the forward and backward directions, depend on the three electromagnetic form factors 
\begin{eqnarray}
A(Q^2) &&= A(G_C)+A(G_M)+A(G_Q)
\nonumber\\
&&= G_C^2(Q^2) + {{8}\over{9}} \eta^2 G_Q^2(Q^2) + 
 {{2}\over{3}} \eta G_M^2(Q^2)\nonumber\\
B(Q^2) &&= {{4}\over{3}} \eta(1+\eta) G_M^2(Q^2)\, ,
\label{AandB}
\end{eqnarray}
where
\bea
\eta=\frac{Q^2}{4M_d^2}\, . \label{eq:eta}
\eea
To further separate $G_C$ and $G_Q$, the polarization of the outgoing deuteron can be measured in a separate, analyzing scattering.  The quantity most extensively measured is 
\begin{equation}
\widetilde{T}_{20} = - \sqrt{2}\; { {y(2+y)}\over{1 + 2 y^2} }
\label{eq:T20}
\end{equation}
where 
\bea
y = \frac{2 \eta G_Q}{3 G_C}\, .  \label{eq:yT20}
\eea

Note that the structure function $B$ depends only on $G_M$, and $T_{20}$ depends on $y$, both of which are {\it linear\/}  in the the nucleon form factors. However, the structure function $A$ is quadratic in the nucleon form factors. 


\section{Fits to the Deuteron Observables} \label{sec:cal}

\subsection{Introduction}

All results for the deuteron form factors depend on the off-shell nucleon form factors $F_3(Q^2)$ and $F_4(Q^2)$.  However, except for the  sole requirement that $F_3(0)=1$, these form factors are completely unknown, and it is appropriate to use the deuteron form factor data to determine them.  The first step, determining $F_3$ and $F_4$, in done in Sec.~\ref{sec:cal}B below.

I have found that the most efficient way to do this is to use the data from $G_M$ (determined directly from $B$) and $\widetilde T_{20}$ [actually $y$ from Eq.~(\ref{eq:yT20})].  Both $G_M$ and $y$ are linear in $F_3$ and $F_4$, so a solution is straightforward and it is easy to determine the errors in $F_3$ and $F_4$ from the errors in $G_M$ and $y$.  Details are given in Appendix \ref{app:F3F4nextract}.

The data are scattered, and to do this efficiently it would first be necessary find a smooth fit to all the data.  Fortunately, Sick has produced a Global Analysis \cite{Sick:2001rh,Marcucci:2015rca} (referred to as the GA in this paper) where he reanalyzed all of the data  starting from the detailed records.  I will use his GA for a representation of the data.  Once $F_3$ and $F_4$ have been determined, the data (that is, the Sick GA) for $B$ and $\widetilde T_{20}$ is exactly reproduced, as shown in Sec.~\ref{sec:cal}C.  

Note that $\widetilde T_{20}$ determines only the {\it ratio\/} of the independent form factors $G_C$ and $G_Q$, not their size.  The third observable, $A$, can vary even when $B$ and $\widetilde T_{20}$ are fixed.   This is studied in Sec.~\ref{sec:cal}D, where it is shown that $G_{En}$ can be adjusted to give the correct $A$ (fortunately, $B$ and $\widetilde T_{20}$  are very insensitive to $G_{En}$, so that this determination of $G_{En}$ does not alter the fits to $B$ and $\widetilde T_{20}$).  The predictions for $G_{En}$ made by each model  is discussed in Sec.~\ref{sec:cal}D, where it is shown that  model WJC1 fails at this point, but model WJC2 works very well.  Finally, using the predicted $F_3$ and $F_4$ and various models of $G_{En}$, Sec.~\ref{sec:cal}E presents the deuteron form factors and compares them to Sick's GA.

In order to keep the number of figures to a minimum, the reader is warned that some of the early figures will show results for models that will not be introduced until later in the discussion.  To help with this, all of the models that will be used are summarized in Table \ref{tab:models}.  Models 1A and 2A are a starting point; their input is the same as the successful VODG calculation (except I never have any $\rho\pi\gamma$ interaction current).  Model VODG used the GK05 prediction of $G_{En}$, and in the absence of any previous knowledge, assumed a standard dipole for $F_3$ and $F_4=0$.  Then, models 1B and 2B replace $F_3$ and $F_4$ by the solutions found in Sec.~\ref{sec:cal}B, giving precise fits to $B$ and $\widetilde T_{20}$.  Finally, models 2C and 2D show the results of using models for $G_{En}$ based on the predictions given in Sec.~\ref{sec:cal}D, and cannot be understood until that section is studied. 

\begin{table}
\begin{minipage}{3.4in}
\caption{Summary of theoretical models discussed in this paper.  All models listed in the table have {\it no\/} $\rho\pi\gamma$ exchange current, except for  model VODG.  The model GK05 is  discussed in Ref.~\cite{Lomon:2002jx} and shown in Fig.~\ref{fig:GEn}.   }
\label{tab:models}
\begin{ruledtabular}
\begin{tabular}{lcccccc}
Name& Deuteron & $G_{En}$ & $F_3$ & $F_4$ & color & line\\[0.05in]
VODG* & IIB & GK05& dipole &0 & black & long-dash \\[0.05in]
VODG0 & IIB & GK05& dipole &0 & black & 2 dash-2 dot\\[0.05in]
1A & WJC1 & GK05& dipole & 0 & blue & short-dash\\[0.05in]
2A & WJC2 & GK05& dipole & 0 & red & short-dash\\[0.05in]
1B & WJC1 & GK05 & $F_3(1)$ & $F_4(1)$  & blue & 2 dash-2 dot\\[0.05in]
2B & WJC2 & GK05& $F_3(2)$ & $F_4(2)$ &red & 2 dash-2 dot \\[0.05in]
2C & WJC2 & CST2 & $F_3(2)$ & $F_4(2)$ &red & long dash-dot \\[0.05in]
2D & WJC2 & CST1 & $F_3(2)$ & $F_4(2)$ & red & thick solid\\[0.05in]
\end{tabular}
\end{ruledtabular}

\flushleft{*Model  VODG includes  a $\rho\pi\gamma$ exchange current calculated using the Rome 2 $\rho\pi\gamma$ form factor \cite{Cardarelli:1995ap}.}
\end{minipage}
\end{table}
%
%

\begin{figure*}
\centerline{
\mbox{
\includegraphics[width=3.3in]{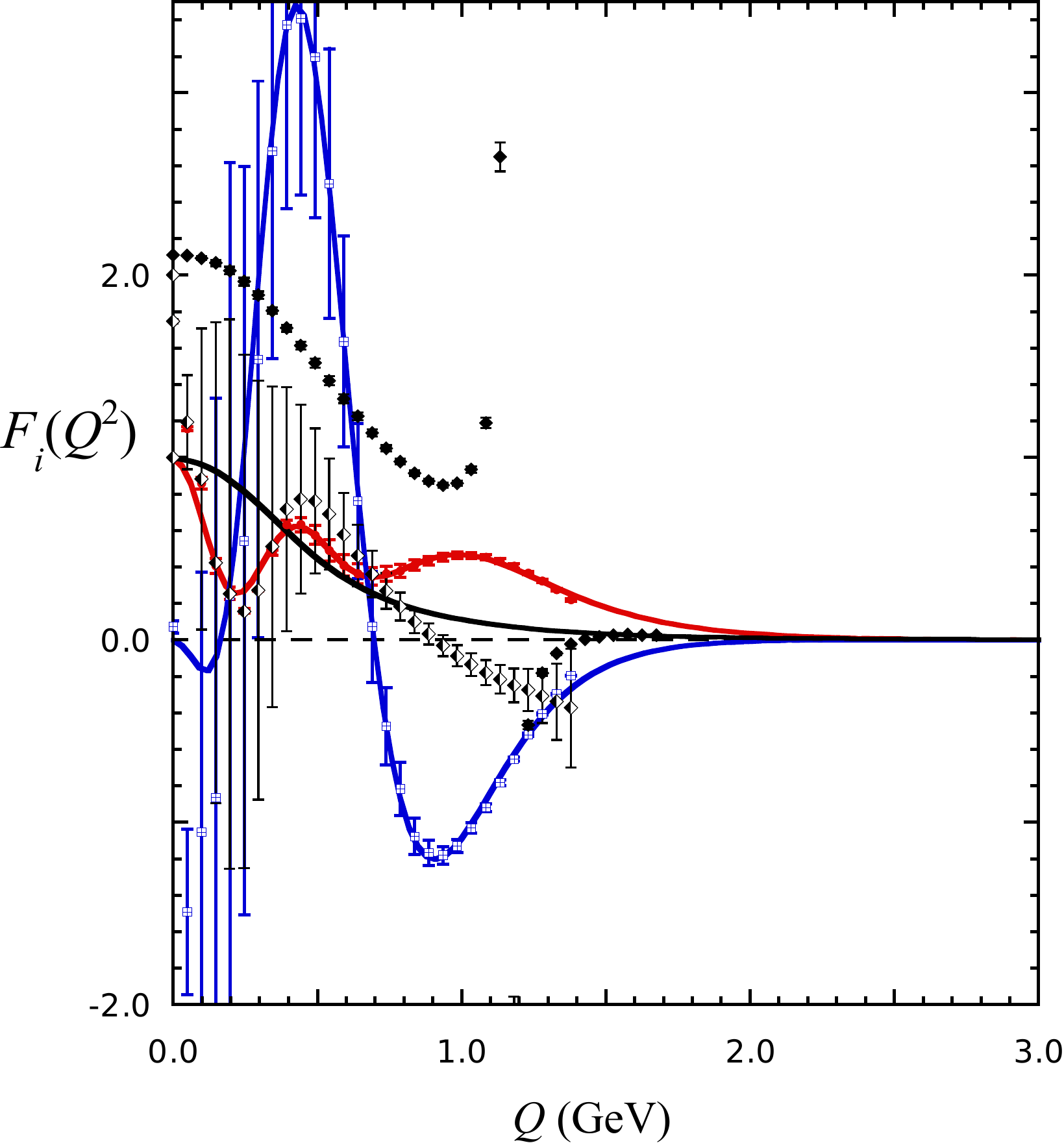} 
\includegraphics[width=3.3in]{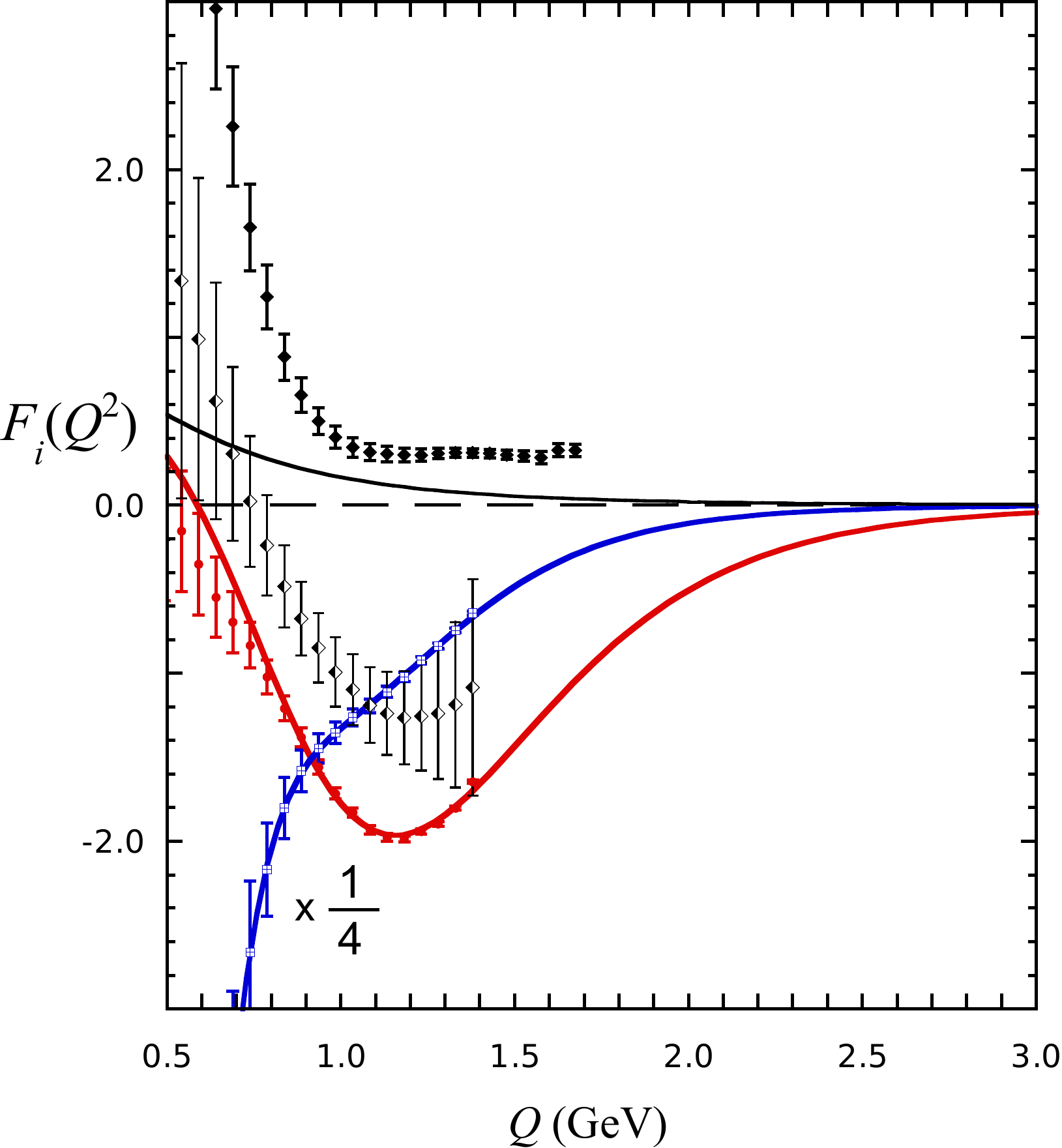} 
}
}
\caption{\footnotesize\baselineskip=10pt Results for Model WJC1 (left panel) and Model WJC2 (right panel).  Both panels show $F_3(Q^2)$ (small red circles) and $F_4(Q^2)$ (small blue squares) obtained by simultaneously fitting  to Sick's GA for $G_M$ and $T_{20}$.  Only the error bars obtained from the errors in Sick's $G_M$ are shown, reflecting the fact that the requirements to fit $G_M$ are far more stringent than those necessary to fit $T_{20}$. (Note that $F_4$ for Model WJC2 is four times larger than shown in the figure.)  The panels also show the results for $F_3$ fitted to $G_M$ (solid black diamonds), or to $T_{20}$ (half filled black diamonds) when $F_4=0$.  The smooth black curve is the dipole model and the red and blue curves are the fits discussed in the text.
}
\label{fig:F3-F4}
\end{figure*} 

\begin{figure}
\centerline{
\mbox{
\includegraphics[width=3.3in]{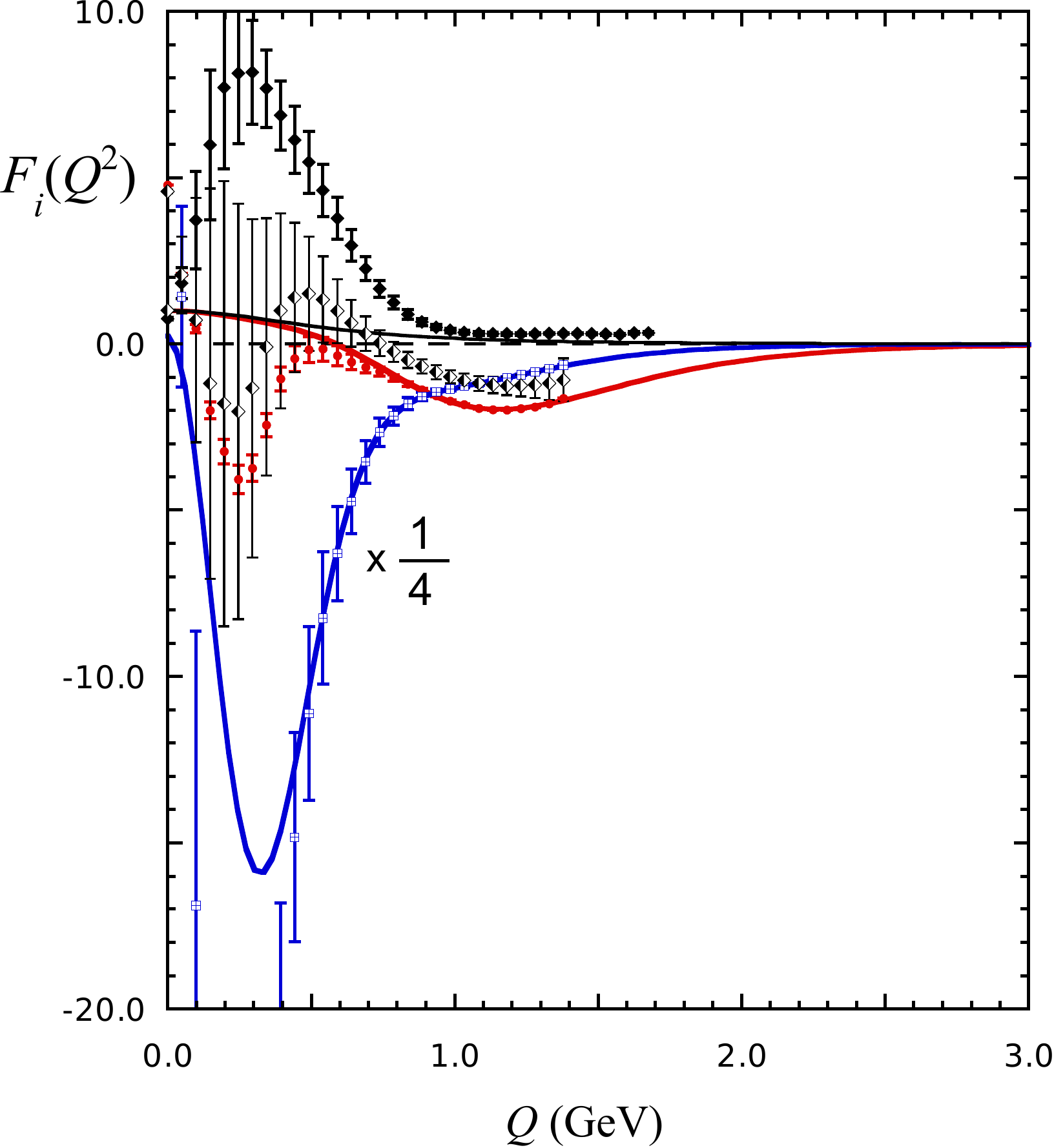}
}
}
\caption{\footnotesize\baselineskip=10pt Same results as shown in the right panel of Fig.~\ref{fig:F3-F4}, but on a bigger scale which allows a fuller picture of the form factors at smaller $Q$.  (Note that $F_4$ is four times larger than shown in the figure.)
}
\label{fig:F4}
\end{figure} 

\subsection{Predictions for the off-shell nucleon form factors} \label{sec:4}

 As mentioned in the Introduction, the off-shell form factors can be found by simultaneously fitting them to the GA data points for $G_M$ and $T_{20}$ (which is independent of $G_M$).  Each GA point has its own error that I use to estimate the errors in the fitted values of the form factors.   The results obtained from Models WJC1 and WJC2 are shown in Figs.~\ref{fig:F3-F4} and \ref{fig:F4}.   Each red and blue point in the figure is the (simultaneous) solution for $F_3$ and $F_4$ at each GA point, which extend out to $Q$=  7 (fm)$^{-1}$ = 1.379 (GeV) (limited my the measurements of $T_{20}$). 

On the same figures I also show the values obtained by fitting $F_3$ separately to $G_M$ (solid black diamonds) or $T_{20}$ (half filled black diamonds) {\it under the assumption\/} that $F_4=0$.  The fact that these fits differer substantially shows that it is not possible to obtain a good fit to the GA data without including a nonzero $F_4$.

\begin{table}[b]
\begin{minipage}{3in}
\caption{Parameters for the fits to $F_3$ and $F_4$ with $Q$ in GeV.  Here $F_i$(X) $\equiv F_i$(WJCX).}
\label{tab:F3models}
\begin{ruledtabular}
\begin{tabular}{crrrr}
   & $F_3$(1) & $F_3$(2)  & $F_4$(1) & $F_4$(2)\\[0.05in]
$a$ & $-$30.905  & 1.3508  & $-$29.467 & $-$1747.8\\[0.05in]
$b$ & 457.66  & 4.0568  & 1141.0&  2395.0\\[0.05in]
$c$ & $-$1401.7 & 0 & $-$2422.0&0 \\[0.05in]
$d$ & 1618.9 &$-$137.69  &404.21 &$-$3370.4 \\[0.05in]
$e$ & 1.2323 & 0.6131 &1.1115 &1.0004 \\[0.05in]
\end{tabular}
\end{ruledtabular}
\end{minipage}
\end{table}
%

\begin{figure}
\centerline{
\mbox{
\includegraphics[width=3.4in]{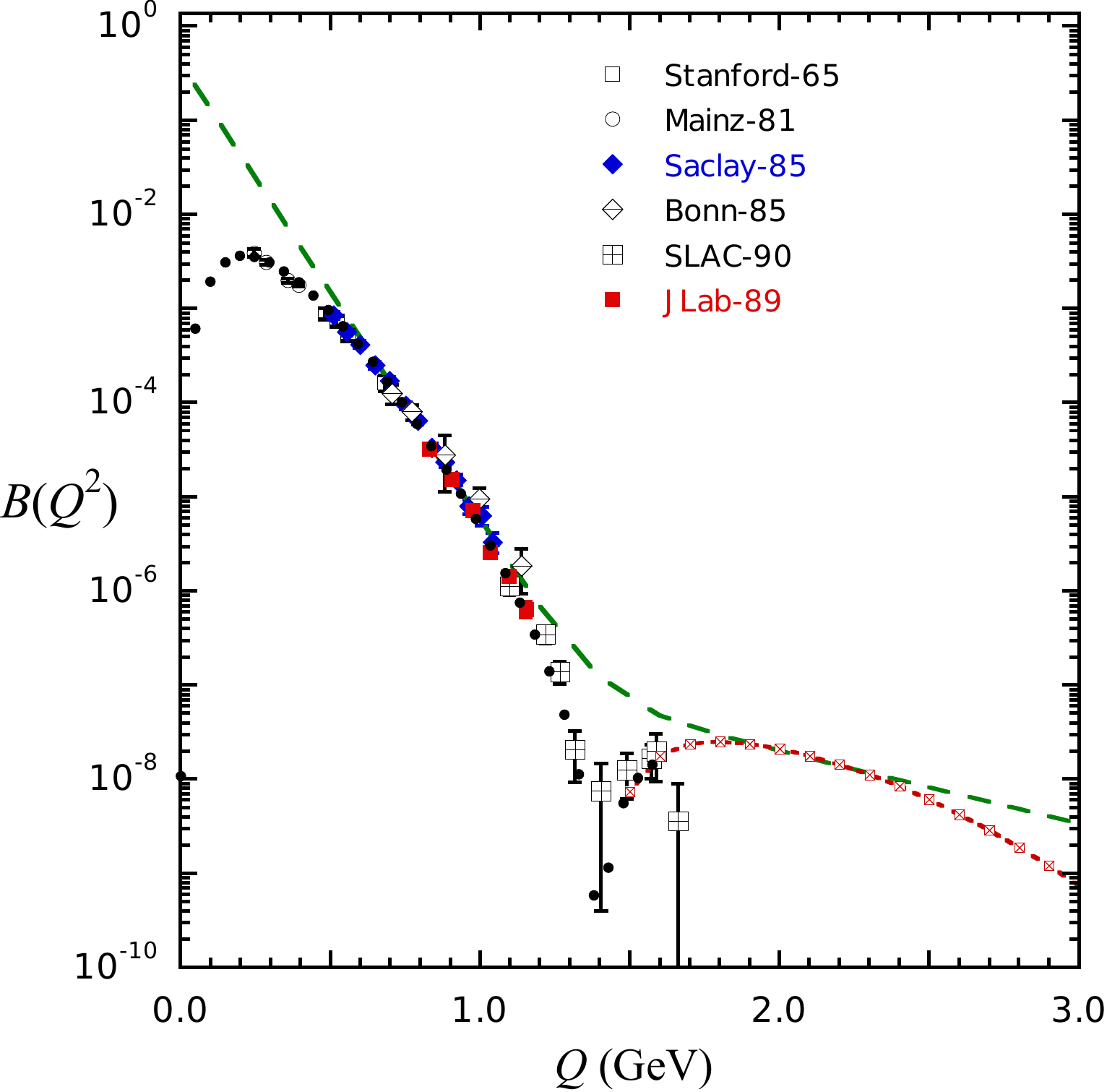}
}
}
\caption{\footnotesize\baselineskip=10pt Data for the magnetic structure function $B(Q^2)$ compared to Sick's GA (small solid black circles).  The function fit{\it B\/} (dashed green line) is shown for comparison.  The data are Stanford-65 \cite{Buchanan:1965zz}, Mainz-81 \cite{Simon:1981br}, Saclay-85 \cite{Auffret:1985tg},  Bonn-85 \cite{Cramer:1986kv}, SLAC-90 \cite{Arnold:1986jda,Bosted:1989hy}, and JLab-89 \cite{sulthesis,Petratos:2000rq}.  The high Q tail of  Model 2B (c.f.~Table \ref{tab:models}), used to construct the high Q tail of fit{\it B\/}, is the red dashed line connected to red squares. 
}
\label{fig:BallQ}
\end{figure} 

\begin{figure*}
\centerline{
\mbox{
\includegraphics[width=3.3in]{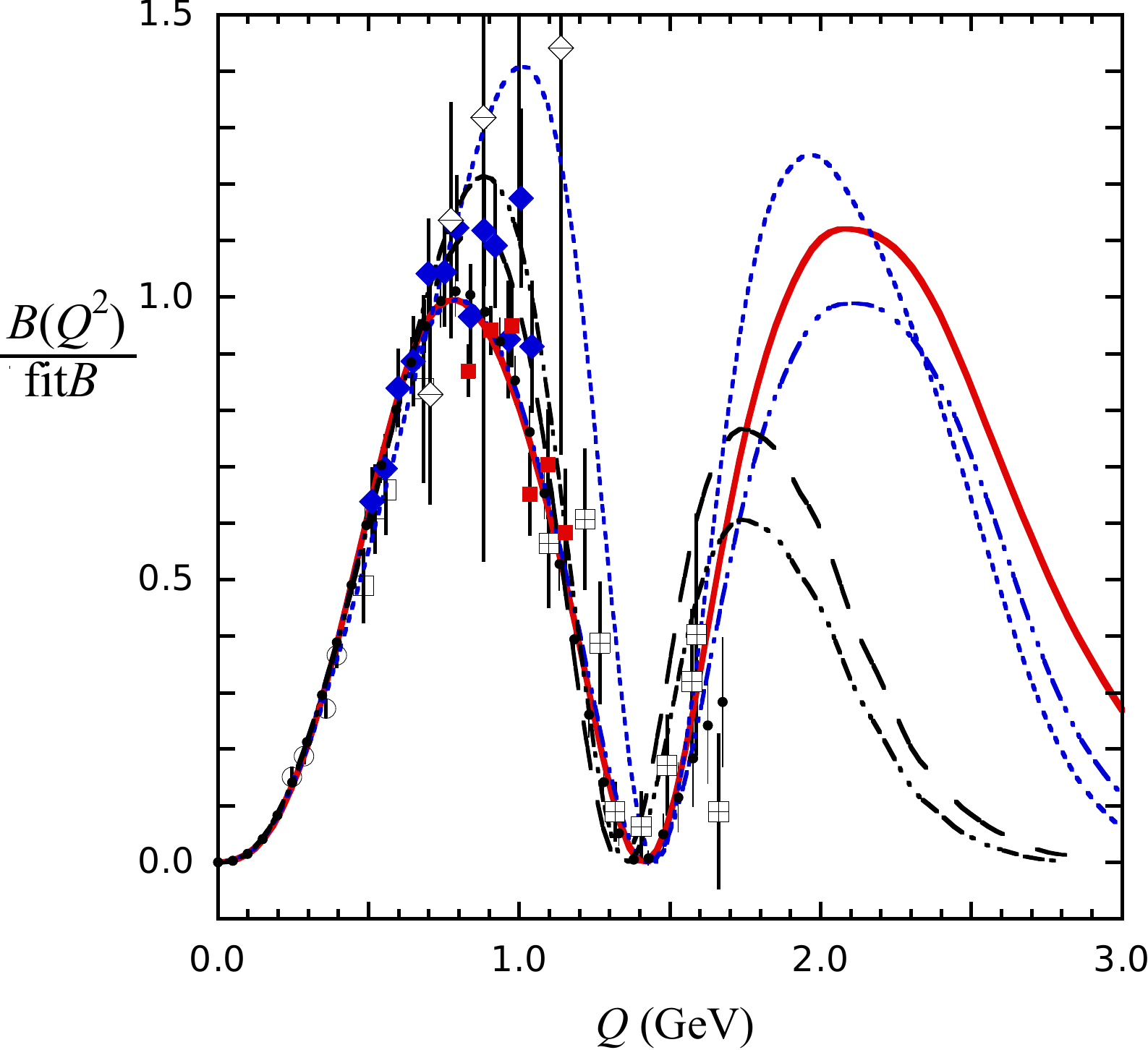} 
\includegraphics[width=3.3in]{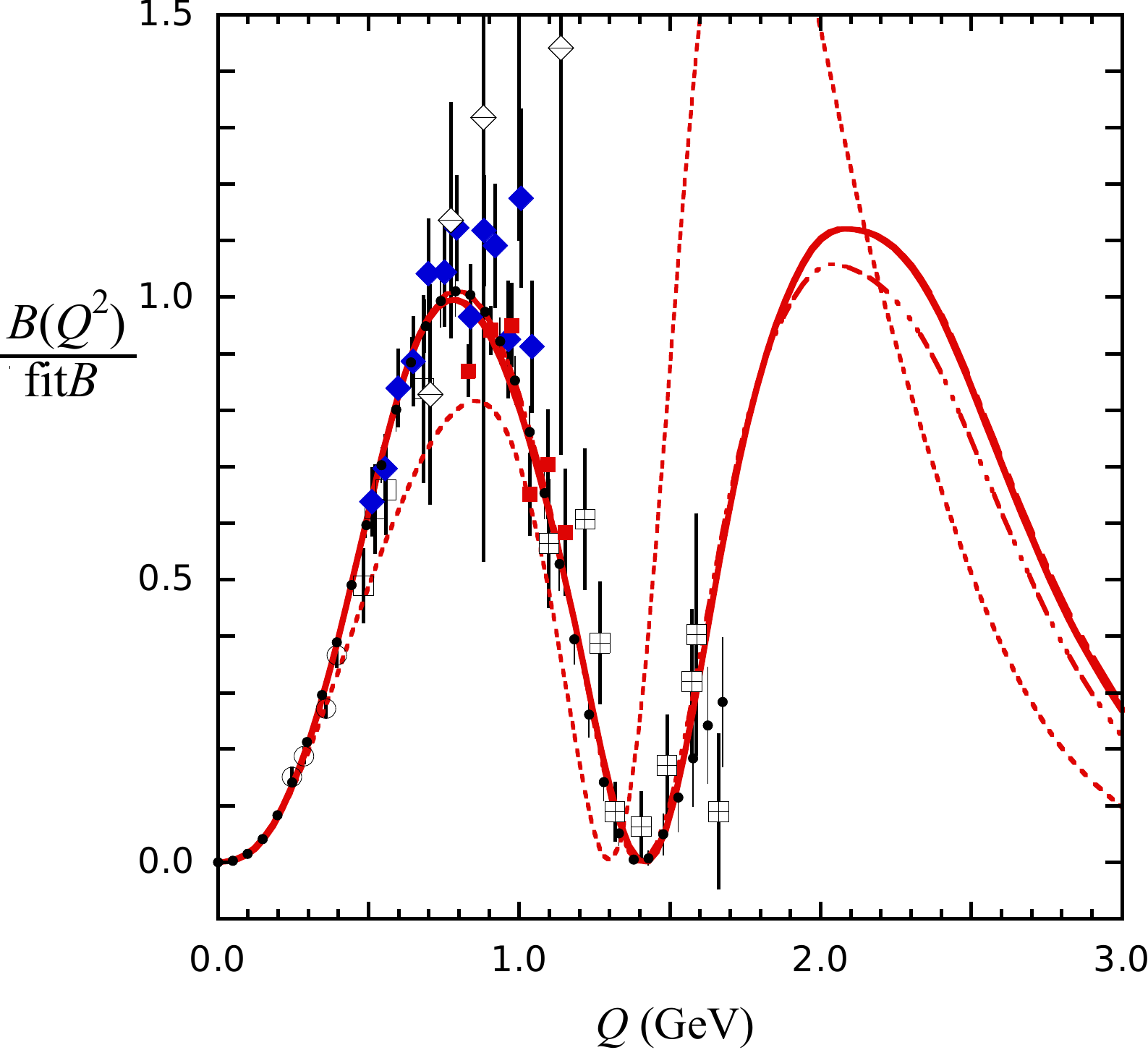} 
}
}
\caption{\footnotesize\baselineskip=10pt The same data for the magnetic structure function $B(Q^2)$ shown in Fig.~\ref{fig:BallQ}, Sick's GA, and various theoretical predictions all scaled by the function fit{\it B\/}, Eq.~(\ref{eq:fitb}).  
Both panels show the data, the GA, and Model 2D (thick red solid line).  The left panel also shows Models 1A (blue short-dashed line),1B (blue double dashed-dotted line),  VODG (black long dashed line) and VODG0 (black double dashed-dotted line).   The  right panel shows Model 2A (red short-dashed line), 2B (red double dashed-dotted line), 2C (red long dashed-dotted line) and 2D (thick red solid line).  Models 2C and 2D are nearly indistinguishable (see some differences in the other plots).
}
\label{fig:BallQ-scaled}
\end{figure*} 

Note that the form factors are largely undetermined at $Q \gtrsim1.4$ GeV, and also at small $Q$ where the errors in the fitted form factors are large.  In order to have results for all $Q$, and especially  beyond the range where data for $T_{20}$ exists, I chose smooth curves that fit the points in the range 0.5 $\gtrsim Q \gtrsim 1.3$ GeV, where they are well constrained.  The generic models used for $F_3$ and $F_4$ are
\bea
F_3(Q^2)&=&\frac{1+aQ^2+bQ^4+cQ^6+dQ^8}{(1+eQ^2)^n}  
\nonumber\\
F_4(Q^2)&=&\frac{aQ^2+bQ^4+cQ^6+dQ^8}{(1+eQ^2)^9}
\eea
where $n=7$ for WJC1, $n=9$ for WJC2, and the other parameters are given in Table \ref{tab:F3models}.  The asymptotic limits of these form factors are 
\bea
\lim_{Q^2\to\infty}F_3&=&\frac{\rm const}{Q^{2n-8}} \sim \begin{cases} Q^{-6} & {\rm WJC1} \cr
Q^{-10} & {\rm WJC2} \end{cases}
\nonumber\\
\lim_{Q^2\to\infty}F_4&=&\frac{\rm const}{Q^{10}}\, ,
\eea
and note that I have constrained 
\bea
\lim_{Q\to0}F_4 = aQ^2\, .
\eea 
Finally, I point out that the models for those form factors are real analytic functions with cuts in the complex $q^2$ plane along the positive real axis.  For the model WJC2, these cuts start at 
\bea
q^2=-Q^2 = 1/e = \begin{cases} 1.63 \;{\rm GeV}^2 & F_3(2) \cr  1.00\;{\rm GeV}^2 & F_4(2)\, . \cr \end{cases}
\eea
Both cuts start near or above the the 7$m_\pi$ threshold, confirming that they are short range effects.  Similar results hold for $F_i(1)$.  I have not investigated the dispersion relations that these functions satisfy.

\begin{figure*}
\centerline{
\mbox{
\includegraphics[width=3.3in]{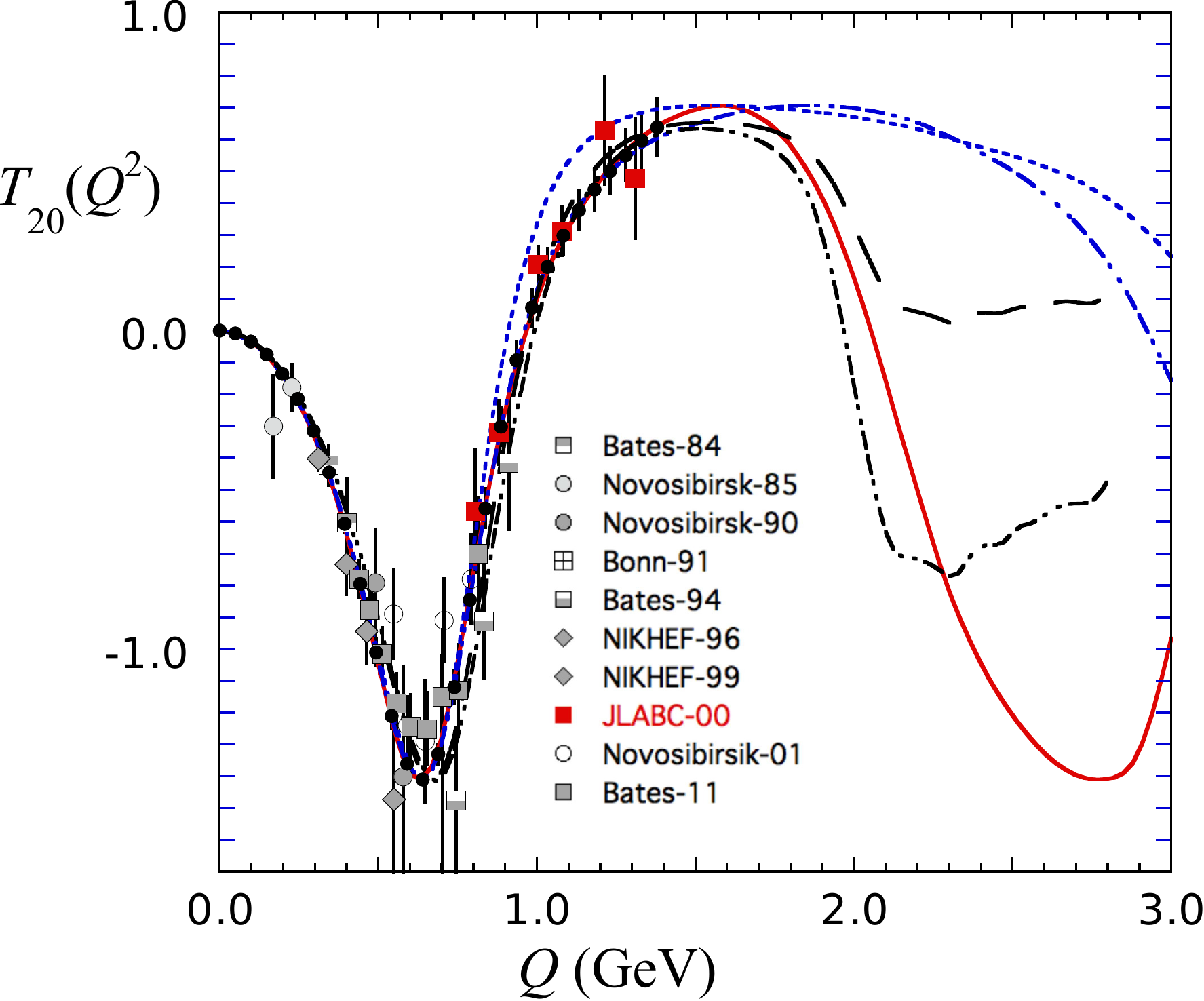} 
\includegraphics[width=3.3in]{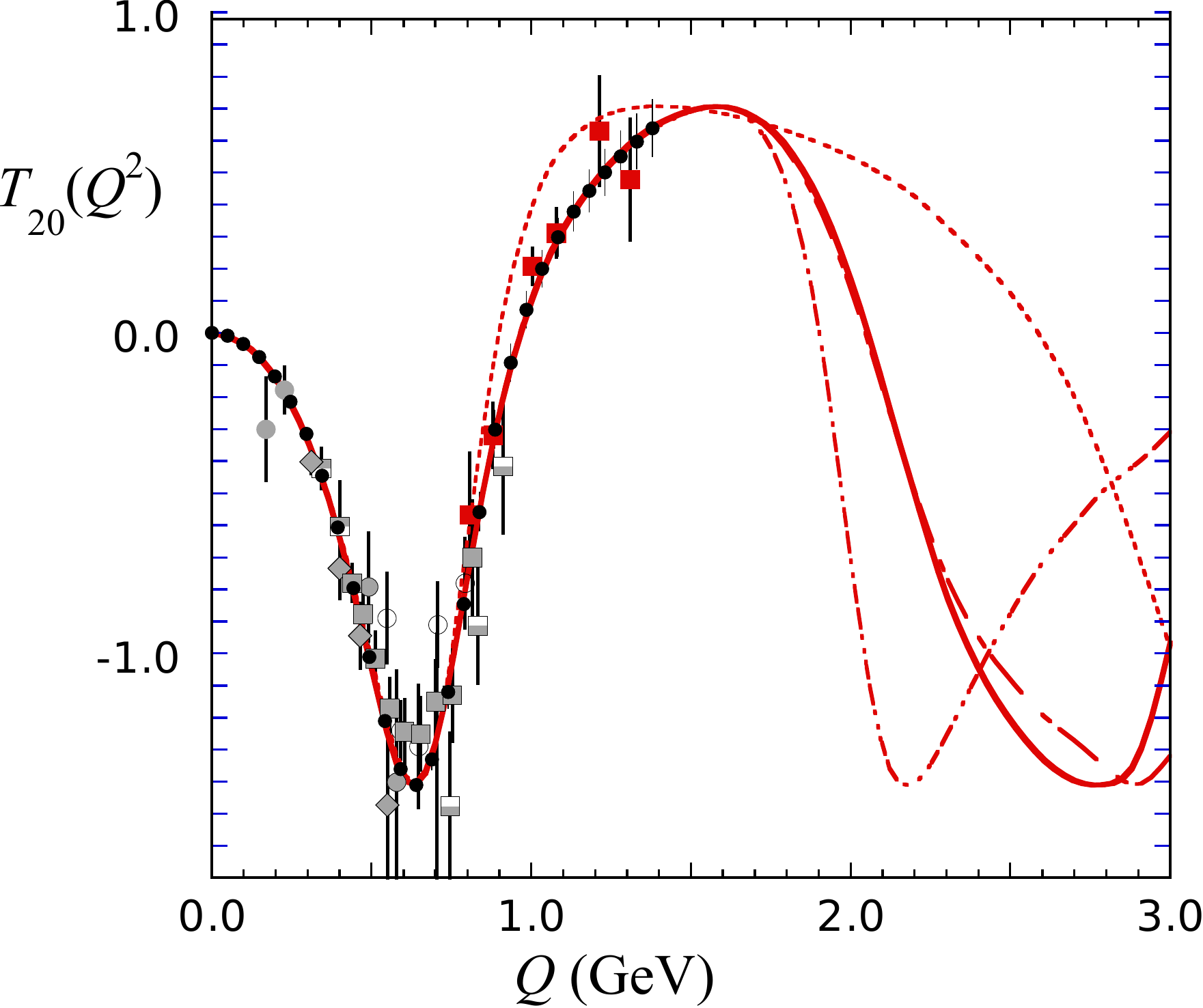} 
}
}
\caption{\footnotesize\baselineskip=10pt The data for $T_{20}$ compared to Sick's GA (black dots) and the same theoretical models labeled as they were in Fig.~\ref{fig:BallQ-scaled} and Table \ref{tab:models}.  The data are Bates-84 \cite{Schulze:1984ms}, Novosibirisk-85 \cite{Dmitriev:1985us,Wojt96},  Novosibirisk-90 \cite{Gilman:1990vg}, Bonn-91 \cite{Boden:1990una}, Bates-94 \cite{The:1991eg,Garcon:1993vm}, NIKHEF-96 \cite{FerroLuzzi:1996dg}, NIKHEF-99 \cite{Bouwhuis:1998jj}, JLabC-00 \cite{Abbott:2000fg}, Novosibirisk-01\cite{Nikolenko:2001zu}, and Bates-11 \cite{Zhang:2011zu}.
}
\label{fig:T20}
\end{figure*} 

\subsection{Fits to $B(Q^2)$ and $T_{20}(Q^2)$}

With the off-shell form factors determined, I now confirm that the fits to $B(Q^2)$ and $T_{20}(Q^2)$ do indeed agree with the Sick GA.  (The fits to $G_M$, related to $B$, will be shown later when the other form factors are discussed.)  This is also an opportunity to compare the results for Models WJC1 and WJC2 with the previous successful calculation of Van Orden \etal \cite{VanOrden:1995eg}, which is refered to as VODG.  The various models under discussion in this and the following sections are defined in Table \ref{tab:models}, and will be referred to by the simple names given in the table.



I begin by showing the data for $B(Q^2)$ in Fig. \ref{fig:BallQ}.
The rapid variation of $B$ with $Q$ makes it difficult to see how the theory compares with data, so I have scaled everything by the simple fit function
\bea
{\rm fit}B=&&0.4 \exp(-2.2\, Q/0.197) 
\nonumber\\
&&+ 0.7 \times 10^{-6}\exp(-0.35\, Q/0.197)  \label{eq:fitb}
\eea 
where $Q$ is measured in GeV, and the tail was adjusted to be near an expected secondary maximum in $B$. The results of dividing both data and predictions by this function are shown in Fig.~\ref{fig:BallQ-scaled}.  This figure also shows how the various theoretical models shown in Table \ref{tab:models} compare with the experimental data and the Sick GA.  Fig.~\ref{fig:T20} shows how the models and Sick GA compare with the experimental data for $T_{20}$.

Study of the curves in Figs.~\ref{fig:BallQ-scaled} and \ref{fig:T20}  show that models 1A and 2A, with the same assumptions as VODG (standard dipole for $F_3$, 
$F_4=0$, and GK05 nucleon form factors) are both successful at low $Q\lesssim 1$ GeV, but 1A seriously overshoots B at higher $Q$ and 2A undershoots $B$ already at about $Q\simeq0.5$.
Model VODG overshoots $B$ a little near $Q\sim 1$ GeV (this will be more clearly displayed when we show $G_M$ below), but the discrepancy is smaller than either Models 1A or 2A.  VODG gives a better explanation than either models 1A and 2A.


As expected models 1B and 2B, that use the appropriate $F_3$ and $F_4$ for each model, do indeed give excellent agreement with the $B(Q^2)$ structure function and $T_{20}(Q^2)$ over the entire region where the GA exists.   
Note that the size of the $\rho\pi\gamma$ exchange current  used by VODG can be inferred from the differences between VODG0 and VODG and is smaller than the effects arising from $F_3$ and $F_4$, particularly for model WJC2.

Finally, the figures show the important result that the best models, 2C and 2D that have not yet been introduced, are practically indistinguishable from 2B in the region of the GA fit.  Their significance will be discussed in the next section.


\begin{figure}
\centerline{
\mbox{
\includegraphics[width=3.6in]{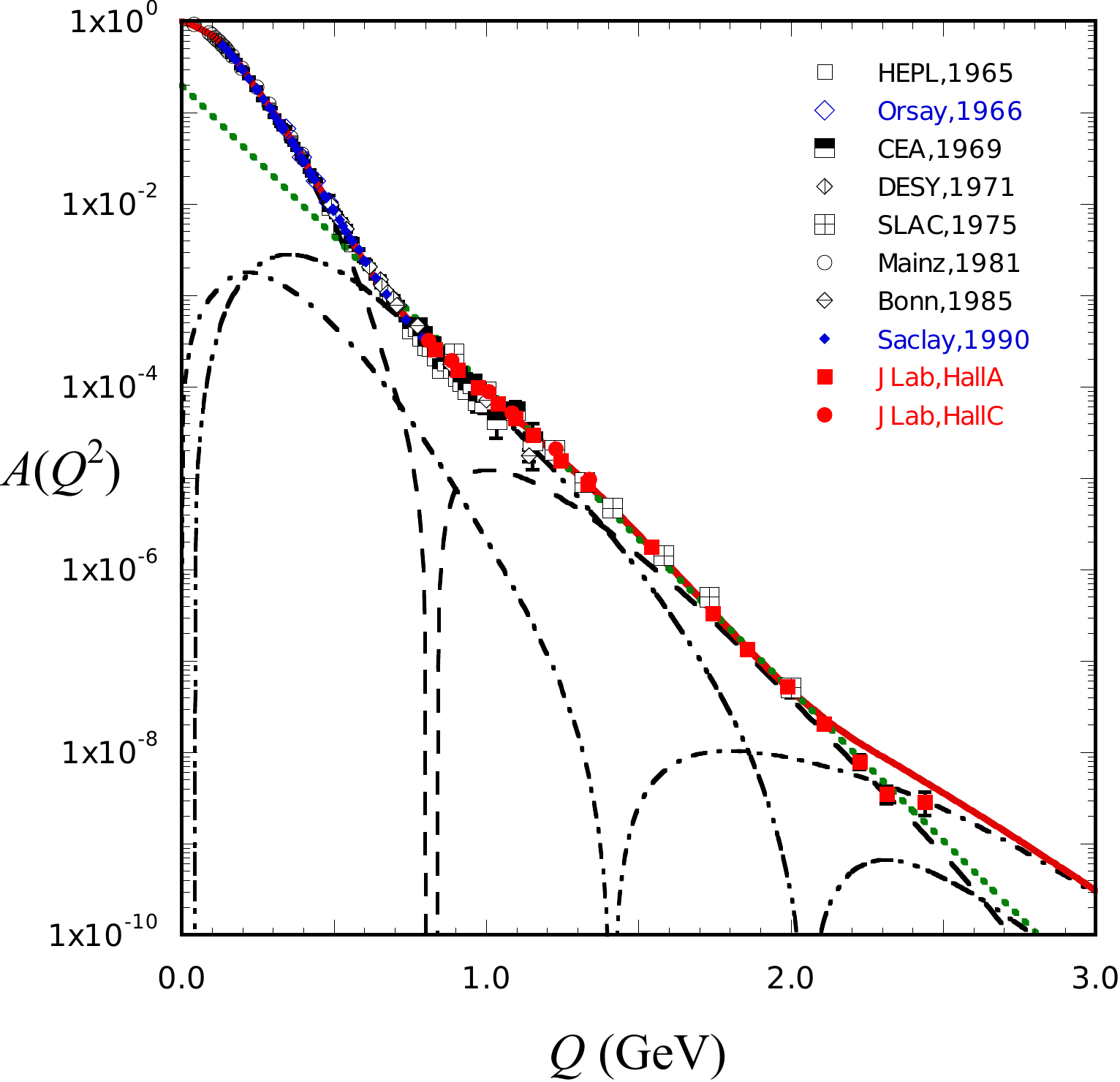}
}
}
\caption{\footnotesize\baselineskip=10pt Data for the structure function $A(Q^2)$.  The function fit{\it A\/} (linear green dotted line) is shown for comparison.  The theoretical curves are model 2D (red line and three black lines) as discussed in the text. The data are HEPL-65 \cite{Buchanan:1965zz}, Orsay-66 \cite{Benaksas:1966zz}, CEA-69 \cite{Elias:1969mi},  DESY-71 \cite{Galster:1971kv}, SLAC-75 \cite{Arnold:1975dd}, Mainz-81 \cite{Simon:1981br}, Bonn-85 \cite{Cramer:1986kv}, Saclay-90 \cite{Platchkov:1989ch}, JLabA-99 \cite{Alexa:1998fe}, and JLabC-99 \cite{Abbott:1998sp}.
}
\label{fig:AallQ}
\end{figure} 

\begin{figure*}
\centerline{
\mbox{
\includegraphics[width=3.4in]{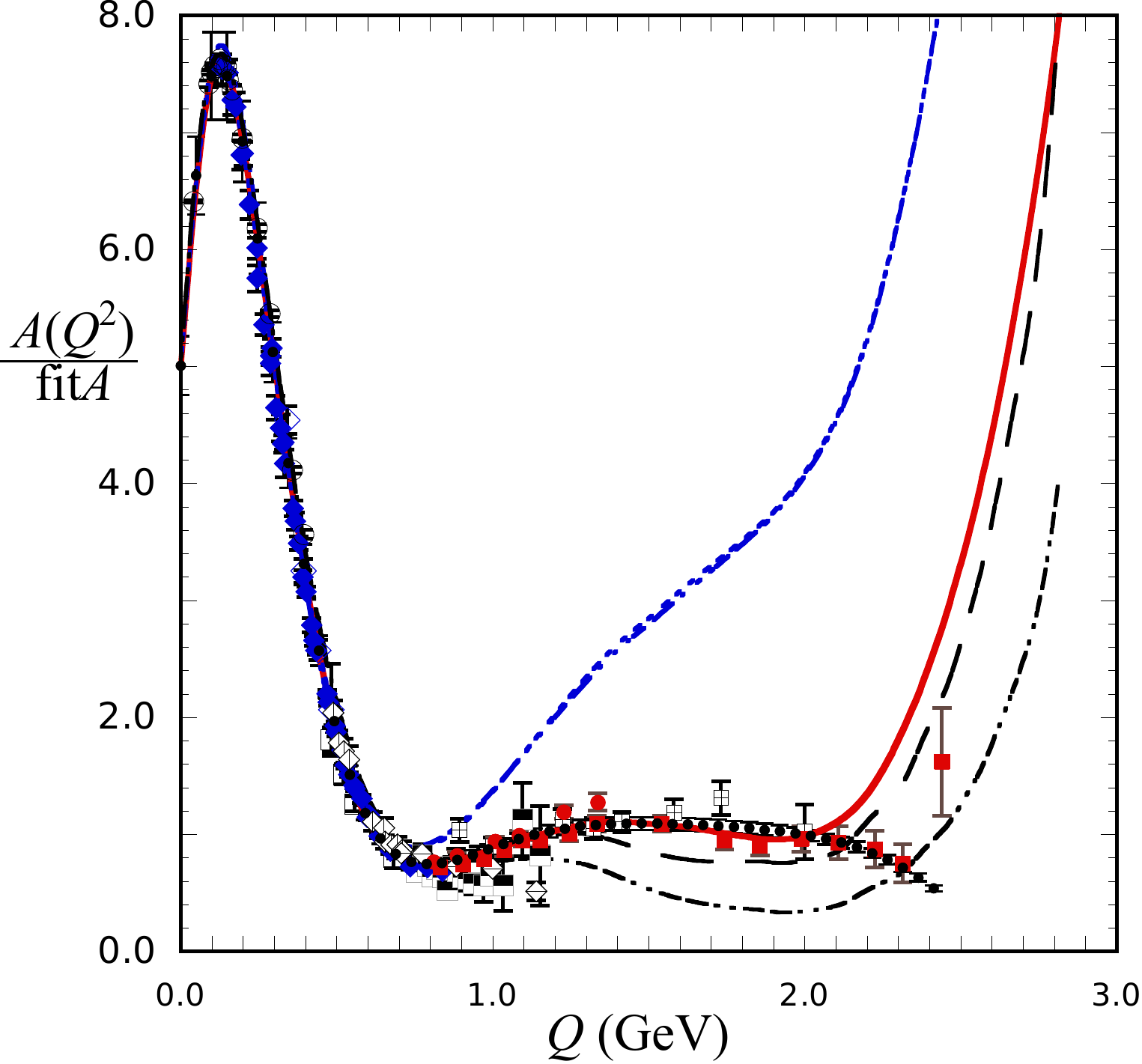} 
\includegraphics[width=3.4in]{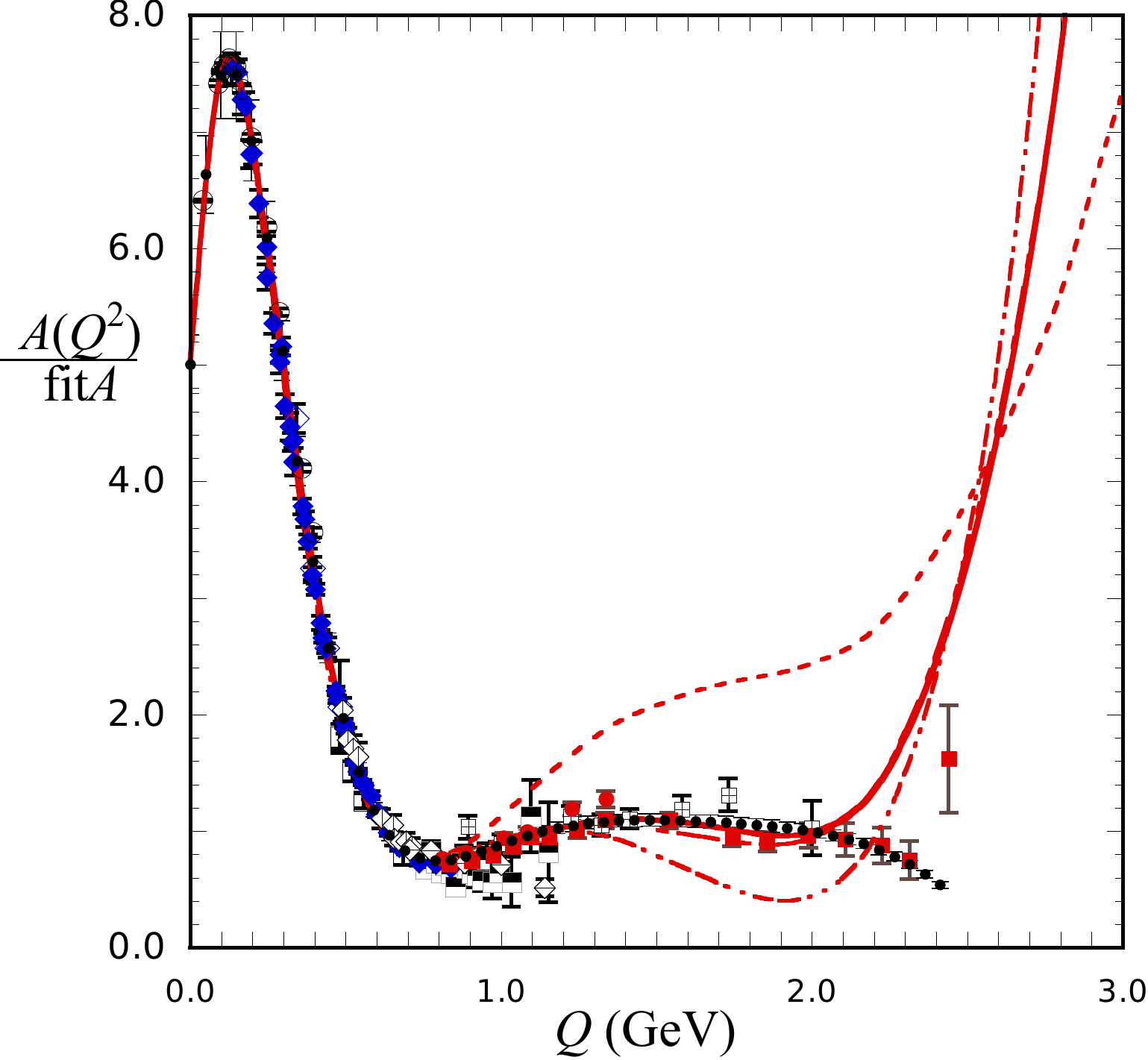} 
}
}
\caption{\footnotesize\baselineskip=10pt The same data for the structure function $A(Q^2)$ shown in Fig.~\ref{fig:AallQ}, Sick's GA (small black solid circles), and various theoretical predictions all scaled by the function fit{\it A\/}, Eq.~(\ref{eq:fitA}) (and all drawn with the same line style used in Figs.~\ref{fig:BallQ-scaled}, \ref{fig:T20} and Table \ref{tab:models}).  
Both panels show Model 2D.  The left panel shows Models 1A and 1B, almost indistinguishable from each other, and VODG and VODG0.   The  right panel shows Model 2A , 2B, 2C and 2D . }
\label{fig:AallQ-scaled}
\end{figure*} 

\subsection{Predictions for $A(Q^2)$ and $G_{En}(Q^2)$}  \label{sec:GEn}

\begin{figure}
\centerline{
\mbox{
\includegraphics[width=3in]{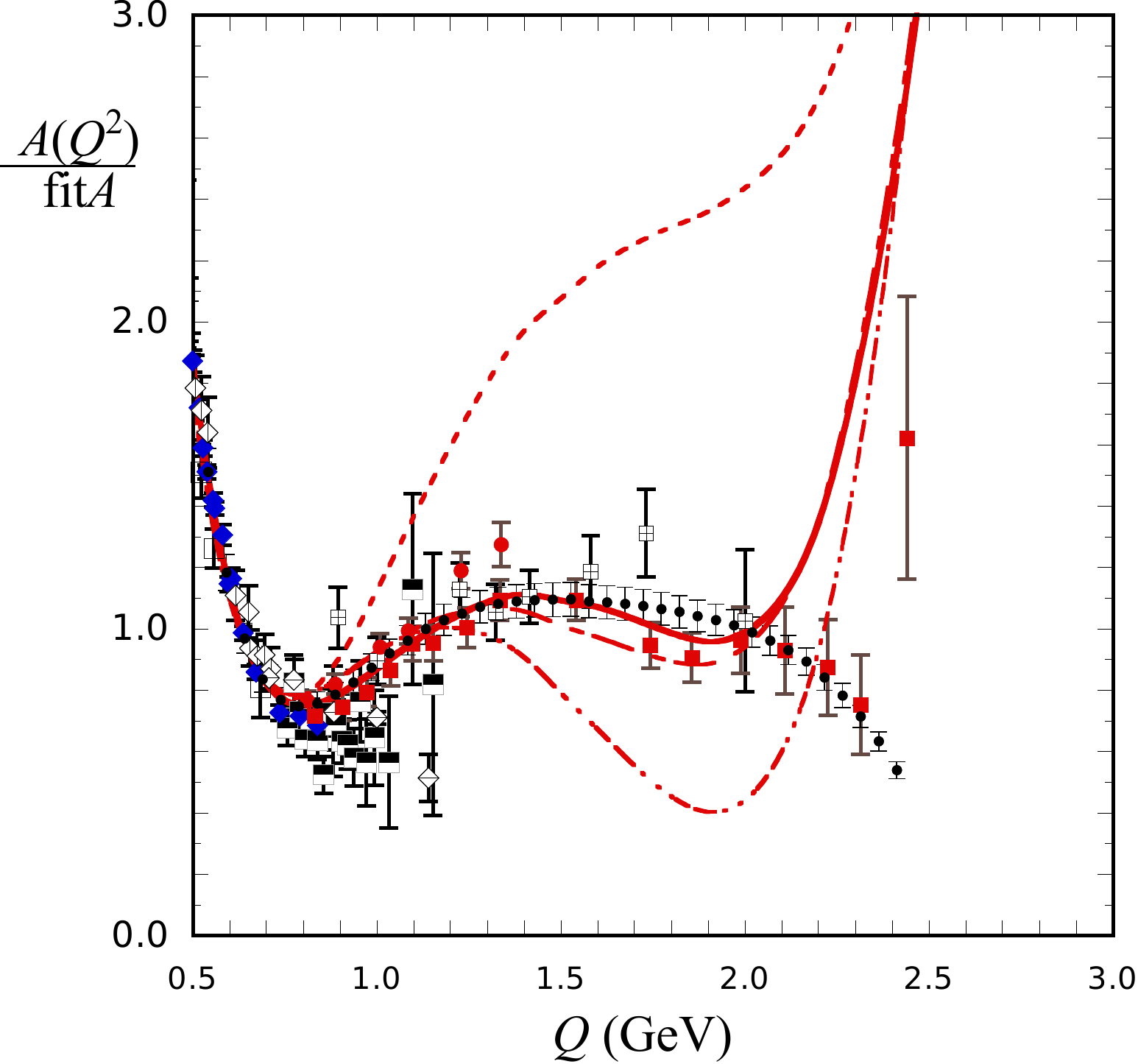} 
}
}
\caption{\footnotesize\baselineskip=10pt The right panel of Fig.~\ref{fig:AallQ-scaled} with expanded scales.
}
\label{fig:AhighQ-scaled}
\end{figure}

To complete the picture, Fig.~\ref{fig:AallQ} shows the data and predictions for $A(Q^2)$ similar to those shown for $B(Q^2)$ in Fig.~\ref{fig:BallQ}.   This figure shows nicely how $A(Q^2)$ falls as an exponential over many decades.  As was the case or $B$, comparing theory to data on such a curve obscures all but huge differences.  To see differences of a factor of 2 or 3, the $A$ structure function is scaled by the simple function
\bea
{\rm fit}A=0.2 \exp(-1.5\, Q/0.197) \label{eq:fitA}
\eea
(where Q is measured in GeV), and Figs.~\ref{fig:AallQ-scaled} and \ref{fig:AhighQ-scaled} show these scaled results, which play a role in our discussion of $A$ similar to that played by Fig.~\ref{fig:BallQ-scaled} in our discussion of $B$.  To emphasize the differences at large $Q$, Fig.~\ref{fig:AhighQ-scaled} is the same as Fig.~\ref{fig:AallQ-scaled}, but with the scales expanded.  

Figs.~\ref{fig:AallQ-scaled} and \ref{fig:AhighQ-scaled} show that
all theoretical models give an excellent description of $A$ at $Q\lesssim 0.7$ GeV. 
However (excluding models 2C and 2D for now) none of the models do very well describing the GA significantly above $Q\sim 0.7$ GeV.  VODG does the best (with the $\rho\pi\gamma$ exchange current playing a decisive role), model 2B is not far off, but  models 1A and B, and 2A all depart substantially from the GA and are clearly unacceptable.  This means that even when the two unknown off-shell form factors $F_3$ and $F_4$ are adjusted to fit $B$ and $T_{20}$, Model WJC1 disagrees with the data for $A$ by such a large amount that it cannot be repaired, as discussed in the next section.

Note that model 2B does well out to $Q\sim 1.4$ GeV, but dips below the data in the region from $1.5\lesssim Q \lesssim 2.2$ GeV.  This is a region where the nucleon charge form factor, $G_{En}$ is unknown, and hence this calculation can be used to {\it predict\/}  $G_{En}$ in this region.

Models 2C and 2D will be discussed below,  
and the failure of any of the models to describe $A$ at the highest values of $Q$  will be discussed in the Conclusions.



\begin{figure}
\centerline{
\mbox{
\includegraphics[width=3.6in]{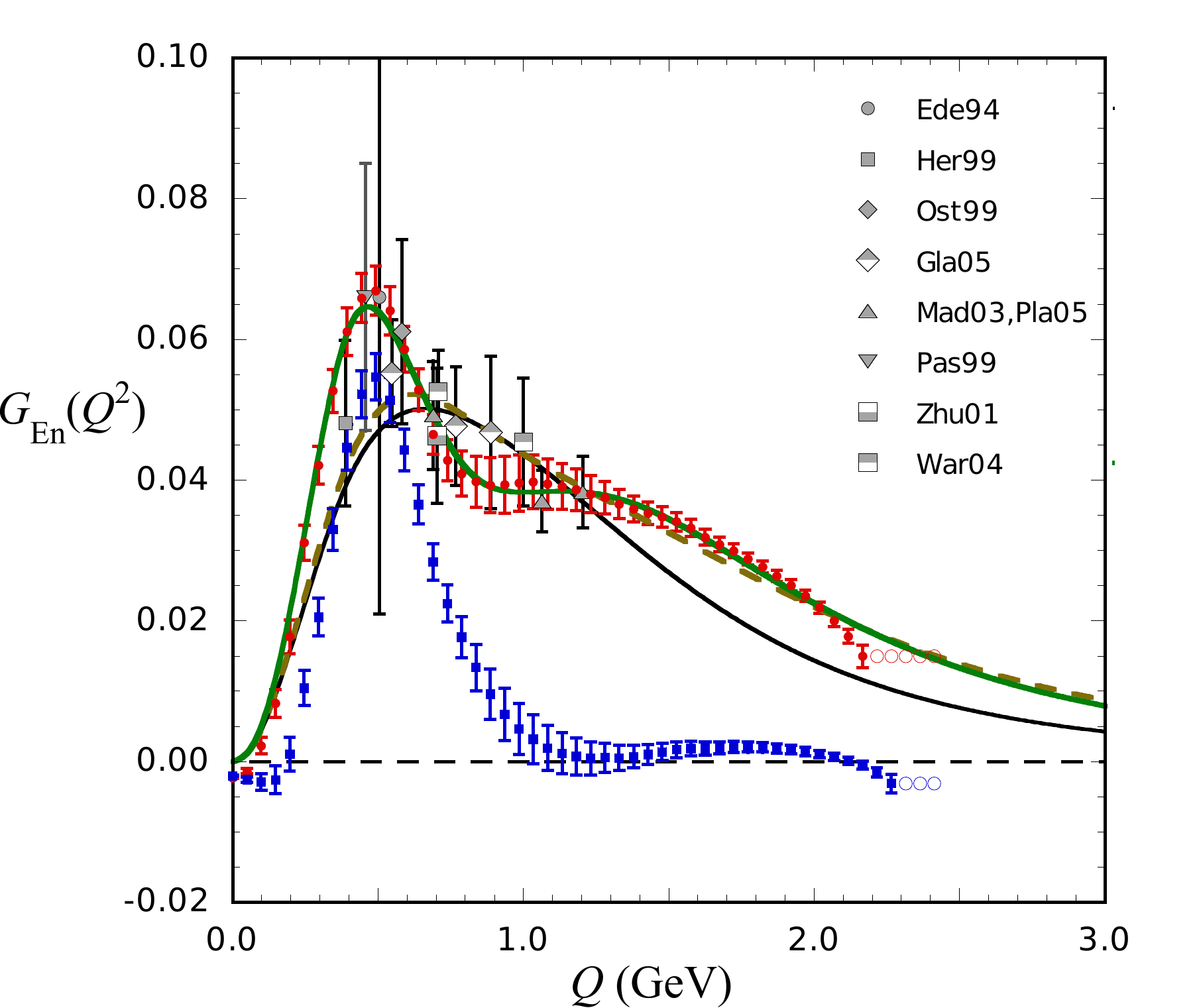}
}
}
\caption{\footnotesize\baselineskip=10pt Experimental data for $G_{En}$ and  "data," with errors, for $G_{En}$ determined by fitting to Sick's GA points for $A$.  The errors were obtained from the errors in $A$ quoted by Sick.  
Solutions WJC1 are the blue points with smaller values of $G_{En}$;  those for WJC2 are the red points.   The small open circles at the highest momenta are points at which there is {\it no\/} solution for $G_{En}$.  The models shown are GK05 (solid black line), CST1 (solid green line with the bump at at small $Q$), and  CST2 (brown dashed line following GK05 at small $Q$ and CST1 at larger $Q$).   The experimental data are Ede94 \cite{Eden:1994ji}, Her99 \cite{Herberg:1999ud}, Ost99 \cite{Ostrick:1999xa}, Gla05 \cite{Glazier:2004ny}, Mad03 \cite{Madey:2003av}, Pla05 \cite{Plaster:2005cx}, Pas99 \cite{Passchier:1999cj}, Zhu01 \cite{Zhu:2001md}, and War04 \cite{Warren:2003ma}.     
}
\label{fig:GEn}
\end{figure} 

The values of $G_{En}$ required to bring each model into agreement with the GA points for $A(Q^2)$  are shown in Fig.~\ref{fig:GEn}.  The GA fits to $A(Q^2)$, shown Figs.~\ref{fig:AallQ-scaled} and \ref{fig:AhighQ-scaled}, extend out to $Q\gtrsim1.576$ (GeV) = 8 (fm)$^{-1}$, well beyond the region where $B(Q^2)$ is known.   However, since $A$ is quadratic in $G_{En}$ (but only one root is acceptable; see the discussion in Appendix \ref{app:GEnextract}),  there is no guarantee that a real solution can be found at each point.  It is remarkable that real solutions do exist except at the highest values of $Q$.  I found that there were {\it no solutions\/} for Model WJC1  at the  3 highest GA points  ($Q \geq 2.319$ GeV), and for WJC2 at the 5 highest GA points ($Q\geq2.216$ GeV).  The errors shown are determined by the GA errors in $A$ only; at high $Q$ the theory depends on the {\it extrapolations\/} obtained from the fits $F_3(i)$ and $F_4(i)$ (where $i=1,2$), and hence are subject to additional errors I have not tried to estimate.   

Here we have a very different situation from our previous study of $F_3$ and $F_4$.   Measurements of $G_{En}$ from {\it free\/} neutrons using recoil polarization, \cite{Eden:1994ji}--\cite{Plaster:2005cx} are completely independent of any theory of the deuteron, and those from a  polarized deuteron target, \cite{Passchier:1999cj}--\cite{Warren:2003ma}, are almost as clean.  All of these measurements are shown in  Fig.~\ref{fig:GEn}, and I chose to focus only on them because they are insensitive to deuteron theory.   For a recent review of the experimental data, see Ref.~\cite{Perdrisat:2006hj}; many other measurements exist.

Fig.~\ref{fig:GEn} shows that the solution for $G_{En}$ for Model WJC1 is in serious disagreement with the form factor measurements from free neutrons.   There seems to be no way to repair model WJC1; 
for this reason I did not study the predictions for model WJC1 further.  

In contrast, the  solution for $G_{En}$ from model WJC2 is in good agreement with the free data.  To study various possibilities, I decided to represent $G_{En}$ by the general functional form
\bea
G_{En}^{\rm model}(Q^2)=\frac{aQ^2(1+bQ^2+cQ^4+dQ^6)}{\left(1+e Q^2\right)^6}\, ,
\label{eq:GEnmodels}
\eea 
which goes like $Q^{-4}$ at large $Q$, and has cuts only for positive $q^2$, required if it is to be represented by a dispersion relation.  This functional form is so flexible that it can describe GK05 and two additional models of potential interest.  The parameters used for these three models of $G_{En}$ are given in Table \ref{tab:GEnmodels}.     Model CST1 is a very good representation of the solution obtained from $A$, while model CST2 follows GK05 up to the highest $Q^2$ points (Mad03,Pla05) and then tracks CST1 at higher $Q^2$.  All three of these models are shown in Fig.~\ref{fig:GEn}.

It turns out that both $B$ and $T_{20}$ are very insensitive to $G_{En}$, so our choice of a new $G_{En}$ different from GK05 will not disturb our previous fits to $B$ and $T_{20}$ (this can be confirmed by noting that the left panels of Figs.~\ref{fig:BallQ-scaled} and \ref{fig:T20} show almost no differences between models 2B, 2C, and 2D in the regions where they were used to obtain $F_3$ and $F_4$). Hence the only effect of choosing a new $G_{En}$ is an improvement in $A$, and as Fig.~\ref{fig:AhighQ-scaled} showns, model 2D (with $G_{En}$ represented by model CST1) provides an excellent fit to the data (except at the highest points -- to be discussed in the Conclusions), while model 2C (with $G_{En}$ represented by model CST2) is almost as good, and this $G_{En}$ tracks GK05 in the region where $G_{En}$ has been measured.  Final conclusions will be drawn in Sec.~\ref{sec:5}.

Predictions for the three form factors $G_C$, $G_M$, and $G_Q$ are shown in the next subsection.

\begin{table}[t]
\begin{minipage}{3in}
\caption{Parameters for the $G_{En}$ models using Eq.~(\ref{eq:GEnmodels}), with $Q$ in GeV.}
\label{tab:GEnmodels}
\begin{ruledtabular}
\begin{tabular}{lrrr}
  & GK05 &CST1 & CST2  \\[0.05in]
$a$ & 0.4779 & 0.4930 & 0.4930  \\[0.05in]
$b$ &0.5798 & 16.254 & 0.5532  \\[0.05in]
$c$  & 1.8452 & $-$27.849 & 0.6805  \\[0.05in]
$d$ & 0.4045 & 33.710 & 0.6861  \\[0.05in]
$e$  & 0.8628 & 1.5836 & 0.7904 \\[0.05in]
\end{tabular}
\end{ruledtabular}
\end{minipage}
\end{table}
%

\begin{figure*}
\centerline{
\mbox{
\includegraphics[width=2.3in]{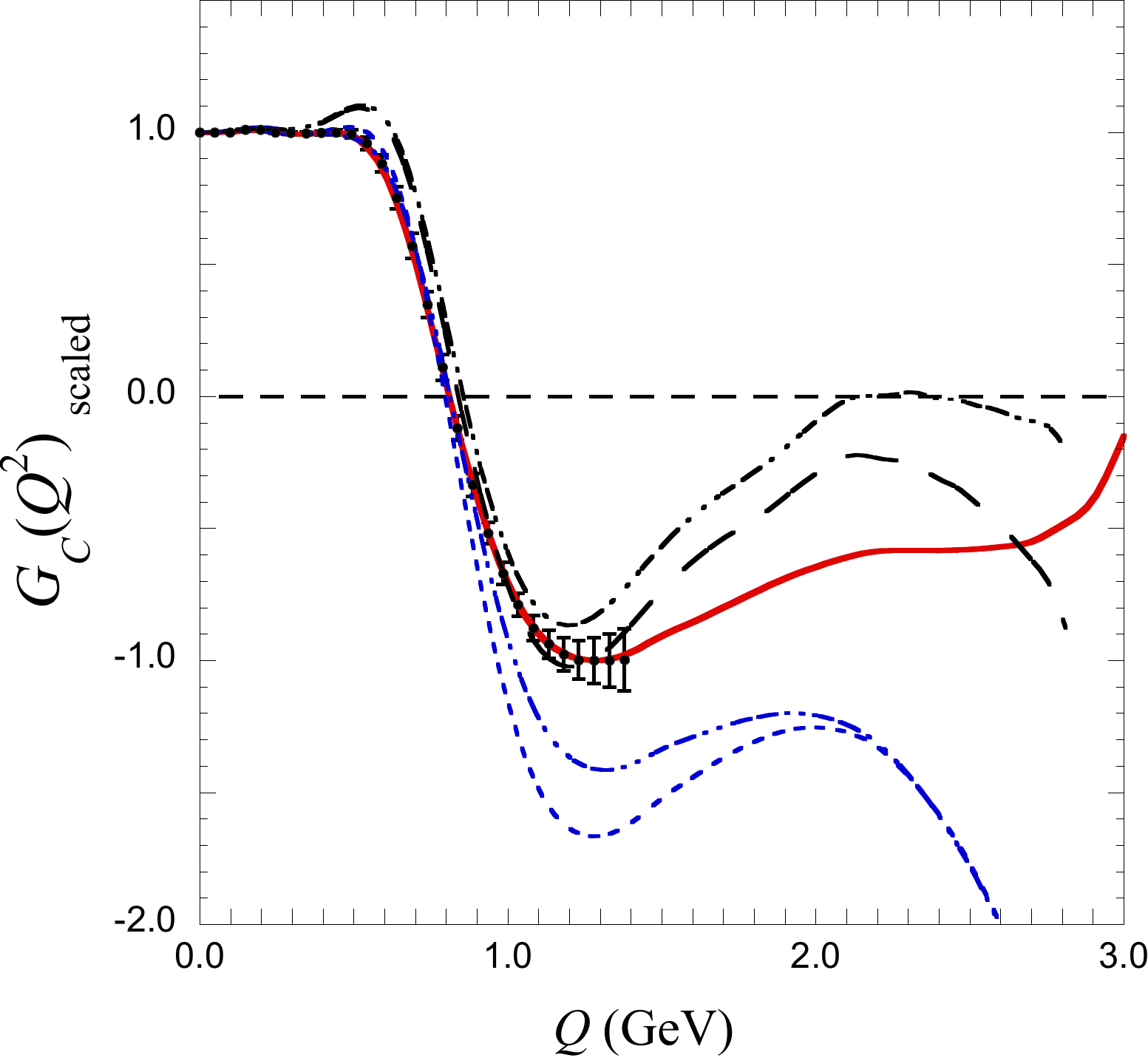} 
\includegraphics[width=2.3in]{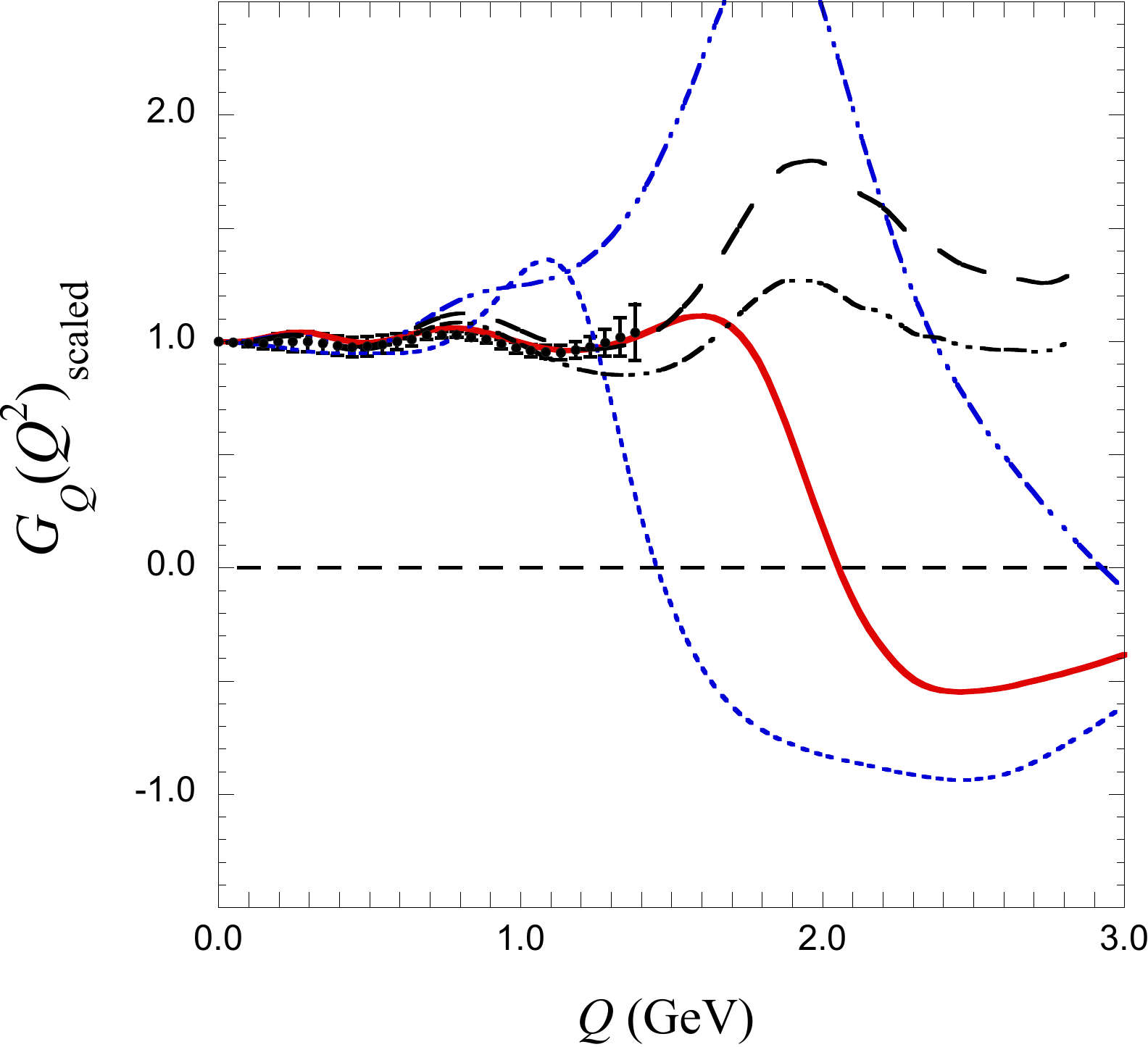} 
\includegraphics[width=2.3in]{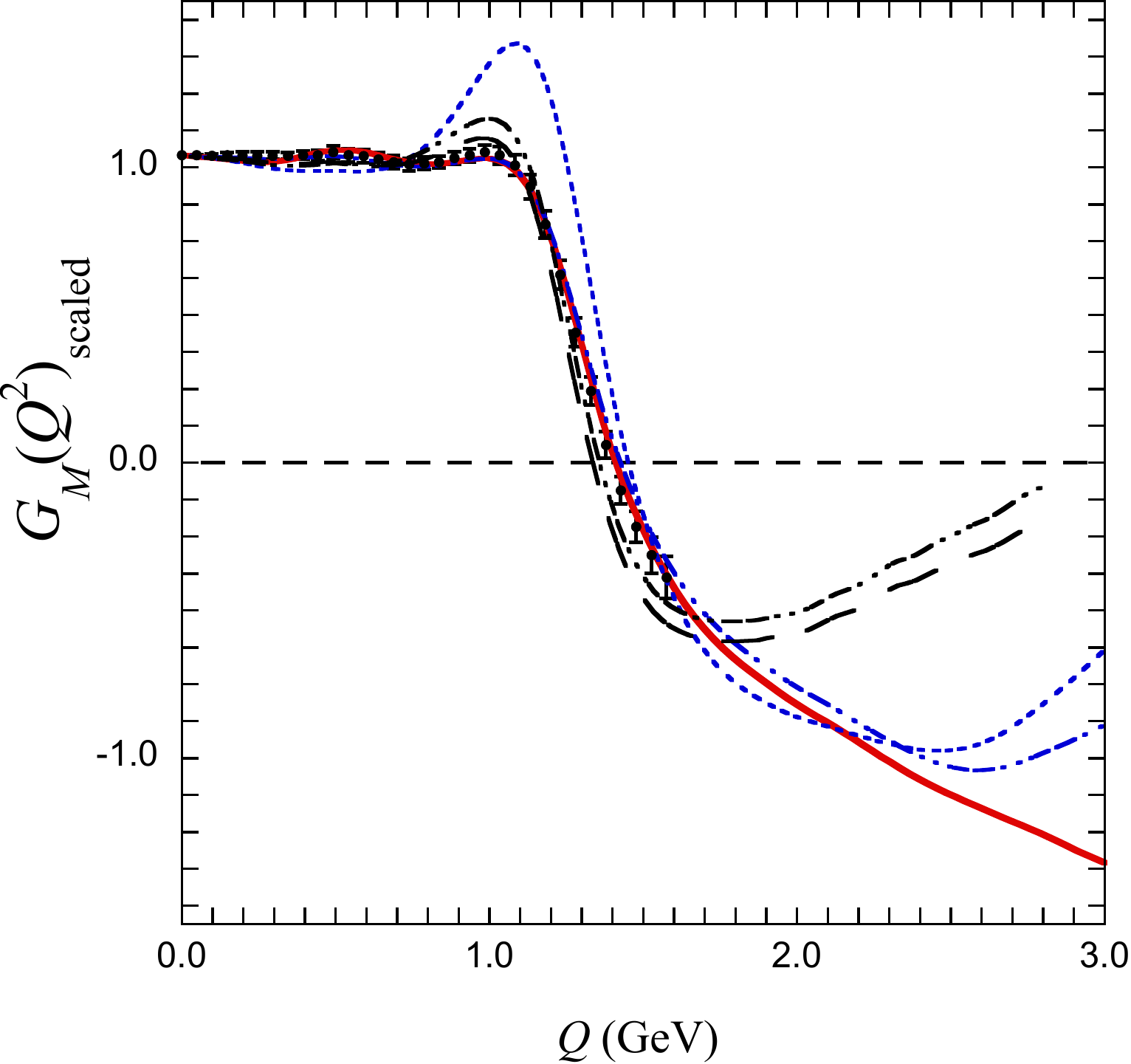} 
}
}
\caption{\footnotesize\baselineskip=10pt Predictions for the three deuteron form factors for models VODG, VODG0, 1A, 1B, and 2D (with lines as in the previous figures) compared to the GA. 
}
\label{fig:GCMQ1}
\end{figure*} 

\begin{figure*}
\centerline{
\mbox{
\includegraphics[width=2.3in]{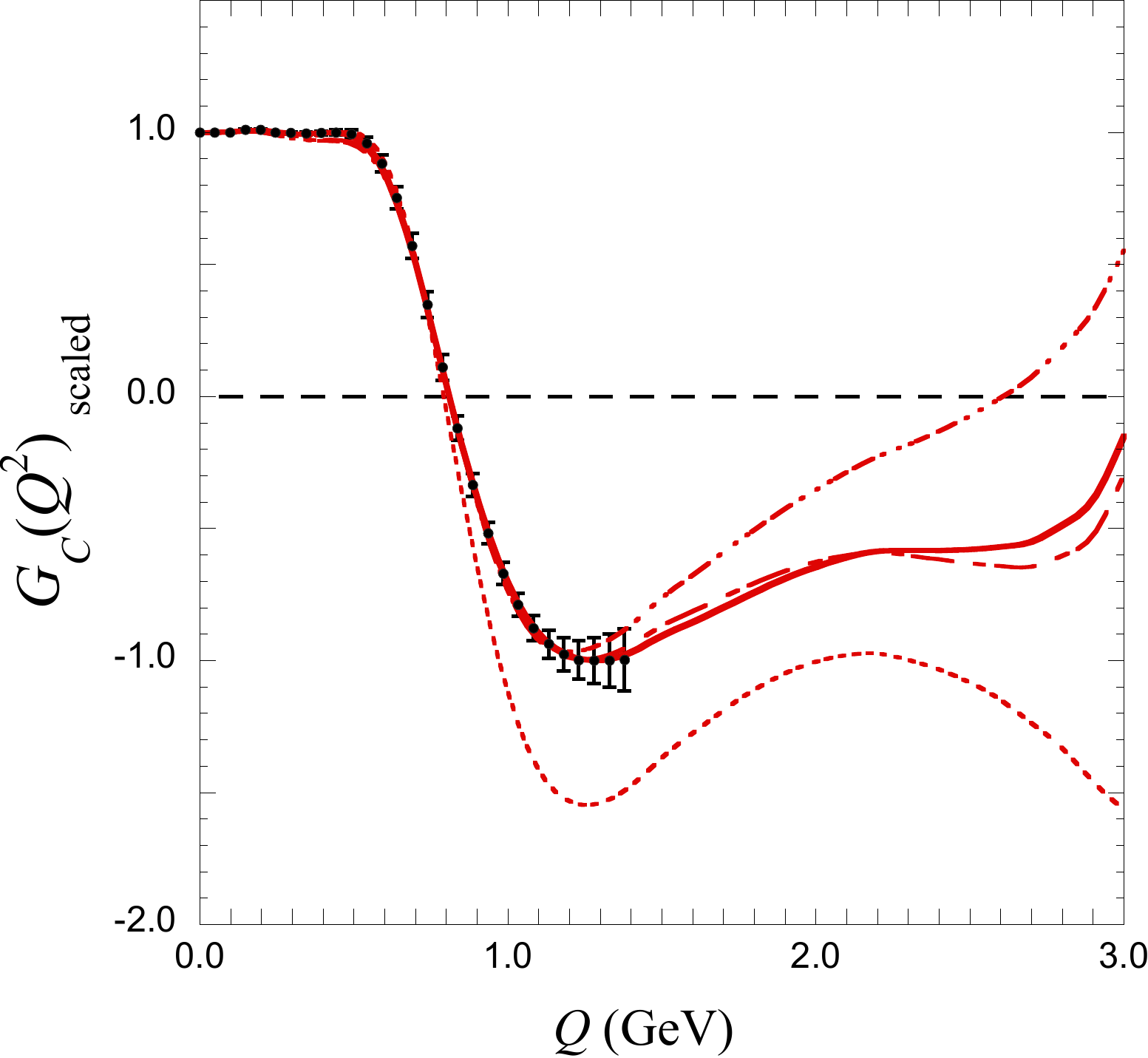} 
\includegraphics[width=2.3in]{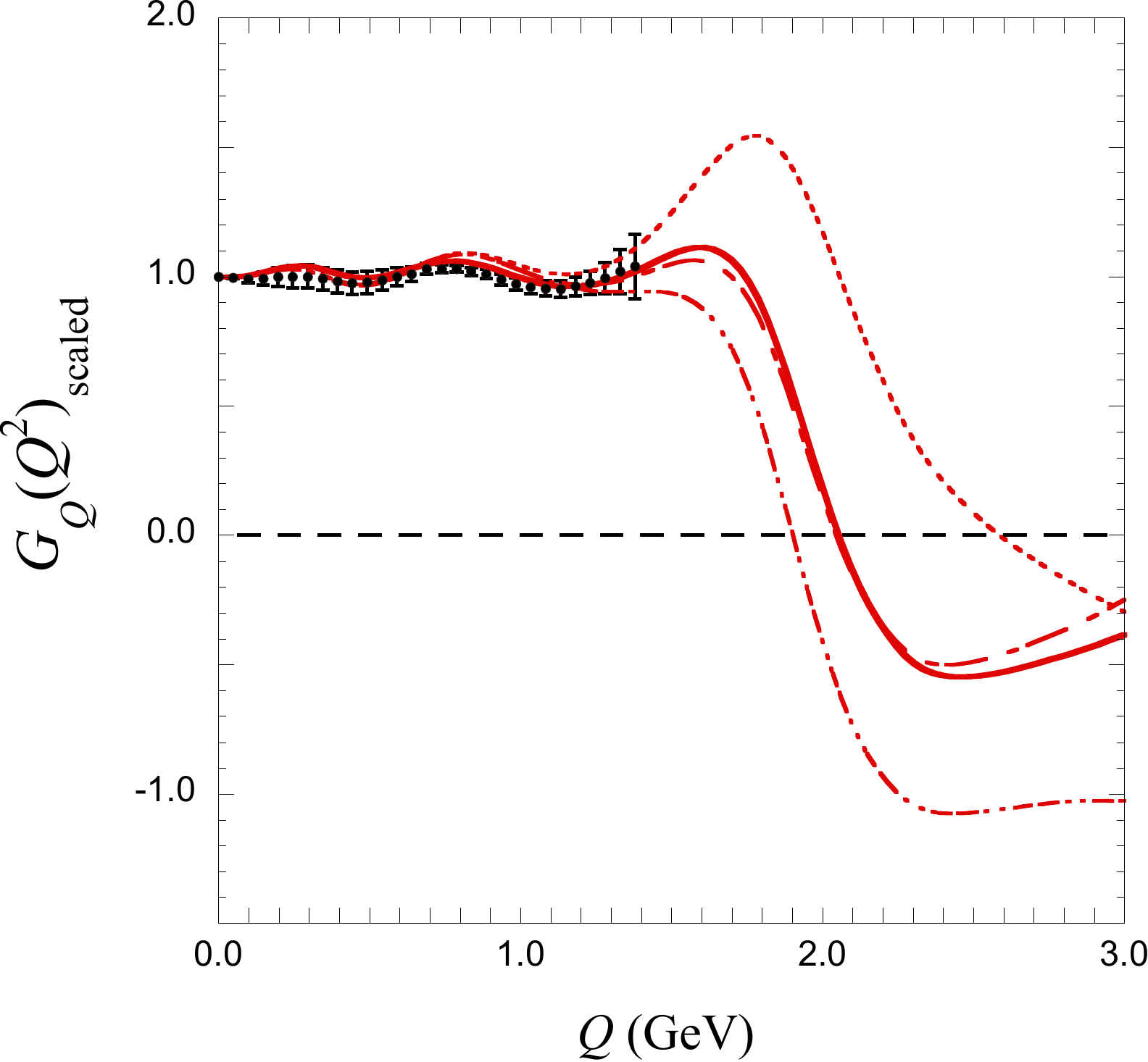} 
\includegraphics[width=2.3in]{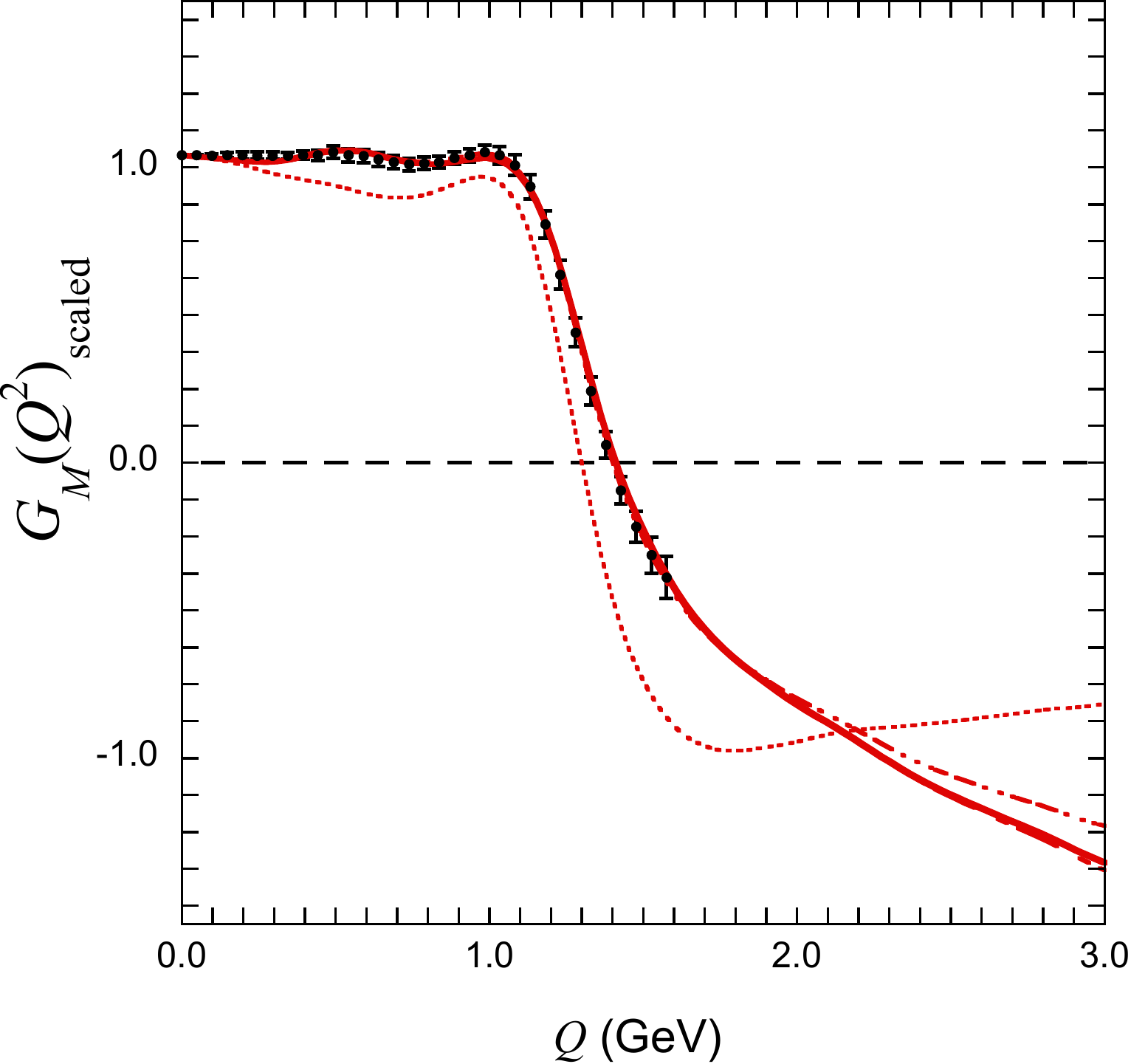} 
}
}
\caption{\footnotesize\baselineskip=10pt Predictions for the three deuteron form factors for models  2A-2D (with lines as in the previous figures) compared to the GA. 
}
\label{fig:GCMQ2}
\end{figure*} 

\begin{table}[t]
\begin{minipage}{3in}
\caption{Parameters for the scaling functions given in Eq.~(\ref{eq:FFscaling}), with $Q^2$ in fm$^{-2}$.}
\label{tab:GXscaling}
\begin{ruledtabular}
\begin{tabular}{lrrr}
  & $G_C (n=1)$ & $G_M (n=2)$ & $G_Q (n=3)$  \\[0.05in]
$a_1(n)$ & 0.6743 & 0.5149 & 0.4980  \\[0.05in]
$a_2(n)$ &0.0693 & 0.2912 & 0.0559  \\[0.05in]
$a_3(n)$  & 0.0084 & 0.0013 & 0.00008  \\[0.05in]
$b_0(n)$ & 1.8478 & 1.8422 & 1.2732  \\[0.05in]
$b_1(n)$  & 0.4185 & 0.5252 & 0.2956 \\[0.05in]
$b_2(n)$  & 0.1557 & 0.1749 & 0.0963 \\[0.05in]
$b_3(n)$  & 0.0321 & 0.0204 & 0.0194 \\[0.05in]
\end{tabular}
\end{ruledtabular}
\end{minipage}
\end{table}
%

\subsection{Predictions for the deuteron form factors}

I now can complete the discussion by presenting the three deuteron form factors and comparing them  to Sick's GA, which has been determined in the region $Q\lesssim 1.4$ GeV.  

In order to better see the details, all form factors are normalized to unity at $Q=0$, and divided by scaling functions with the same functional form as  used in Ref.~\cite{Marcucci:2015rca}:
\bea
{\rm ScaleG}_n&&(Q^2)=\sum_{i=0}^3a_i(n) \exp(-b_i(n) Q^2) \label{eq:FFscaling}
\eea
where
\bea
a_0(n)= 1-a_1(n)-a_2(n)-a_3(n)
\eea
ensuring that ScaleG$_n(0)=1$.  In order to scale the large $Q$ behavior of the form factors, I found it necessary to refit the coefficients, and the values I use in this paper are given in Table \ref{tab:GXscaling}.

\begin{table}[b]
\begin{minipage}{2.8in}
\caption{The $\chi^2/$datum  for the predictions of models 2C and 2D compared to Sick's GA.   The first point at Q=0.001 fm$^{-1}$ has been excluded.
} 
\label{tab:chi2}
\begin{ruledtabular}
\begin{tabular}{lccc}
 & nbr of points  & 2C& 2D\\[0.05in]
$G_C$  & 28 &3.613 & 0.116\\[0.05in]
$G_M$ & 32 &0.713 &0.763 \\[0.05in]
$G_Q$ & 28 & 1.920 &0.446  \\[0.05in]
$A$  & 44 & 6.440 &0.774  \\[0.05in]
$A$tail & 5  & 125.1 & 116.5  \\[0.05in]
$B$  & 34 &1.130 &1.131  \\[0.05in]
$T_{20}$ & 28 & 0.127&0.131  \\[0.05in]
\end{tabular}
\end{ruledtabular}
\end{minipage}
\end{table}
%

\begin{table}
\begin{minipage}{3.3in}
\caption{The $\chi^2/$datum  for the prediction of model 2D compared to the published data for $A, B,$ and $\widetilde T_{20}$.
} 
\label{tab:chi2A}
\begin{ruledtabular}
\begin{tabular}{lcclcc}
 $A(Q^2)$& number & $\chi^2$/d     &$T_{20}(Q^2)$& number & $\chi^2$/d\\[0.05in]
HEPL-65  & 5 & 2.53 & Bates-84& 2&0.13\\[0.05in]
Orsay-66 & 4 &1.63 & Nuovo-85&2& 0.78\\[0.05in]
CEA-69 & 18 & 3.01& Nuovo-90 & 2 & 0.83 \\[0.05in]
DESY-71 & 10 & 0.71 & Bonn-91 & 1 &0.55\\[0.05in]
SLAC-75 & 8 & 1.61  & Bates-94 &  3 & 2.50 \\[0.05in]
Mainz-81  & 18 & 7.36 & NIK-96& 1 & 1.02 \\[0.05in]
Bonn-85 & 5  & 20.18  & NIK-99& 3 & 0.70\\[0.05in]
Saclay-90  & 43 & 2.77 & JLabC-00& 6 &0.86 \\[0.05in]
JLab-A & 16 & 4.59  & Nuovo-01 & 5 & 3.14\\[0.05in]
JLab-C & 6 & 2.87 & Bates-11& 9 &0.94 \\[0.05in]
All & 131 & 3.98 & All & 34 &1.29 \\[0.05in]
\tableline\\[-0.08in]
$A(Q^2)$ ranges & & & $B(Q^2)$ & & \\[0.05in]
$Q \leq 0.6$ GeV & 64 & 3.82& Stan-65 &4 & 1.08 \\[0.05in]
$Q>0.6$ GeV &67 & 4.14  & Mainz-81& 4 & 2.87 \\[0.05in]
3 largest & 3 & 21.83\;& Saclay-85 & 13 & 0.75 \\[0.05in]
&  & & Bonn-85 & 5  &1.07  \\[0.05in]
&  & & JLab-89 & 6 & 2.06  \\[0.05in]
&  & & SLAC-90 & 9 & 2.28  \\[0.05in]
&  & & All & 41 & 1.56  \\[0.05in]
\end{tabular}
\end{ruledtabular}
\end{minipage}
\end{table}
%

\begin{table}
\begin{minipage}{3.4in}
\caption{Physical quantities that enter into a calculation of the deuteron form factors
} 
\label{tab:ingredients}
\begin{ruledtabular}
\begin{tabular}{ll}
$\Gamma=S^{-1}\Psi$ & Relativistic vertex function (with particle 1\cr
&   on-shell); contributes to all diagrams shown in  \cr & 
Fig.~\ref{Fig2}; solution of a two-nucleon CST equation 
 \cr
& using the OBE kernel \\[0.05in]
$\Gamma^{(2)}=S^{-1}\Psi^{(2)}$ & Relativistic wave function (with particle 1 \cr & 
on-shell); generated by interaction currents \cr  &
of type $V^{(2)}$ which arise 
 from the momentum \cr 
& dependence of the boson couplings to particle  \\& 2; calculated by iterating 
the CST  equation  \cr
& once using the kernel $V^{(2)}$; diagram \ref{Fig2}(A$^{(2)}$)  \\[0.05in]
$\widehat \Gamma_{\rm BS}=\widetilde \Gamma
-\widetilde \Gamma^{(1)}$
&  Subtracted vertex function (with {\it both\/} particles \cr 
 & off-shell); the $\widetilde \Gamma^{(1)}_{\rm BS}$ subtraction arises from the \\& 
 interaction currents $V^{(1)}$ coming  from the \cr 
 & momentum dependence of the  boson couplings 
 \cr & to particle 1;   calculated by iterating the CST\cr
 & equation  once using the subtracted kernel \cr & $V-V^{(1)}$ 
  with {\it both\/} particles off-shell in the \cr
  & final state; diagram \ref{Fig2}(B) \\[0.05in]
$F_3$ and $F_4$ & Form factors describing the off-shell nucleon\cr
& current; diagram \ref{Fig2}(A) \\[0.05in]
\end{tabular}
\end{ruledtabular}
\end{minipage}
\end{table}
%

The scaled deuteron form factors are shown in Figs.~\ref{fig:GCMQ1} and \ref{fig:GCMQ2}.  In the figures I display all of the cases studied in the previous sections, even though the models 2C and 2D are the only ones that are in quantitative agreement with the Sick GA.

Note that model VODG predicts all of the form factors within  1-2 standard deviations over the entire range.  Model 2B, designed to agree precisely with $B$ and $T_{20}$, gives an exact description of $G_M$ over the  entire range (as expected) but fails to provide a precise explanation of $G_C$ and $G_Q$. 
In the region $1\lesssim Q \lesssim 1.4$, both $-G_C$ and $G_Q$ are too large, so that their ratio, measured in $T_{20}$, is correct.  Only  models 2C and 2D give a precise explanation of all form factors.  

I call attention to the contributions of the $F_3$ and $F_4$ form factors which are easy to see on these plots.  Since $F_3$ cannot be zero (because of the constraint $F_3(0)=1$) the best way to isolate the size of these contributions is to  compare models 1A and 1B, or 2A and 2B, shown respectively by the short dashed and double dash-dot lines (blue for WJC1 and red for WJC2).  The figures show that that there is little difference at $Q\lesssim 0.6$ GeV, except that model 2A fails to describe $G_M$  even at quite small $Q$. 


\begin{figure*}
\centerline{
\mbox{
\includegraphics[width=2.3in]{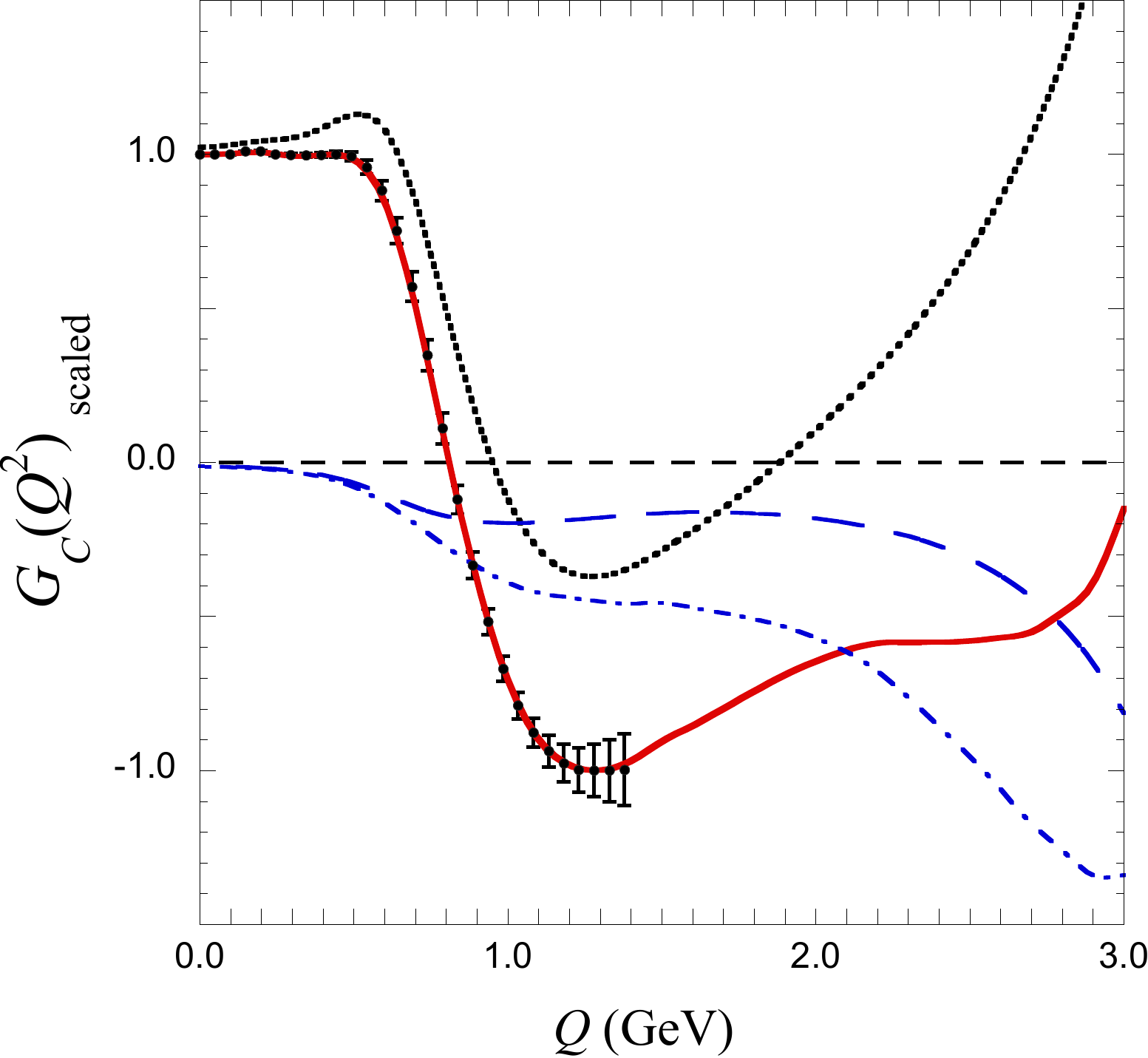} 
\includegraphics[width=2.3in]{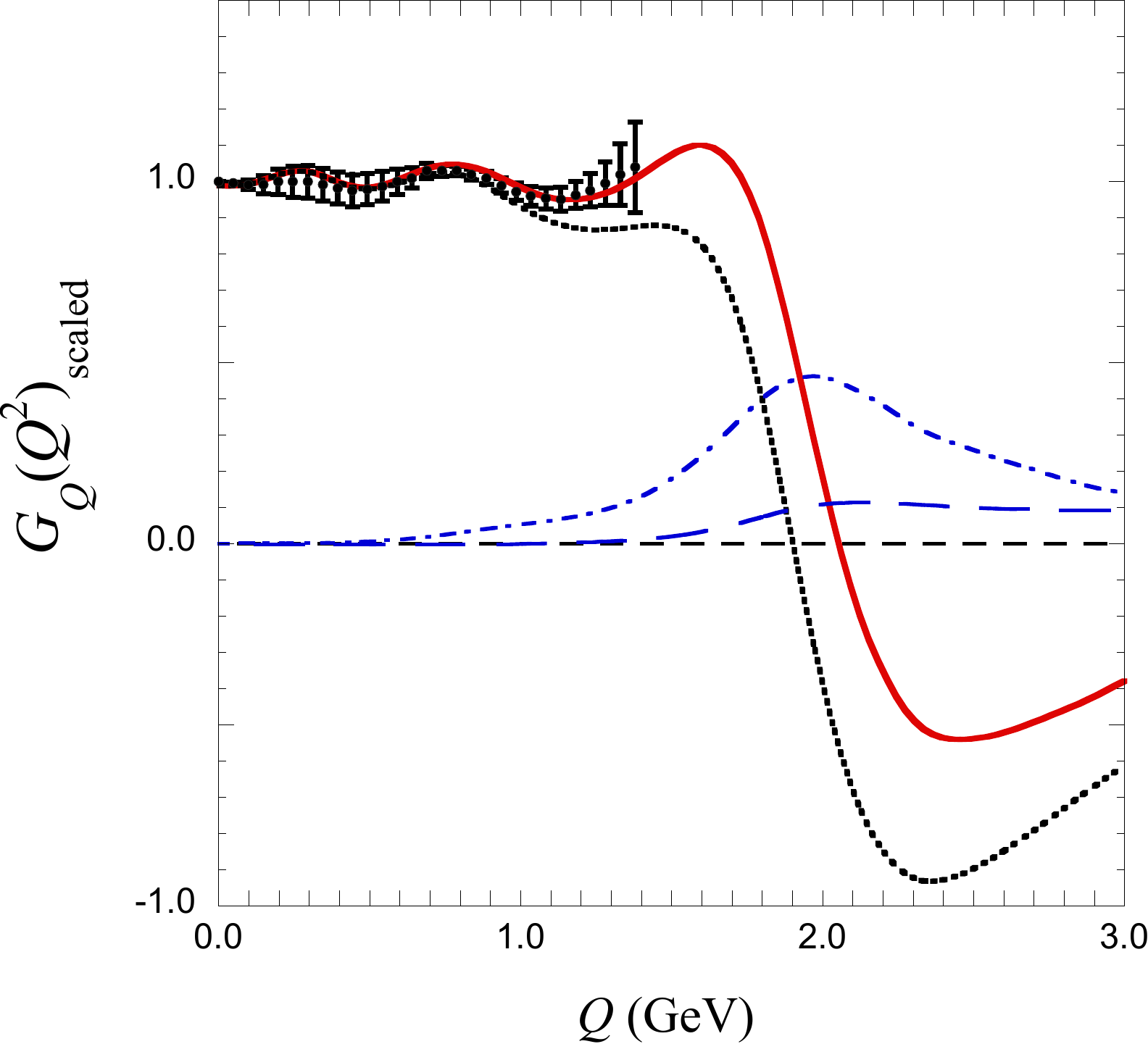} 
\includegraphics[width=2.3in]{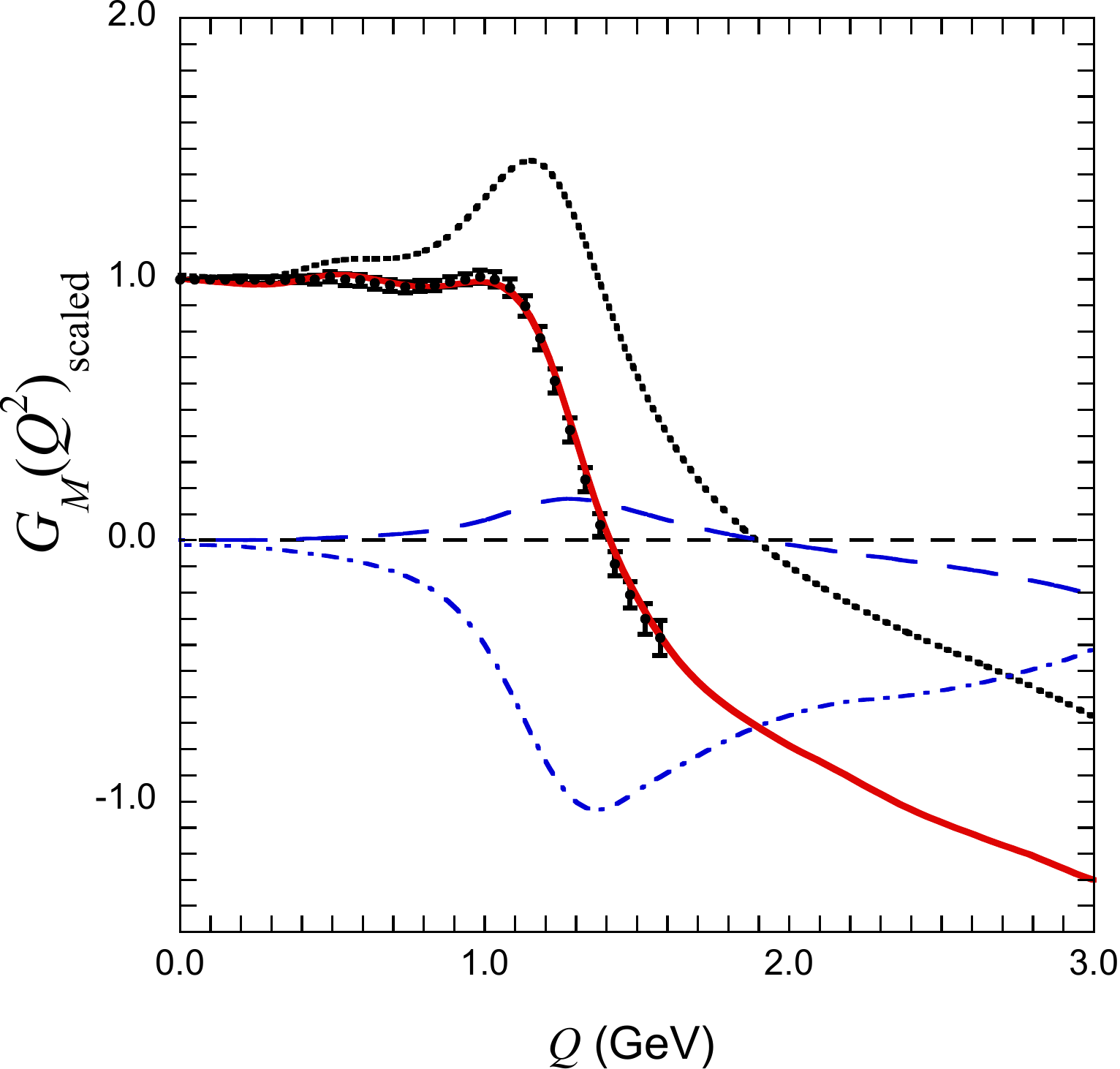} 
}
}
\caption{\footnotesize\baselineskip=10pt  Study of the two interaction currents, $V^{(2)}$ (blue dot-dashed line) and $V^{(1)}$ (blue long dashed line).  Setting both to zero gives model 2D with no IC (thick black dotted line), while restoring both gives the full model2D (thick red line) shown in previous figures.  
}
\label{fig:IC}
\end{figure*} 

Table \ref{tab:chi2} shows how closely models 2C and 2D predict Sick's GA (using Sick's error bars).  Except for few points in $A$ at the highest $Q$ (the tail), the fits are excellent, of comparable quality except of $G_C$, $G_Q$ and $A$, where model 2D provides a more accurate prediction than 2C.

Table \ref{tab:chi2A} shows the $\chi^2$/datum for the published data compared to model 2D.  Note that the CST prediction is in reasonable agreement with the data for $B$ and $T_{20}$, but that there are large discrepancies with the data for $A(Q^2)$, even for $Q\leq0.6$ GeV, and that the measurements at the three largest $Q$ points (JLabA) are in significant disagreement with the perdiction (but the disagreement is not as large as with the Sick GA).  I will discuss this further in the Conclusions

\section{Physical Insights} \label{sec:Rel}

In this section I study the size of various partial contributions to the form factors.  The study is limited to model 2D, which gives the best fit to the Sick GA.  
Before discussing the individual contributions, it is helpful to briefly identify the ingredients of the theory.  


\subsection{Physical quantities of the theory}


The physical quantities that I will focus in in this section  are summarized in Table \ref{tab:ingredients}.  They are:~(i)  vertex functions $\Gamma$ and $\Gamma^{(2)}$ when one particle is on-shell, (ii) the subtracted vertex function $\widehat\Gamma_{\rm BS}$ for {\it both\/} nucleons off-shell, 
and (iii) the new off-shell nucleon form factors $F_3$ and $F_4$ already discussed extensively above.  
To make the presentation simple, I postpone all precise definitions until Appendix~\ref{app:theory}. 


\subsection{Study of the isoscalar interaction currents}

The isoscalar interaction  currents (IC) produce the interaction current vertex function $\Gamma^{(2)}$ generated by $V^{(2)}$ and giving rise to diagram Fig.~\ref{Fig2}(A$^{(2)}$), and the subtraction terms $\tilde \Gamma^{(1)}$ generated by $V^{(1)}$ and 
discussed in Ref.~I.  
The behavior of these terms is shown in Fig.~\ref{fig:IC}.

Fig.~\ref{fig:IC} shows that both IC's make significant contributions to $G_C$, even at low $Q$.  For the other form factors, $G_M$ and $G_Q$, their contributions are quite small at low $Q$, but are still important for $Q\gtrsim0.5$ GeV (for $G_M$) and  $Q\gtrsim1$ GeV (for $G_Q$).   These interaction currents are a significant part to the overall theoretical picture. 

\begin{figure}
\centerline{
\mbox{
\includegraphics[width=2.5in]{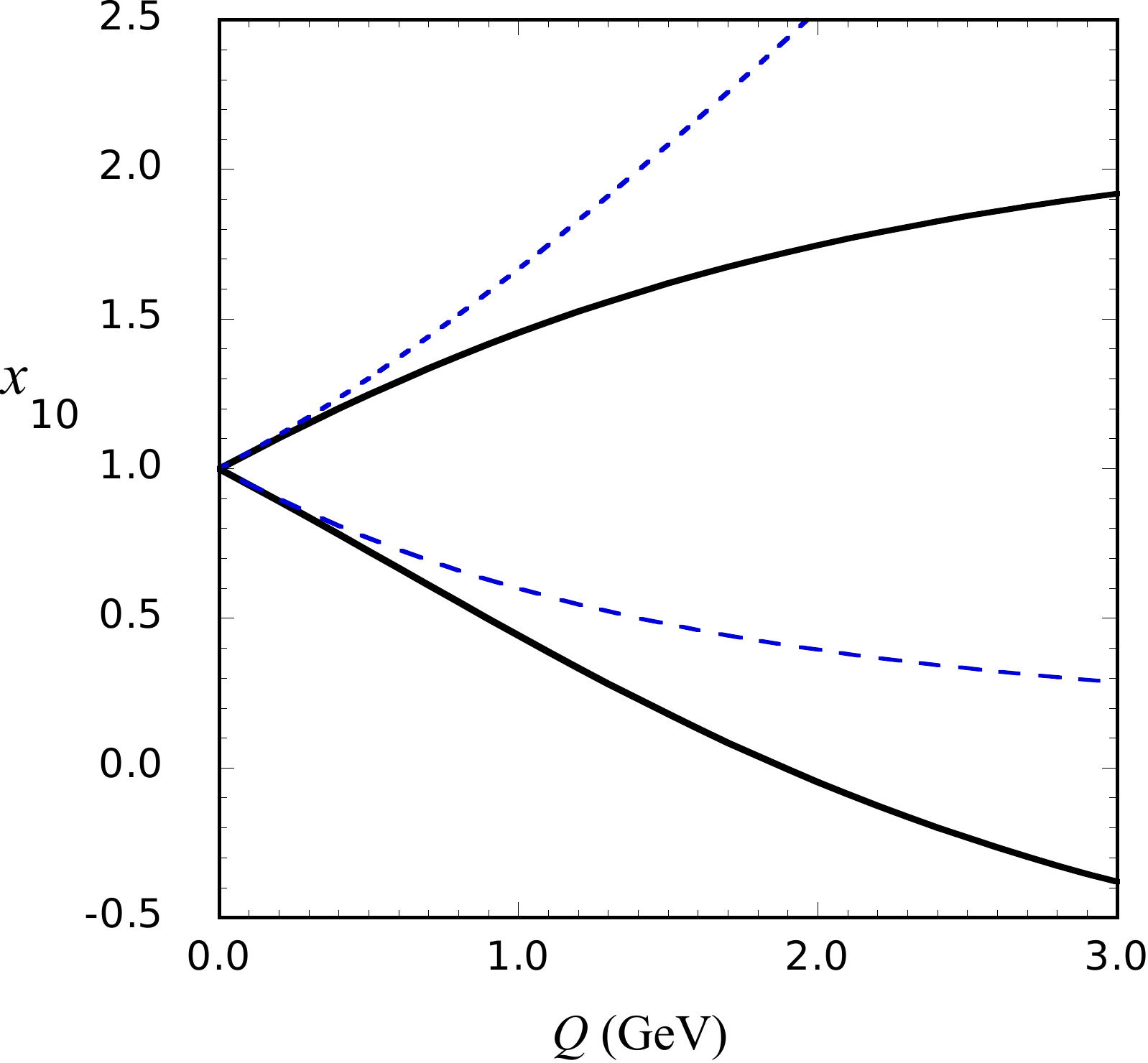}
}
}

\caption{\footnotesize\baselineskip=10pt Plot of the maximum and minimum of $x_{10}$ as a function of $Q$.   
The black solid lines show the maximum and minimum, including boost effects, as obtained from Eq.~(\ref{eq:x10exact}); the blue dashed lines are the maximum and minimum without boost effects as obtained from Eq.~(\ref{eq:x10a}).}
\label{fig:X0}
\end{figure} 

\begin{figure*}
\centerline{
\mbox{
\includegraphics[width=3in]{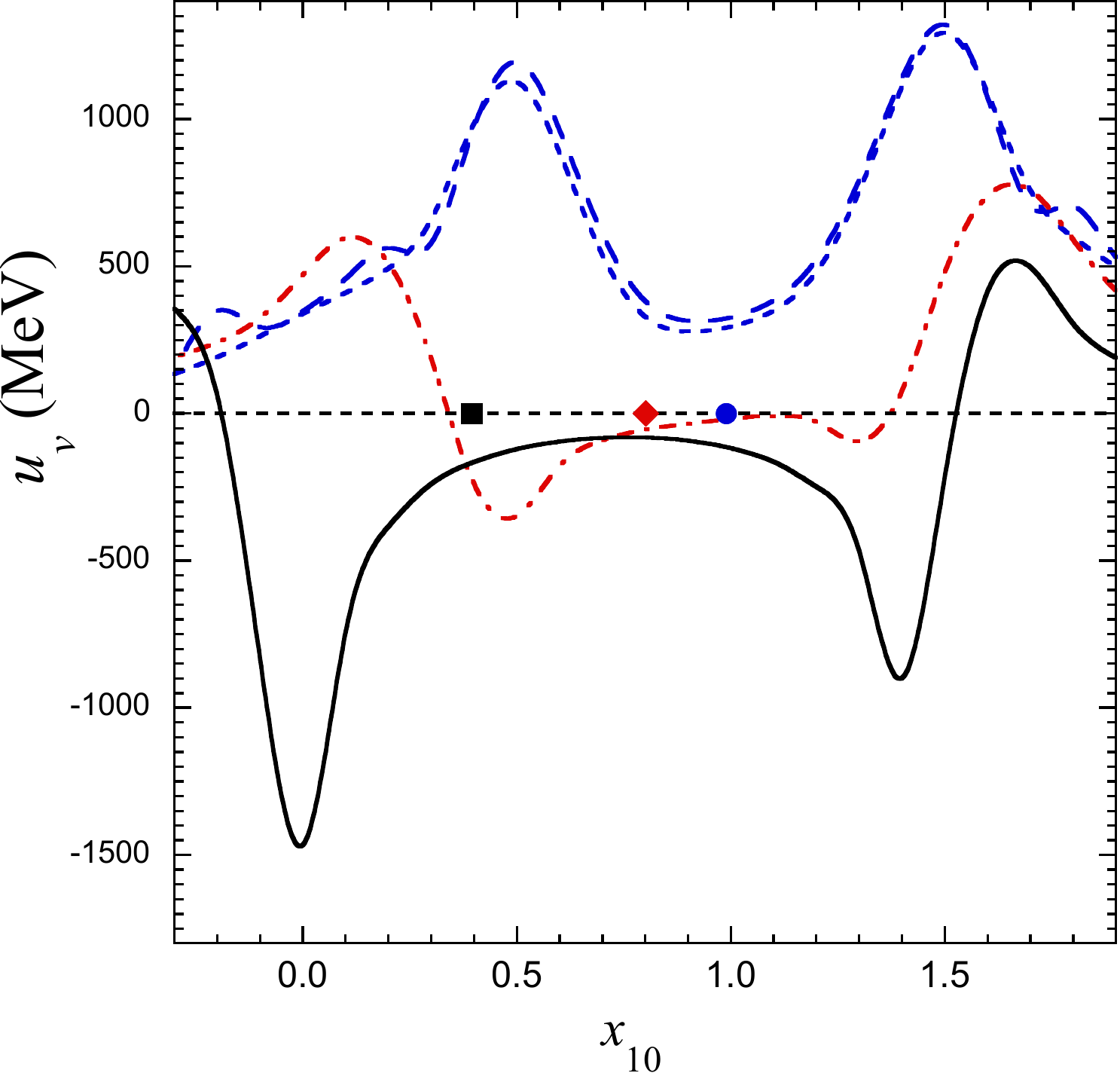}
\hspace{0.5in}
\includegraphics[width=3in]{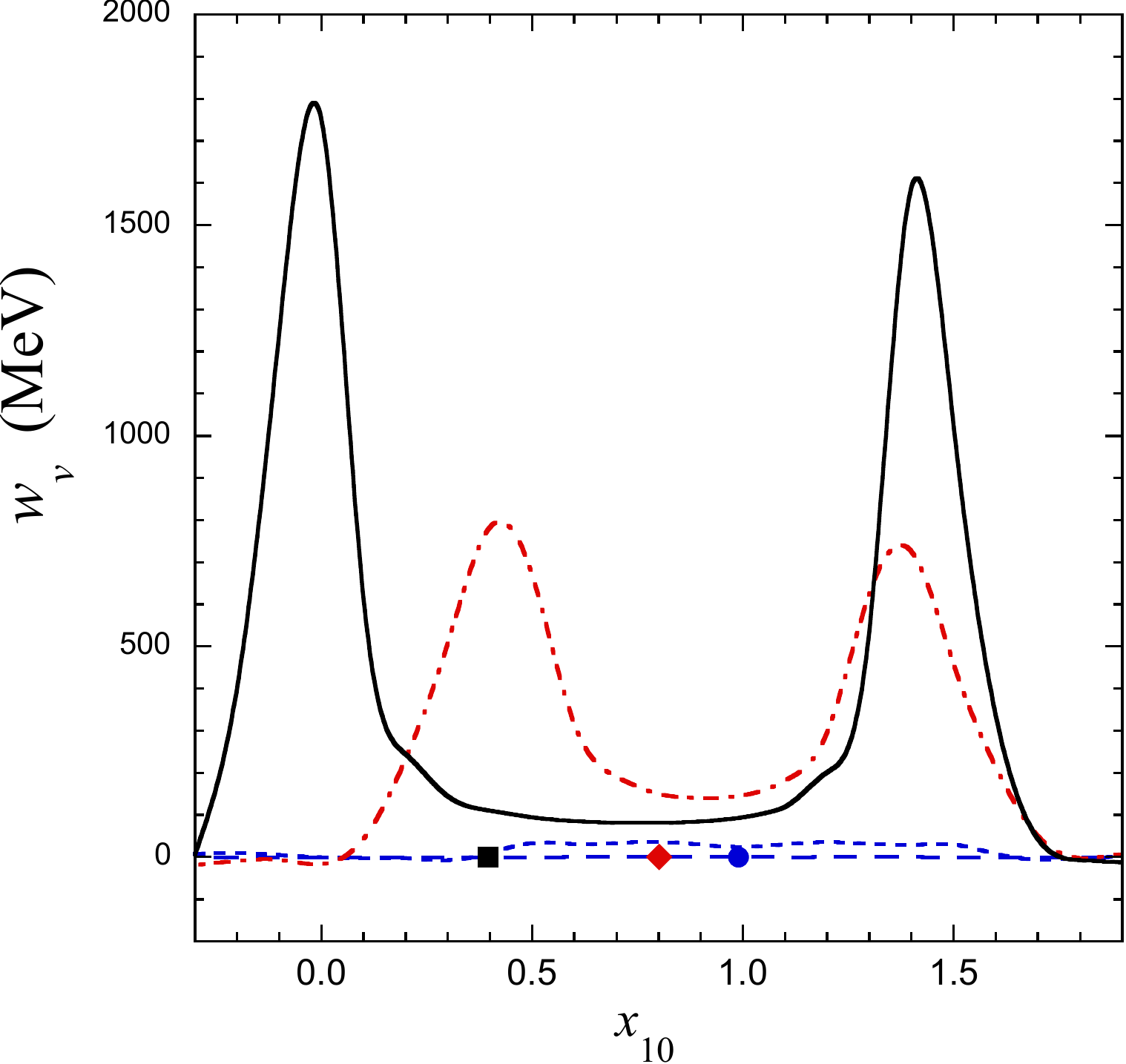}
}
}
\caption{\footnotesize\baselineskip=10pt The invariants $u_v$ and $w_v$ defined in Eq.~(\ref{eq:uvwv}) 
for the favored model WJC2, shown as a function of $x_{10}$ for four fixed momenta:  $k=k(1) \simeq 0.527$ MeV (blue long dashed lines), $k=k(10)\simeq 84.0$ MeV (blue short dashed lines), $k=k(23)\simeq 450.3$ MeV (red dot-dashed line), and $k=k(32)\simeq 960.8$ MeV,  where $k(n)$ is the $nth$ gauss point in the mapped grid of 60 points.   The values of $x_{10}$ when particle 2 is on shell are shown for $k(10)$ (solid blue circle), $k(23)$ (solid red diamond), and $k(32)$ (solid black square).
}
\label{fig:uvwv}
\end{figure*} 

\subsection{Off-shell effects} \label{sec:off-shell}

What are off-shell effects?  This discussion must be approached carefully or serous misunderstandings may emerge.  For example, in the CST one nucleon is always off-shell in intermediate states; this is the way the CST creates virtual intermediate states and, at the same time, preserves  four-momentum conservation.  In conventional quantum mechanics, the particles are always on-shell, but the virtual intermediate states do not conserve the total energy of the particles.  It can be shown that these two approaches are largely equivalent, with the CST having the advantage that it is relativistically covariant, and the disadvantage that it must learn how to describe off-shell particles (with their accompanying antiparticle components).  In the context of the discussion of $NN$ scattering, for example, the role of the virtual antiparticles is an interesting off-shell effect.   However, in the context of $ed$ scattering, I will look only at new effects that did not already arise in $NN$ scattering. 

The unique off-shell effects that are studied here are the contributions that arise  when {\it both\/} nucleons are off-shell.  These are the contributions from the vertex functions $\widehat\Gamma_{\rm BS}$, which take us outside the usual boundaries of the CST.   The need to discuss the physics of two nucleons off-shell does not arise in the discussion of three-nucleon scattering \cite{Stadler:1996ut,Gross:1982ny,Stadler:1995cjp,Stadler:1997iu} but does arise in the discussion of $ed$ scattering and electron-triton scattering \cite{Pinto:2009dh,Pinto:2009jz}.  How should these effects be defined so that they give us useful insight into the physics of this theory? 

\begin{figure*}
\centerline{
\mbox{
\includegraphics[width=2.3in]{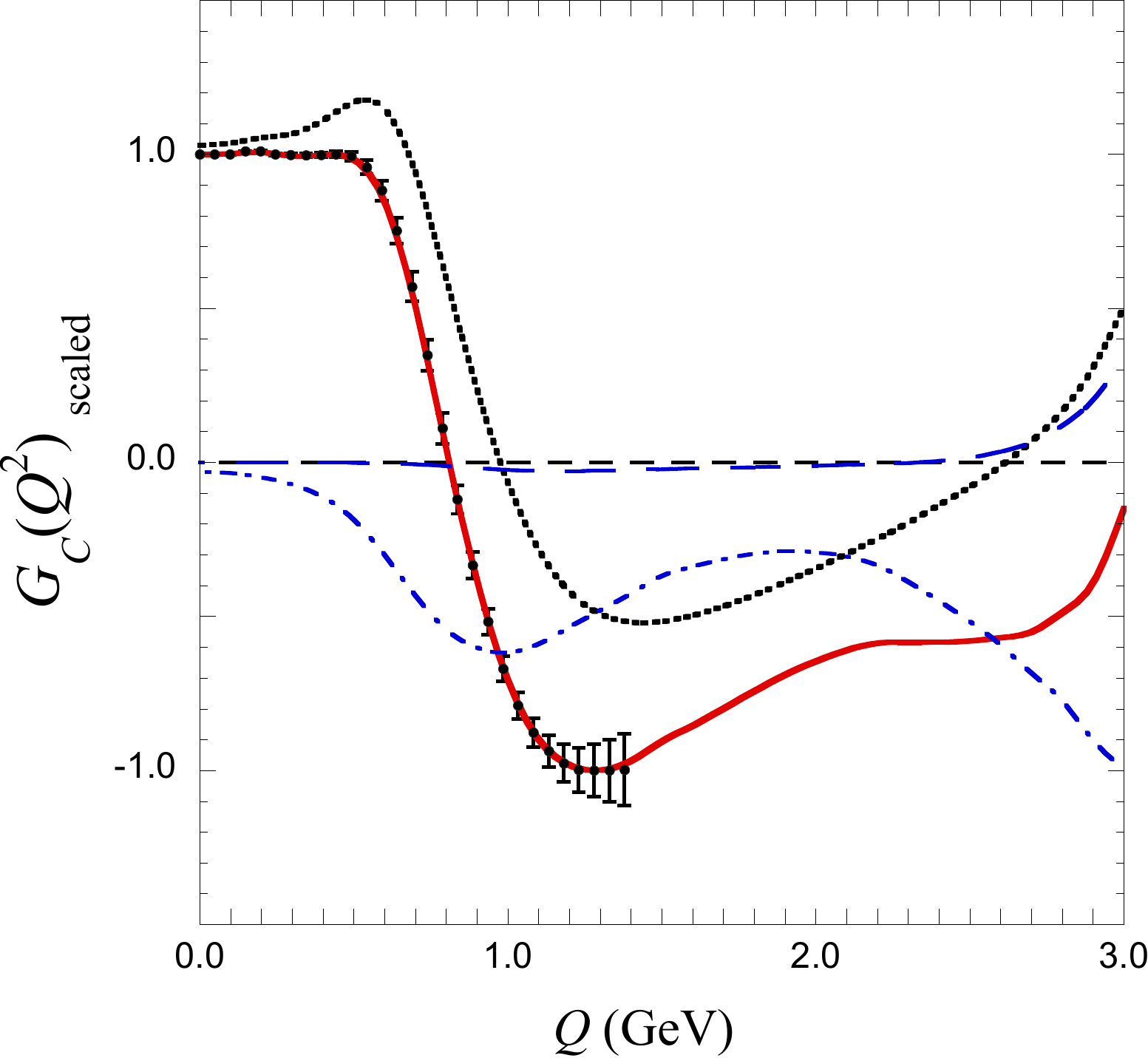} 
\includegraphics[width=2.3in]{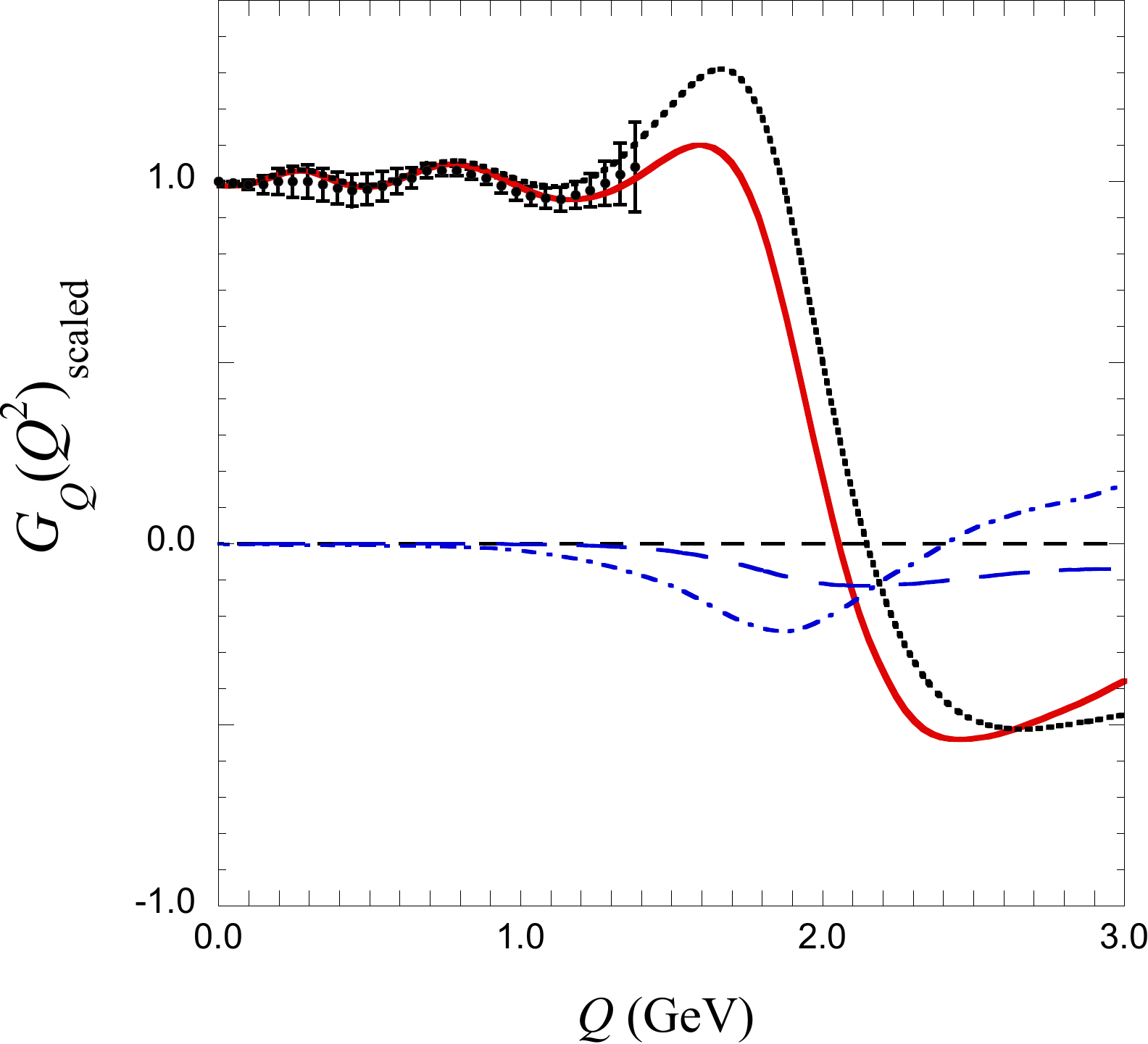} 
\includegraphics[width=2.3in]{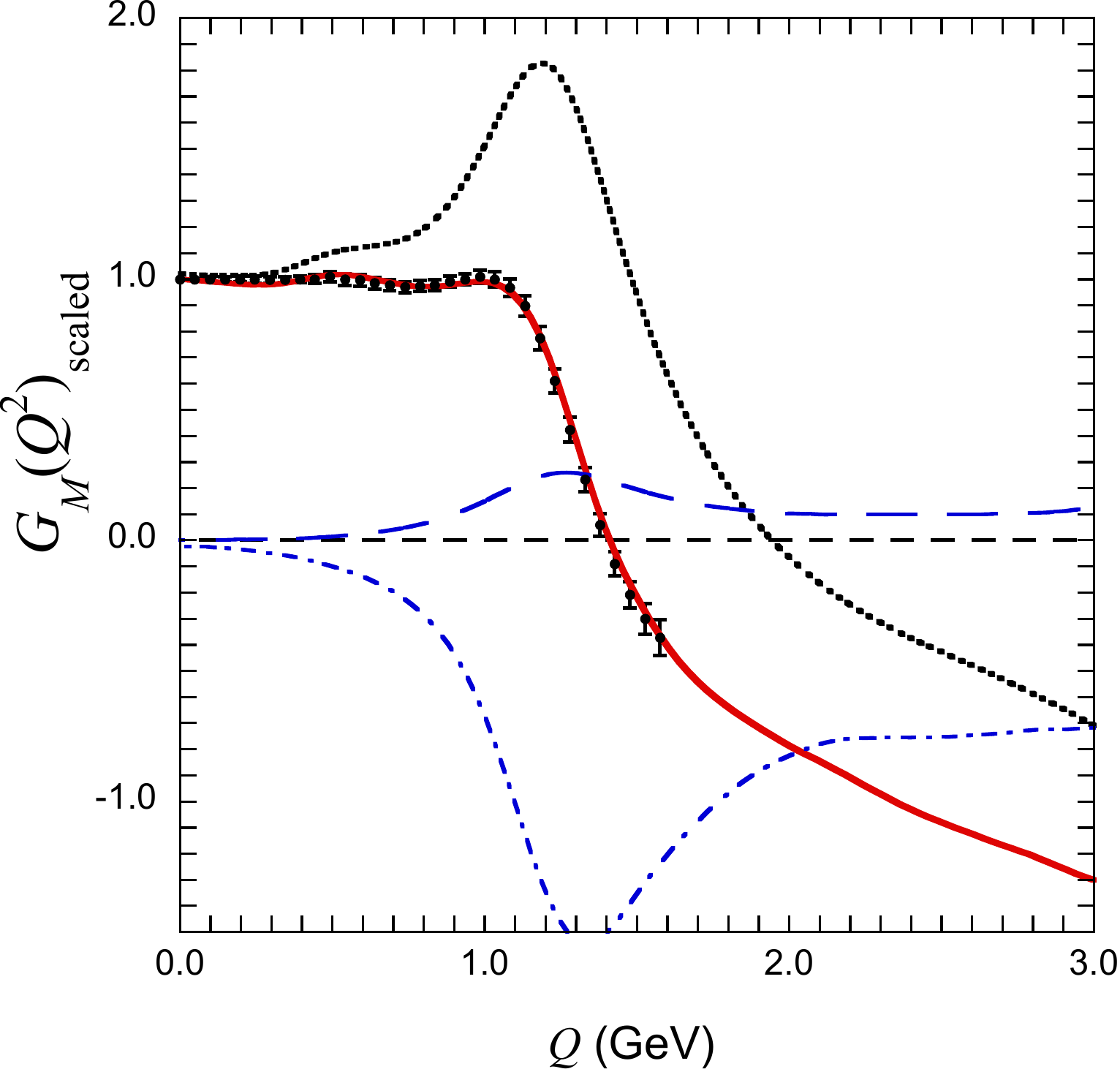} 
}
}
\caption{\footnotesize\baselineskip=10pt  Study of the off-shell effects defined in the text. 
}
\label{fig:OSStudy}
\end{figure*} 

Only diagram \ref{Fig2}(B) requires  particle 1 to be driven off-shell.  In the Breit frame, $P_\pm=\{D_0,\pm \frac12 q_z\}$, with $q_z=Q$.  When the incoming (outgoing) particle 1 is on-shell, the outgoing (incoming) particle 1 will have four-momentum
\bea
\tilde k_\pm=\{\tilde E_\mp, \tilde {\bf k}\pm\frac12 q_z\}\, , \label{eq:ktildepm}
\eea
where 
\bea
\tilde E_\pm=\sqrt{m^2+\big(\tilde{\bf k}\pm\frac12 q_z\big)^2}\, .
\eea
This particle is off-shell with an energy $\tilde k_0^\pm=\tilde E_\mp \ne \tilde E_\pm$.  I find it convenient to describe this extra degree of freedom by the parameter $x_{10}$, which is defined as the ratio of the off-shell energy to the on-shell energy.  In this case the ratio is
%
%
%
\bea
x_{10}^\pm=\frac{\tilde k_0^\pm}{\tilde E_\pm} \to \frac{\tilde E_\mp}{\tilde E_\pm}=\sqrt{\frac{E_{\tilde k}^2\mp \tilde k_z Q+\frac14Q^2}{E_{\tilde k}^2 \pm \tilde k_z Q+\frac14Q^2}}
\equiv \zeta^\pm ,\qquad
\label{eq:x10actual}
\eea
which is always positive.  
The maximum of $x_{10}^-$ ($x_{10}^+$) occurs when $\tilde k_z=\tilde k$  (or $-\tilde k$), $\tilde k=k_{\rm max}$, and solving for $k_{\rm max}$ gives
\bea
x_{10}^{\rm max}&=&\sqrt{\frac{m^2+(k_{\rm max}+\frac12 Q)^2}{m^2+(k_{\rm max}-\frac12 Q)^2}}
\nonumber\\
&\to& \frac1{2m}(\sqrt{4m^2+Q^2}+Q)\, , \label{eq:x10a}
\eea
and the minimum is $1/x_{10}^{\rm max}$.  This shows that as $Q$ increases, the particle 1 (either incoming or outgoing) is forced further and further off-shell.

While this is useful for our understanding, what we really want is the result in the {\it rest\/} system of the deuteron, so (\ref{eq:x10a}) must be transformed to the rest system.  This is discussed in detail in Appendix \ref{sec:trans}.  The results for both (\ref{eq:x10a}) and the relativistically  correct result $x_{10}^{B {\rm max}}$, given in Eq.~(\ref{eq:x10exact}), are shown in Fig.~\ref{fig:X0}.  Note that the boost effects are significant.

The invariants that describe $\widehat\Gamma_{\rm BS}$ depend on the two variables $k$ and $x_{10}$ (with $x_{10}=1$ when particle 1 is on shell).   As shown in Fig.~\ref{fig:X0}, for studies of the form factor below $Q\simeq3$ it is sufficient to know the off-shell dependence of the invariants that describe  $\widehat\Gamma_{\rm BS}$ in the range $1.9 \gtrsim x_{10} \gtrsim -0.3$.  This behavior is shown in  Fig.~\ref{fig:uvwv}, with $u_v$ and $w_v$ related to the largest deuteron wave functions $u$ and $w$ by
\bea
u_v(k,x_{10})&=&[(1+x_{10})E_k-m_d] u(k,x_{10})
\nonumber\\
w_v(k,x_{10})&=&[(1-x_{10})E_k-m_d] w(k,x_{10})\, . \label{eq:uvwv}
\eea  
The other wave functions are much smaller.   

To obtain the off-shell behavior, the wave functions are iterated once using the fully off-shell kernel, as shown in Eq.~(\ref{eq:hatgamma2}).   I found that the resulting wave functions were much smother at low momentum if the small one photon exchange term was removed from the iterating kernel, and all of the results presented in this paper were calculated in this way.  This is partly justified by the observation that keeping the ``last'' one photon exchange could be regarded as  including one higher order effect in $\alpha_\gamma$ while ignoring others, and may not even be consistent.  In any case, it introduces a small inconsistency: when the on-shell wave functions are iterated without the last one photon exchange, the normalization is changed slightly.  To obtain the original normalization, the results for WJC1 are multiplied by 0.9962  and those for WJC2 by 0.9954.

\begin{figure*}
\centerline{
\mbox{
\includegraphics[width=2.3in]{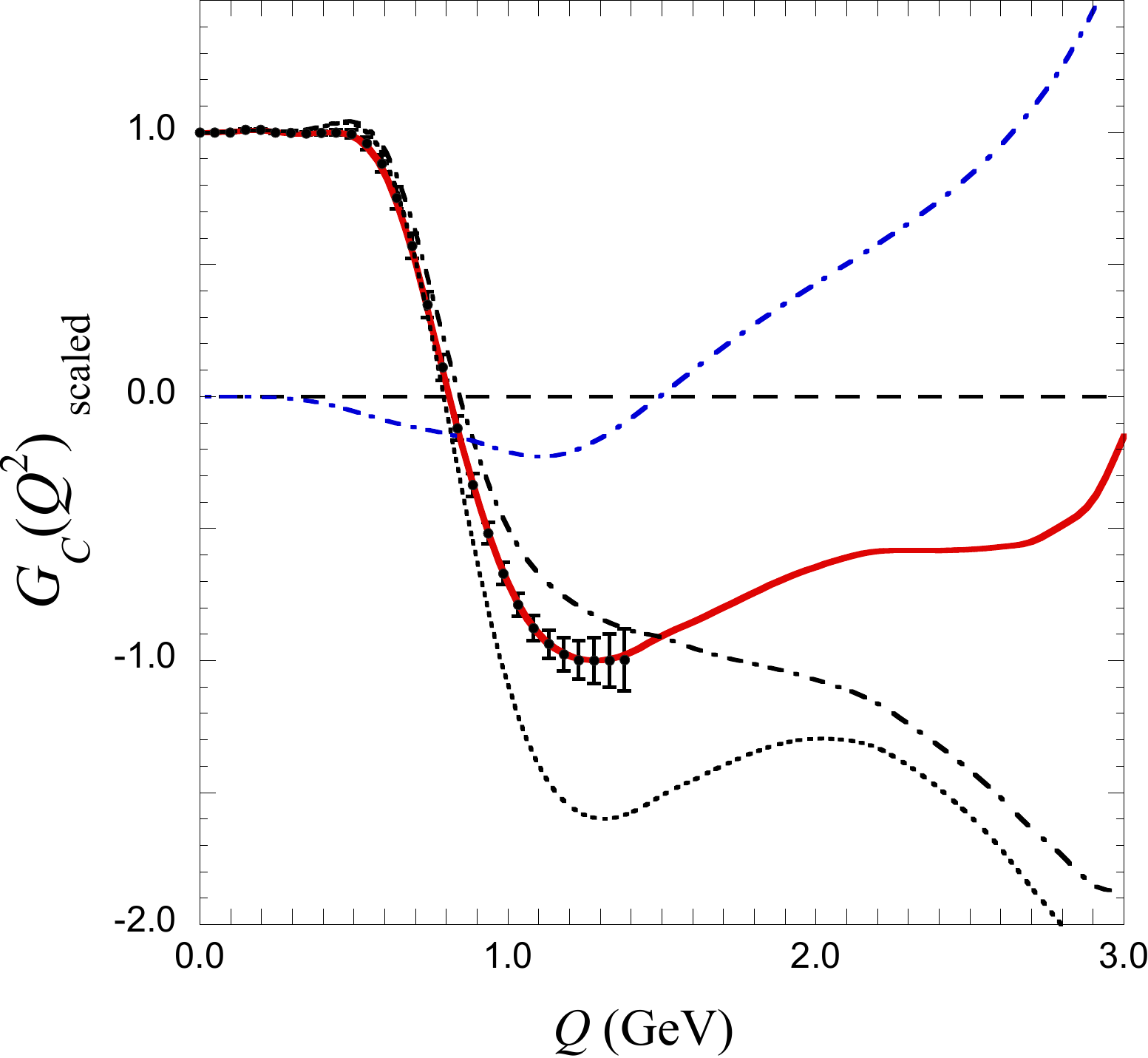} 
\includegraphics[width=2.3in]{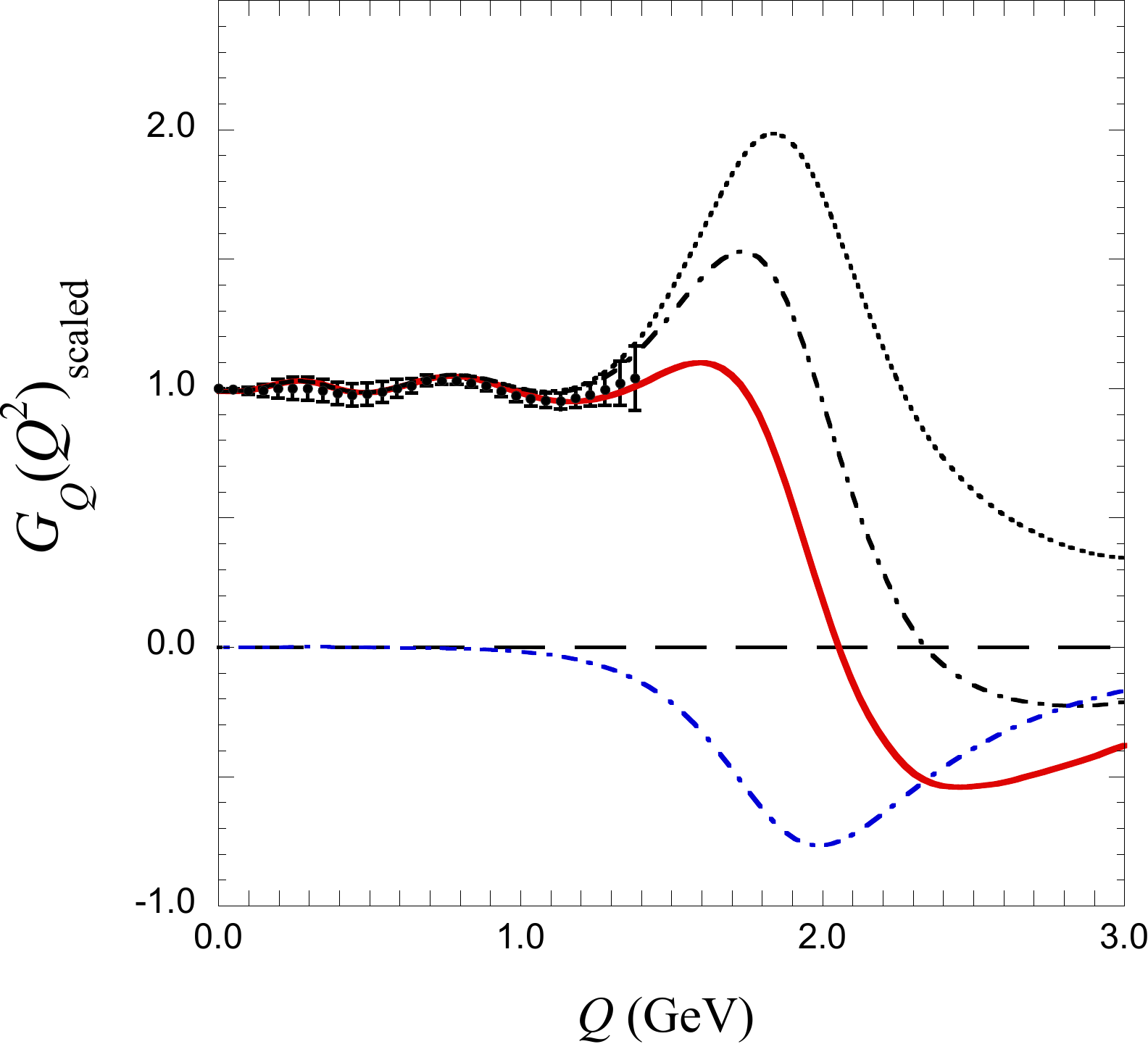} 
\includegraphics[width=2.3in]{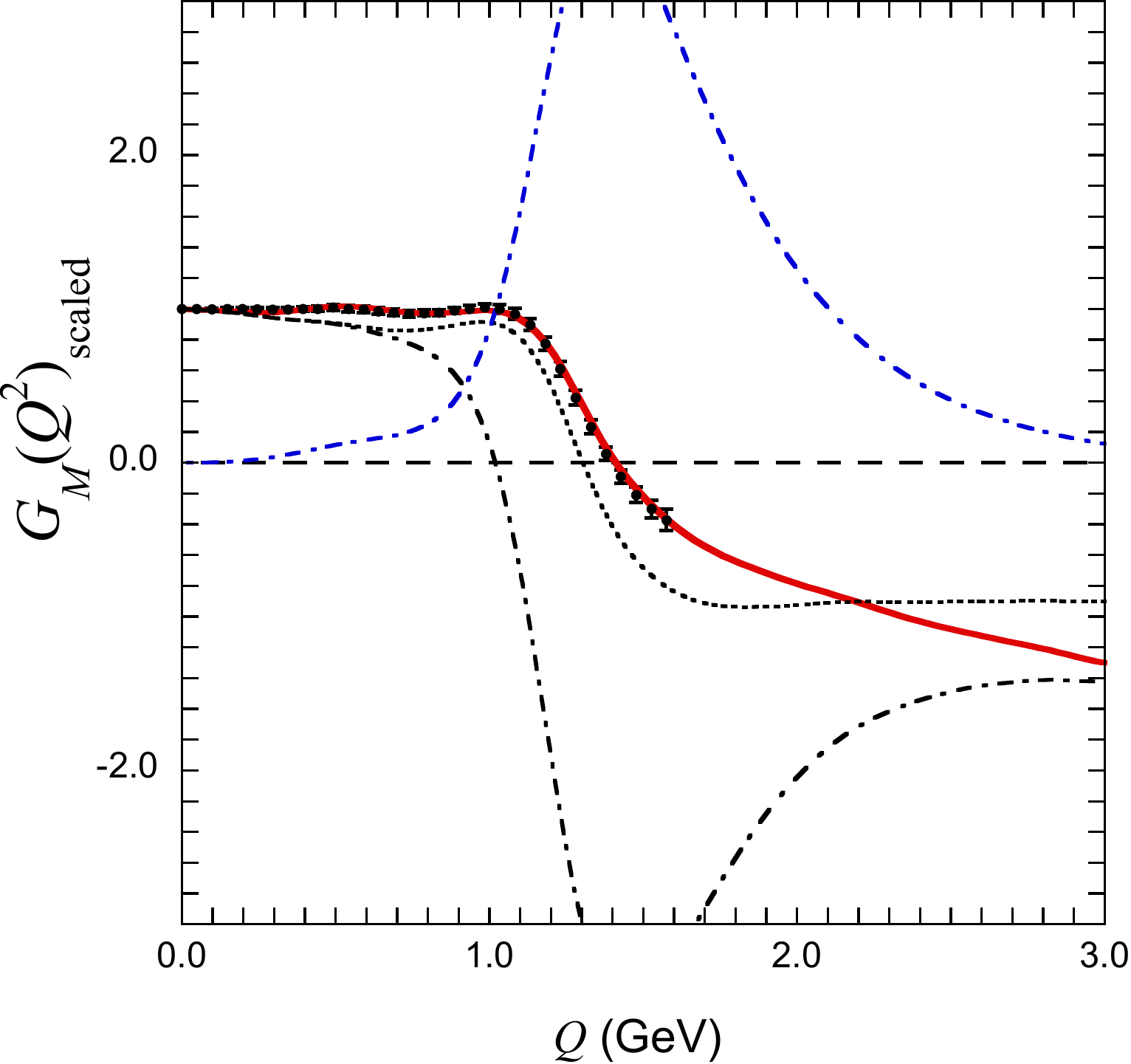} 
}
}
\caption{\footnotesize\baselineskip=10pt  Study of the sensitivity of the form factors $F_3$ and $F_4$.  {\it All\/} curves use the predicted CST1 for $G_{En}$.  Both black curves set $F_4=0$  and $F_3\ne0$: the black dotted curves use a dipole form for $F_3$ while the black dot-dashed lines uses model $F_3(2)$.   The heavy red line is the result of adding the contributions from $F_4(2)$ to the black dot-dashed lines and gives the best model 2D.  The blue dot-dashed lines, which show the size of the $F_4(2)$ contributions by themselves, are given for reference. 
}
\label{fig:F3F4Study}
\end{figure*} 

The point on the curves where particle 2 is on-shell is given by 
\bea
x_{10}^{\rm ex}(k)=\frac{m_d-E_k}{E_k}\, ,
\eea
which depends on $k$.  This point is marked by the small solid black squares (for $k=960.8$ MeV), red diamonds ($k=450.3$ MeV), and blue circles ($k=84$ MeV) along the $x$ axis in the panels of Fig.~\ref{fig:uvwv}.   These points are interesting because the two-body $NN$ bound state equation depends on vertex functions defined only at $x_{10}=1$ and  $x_{10}=x_{10}^{\rm ex}(k)$; values of the vertex functions at all other values of $x_{10}$ have not played {\it any\/} role in previous fits to the $NN$ data..  The off-shell dependence of elastic ed scattering depends on values of the vertex functions determined theoretically, but {\it never tested experimentally\/}.

The size of these effects is shown in Fig.~\ref{fig:OSStudy}.   In each panel the black dotted line is a calculation using the parameters of model 2D with $x_{10}=1$ in the (B) diagrams, and the thick red solid line is the full model 2D with $x_{10}$ free to vary as the kinematics dictates (as shown previously).  The contribution from Fig.~\ref{Fig2}(B) decomposes into a contribution multiplied by the projector $\Theta(-k)$ that vanishes when particle 1 is on-shell (referred to as the C contribution) and a remainder (referred to as the B contribution, distinguished from the total by the absence of the parentheses):
\bea
\widehat\Gamma_{\rm BS}(k,P)&=&\underbrace{\Gamma(k,P)}_\text{B}-\underbrace{\Gamma_{\rm off}(k,P) \,2\Theta(-k)}_\text{C}\, ,
\label{eq:BandCparts} 
\eea
The C contribution (labeled by the blue long dashed lines in the figure) is quite small, but still of great interest because it depends on invariants that do not exist when one of the particles is on-shell.  The largest off-shell contributions come from the B terms (dash-dotted blue lines), which make a significant contribution to all the form factors, especially $G_C$ and $G_M$.  

The calculations are sensitive to off-shell effects at all values of $Q$.

\begin{figure*}
\centerline{
\mbox{
\includegraphics[width=2.3in]{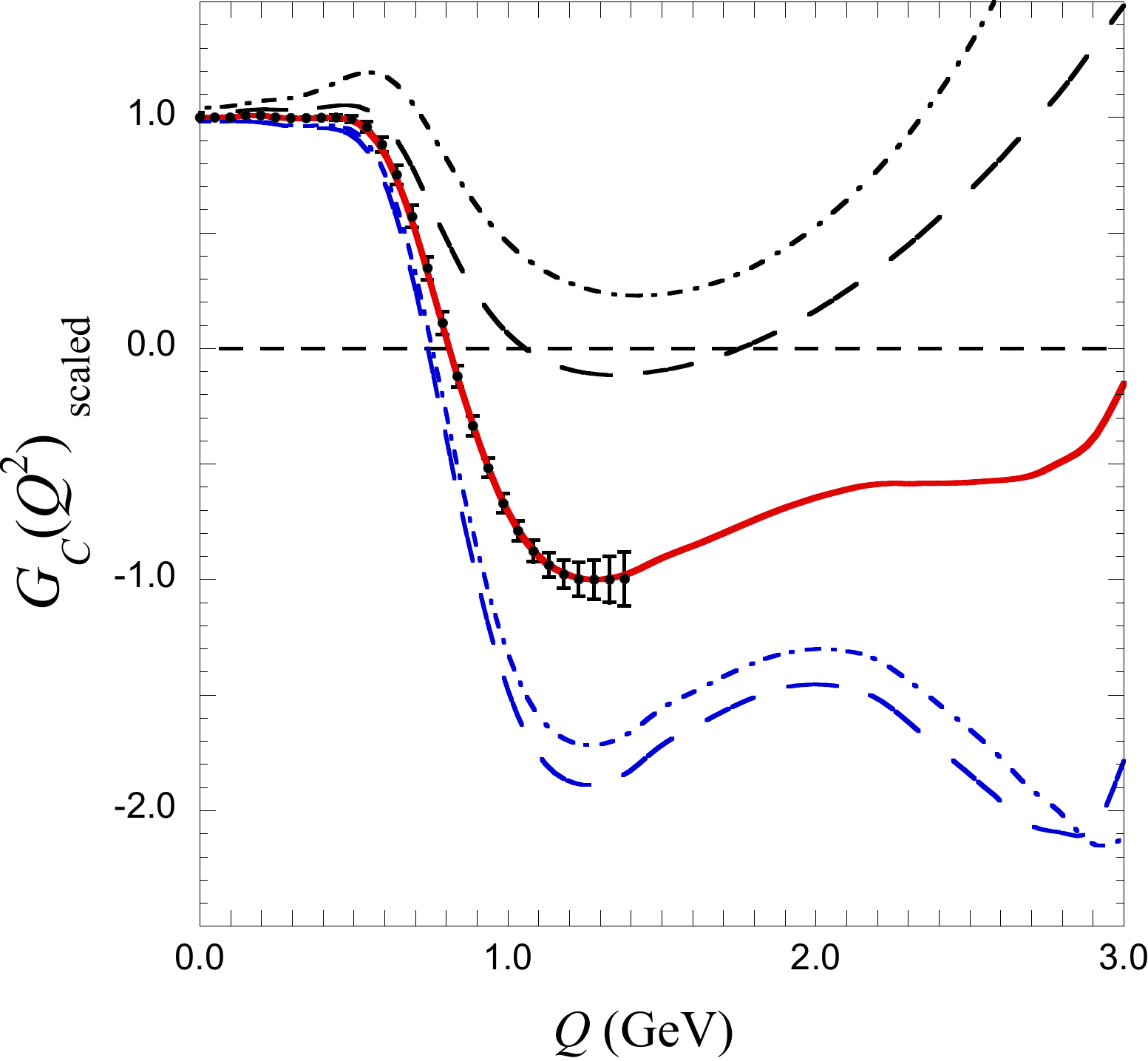} 
\includegraphics[width=2.3in]{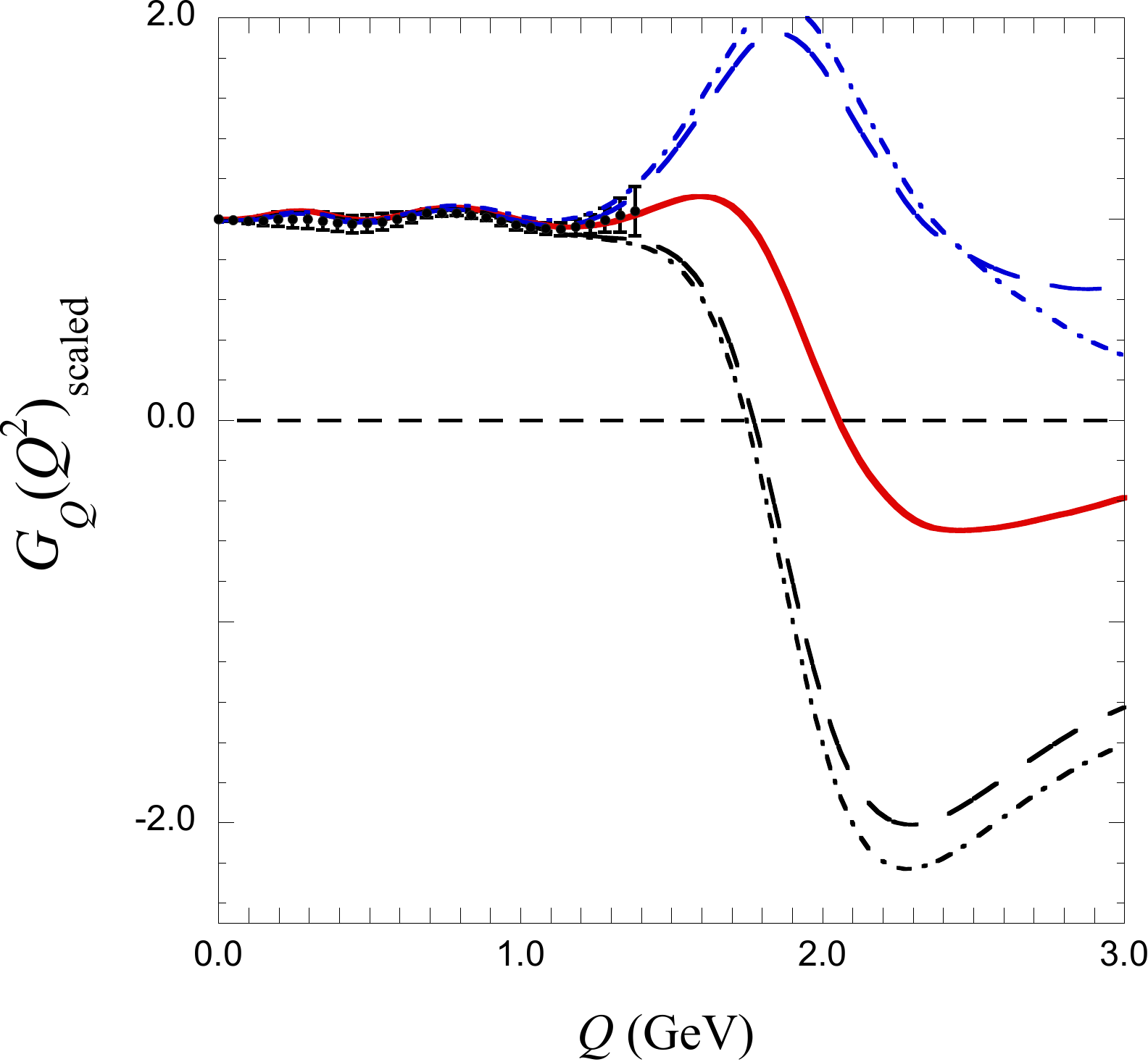} 
\includegraphics[width=2.3in]{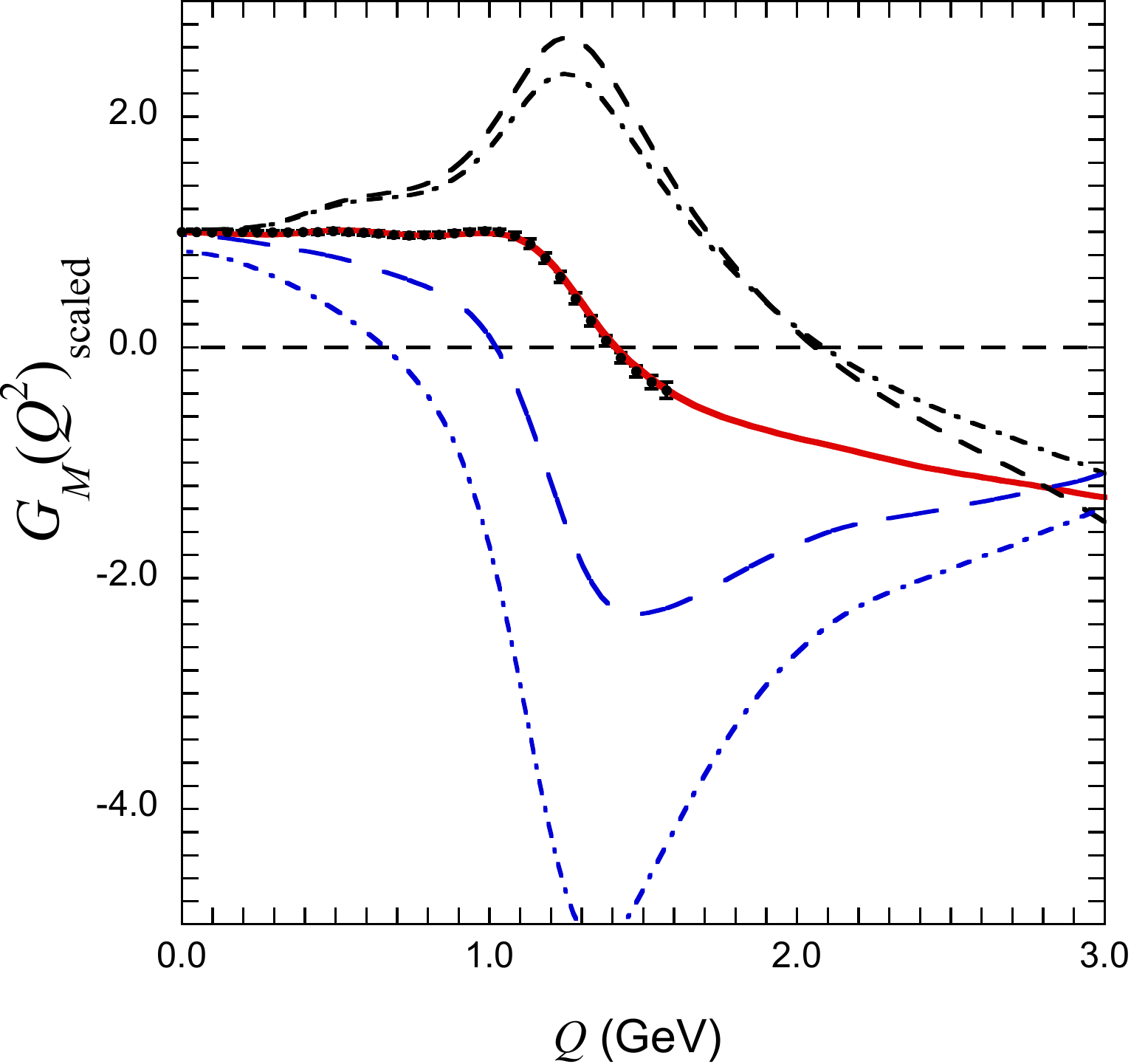} 
}
}
\caption{\footnotesize\baselineskip=10pt  Study of the validity of the RIA.  Contributions from diagram \ref{Fig2}(A) (black dot-dashed line) and \ref{Fig2}(A) + \ref{Fig2}(A$^{(2)}$) (black long dashed line) are compared to the 
contributions from \ref{Fig2}(B) (blue long dashed line).  The difference between the two blue lines is the C contribution to $\Gamma_{\rm BS}$,  defined in Eq.~(\ref{eq:BandCparts}).  The heavy solid red line is model 2D as shown in previous figures.
}
\label{fig:RIA}
\end{figure*} 

\subsection{Size of the $F_3$ and $F_4$ contributions}

The size of the $F_3$ and $F_4$ contributions was  addressed in Fig.~\ref{fig:GCMQ2}; Fig.~\ref{fig:F3F4Study} shows these effects in more detail.  Both $F_3$ and $F_4$ make comparable contributions.  It is interesting to note that the $F_4$ contribution plays a very important role in correcting the failure of model 2A at low $Q$.  In this case the $F_3$ and $F_4$ contributions are individually quite large and tend to cancel each other.

\subsection {Accuracy of the RIA}

In the absence of isoscalar interaction currents, the relativistic impulse approximation (RIA) was originally defined to be twice the contribution from diagram \ref{Fig2}(A).  The interest in this approximation arose from the idea that symmetry (the CST equations are explicitly symmetrized to ensure that $NN$ scattering satisfies the generalized Pauli principle exactly) should allow one to get the full result from the electromagnetic scattering from only one of the nucleons (multiplied by a factor of 2).  If this were true, after adding interaction currents the results from diagrams  \ref{Fig2}(A)+ \ref{Fig2}(A$^{(2)}$) should equal the results from  \ref{Fig2}(B), so that the full result would come from either of these alone, or their average, which emerges if we take 1/2 the sum of the contributions from the lower and upper half plane.

The contributions from diagrams  \ref{Fig2}(A)+ \ref{Fig2}(A$^{(2)}$) and \ref{Fig2}(B) are compared in Fig.~\ref{fig:RIA}.  The contributions from \ref{Fig2}(B) is given in two parts: the B and C contributions discussed in Eq.~(\ref{eq:BandCparts}).  The B contributions (labeled with the blue dot-dashed line) and the explicitly off-shell C contribution.   The sum of these contributions, the total from diagram (B), is the long dashed blue line.  The average of the two long dashed lines (black and blue) is the total result for model 2D.

I conclude from this figure that the RIA disagrees the the magnetic form factor even at low $Q$, but that it works reasonably well at low momentum transfer for the two charge form factors.  In any case, it is not good enough to be a replacement for the full theory, as was hoped at one time.

\begin{figure*}
\centerline{
\mbox{
\includegraphics[width=3in]{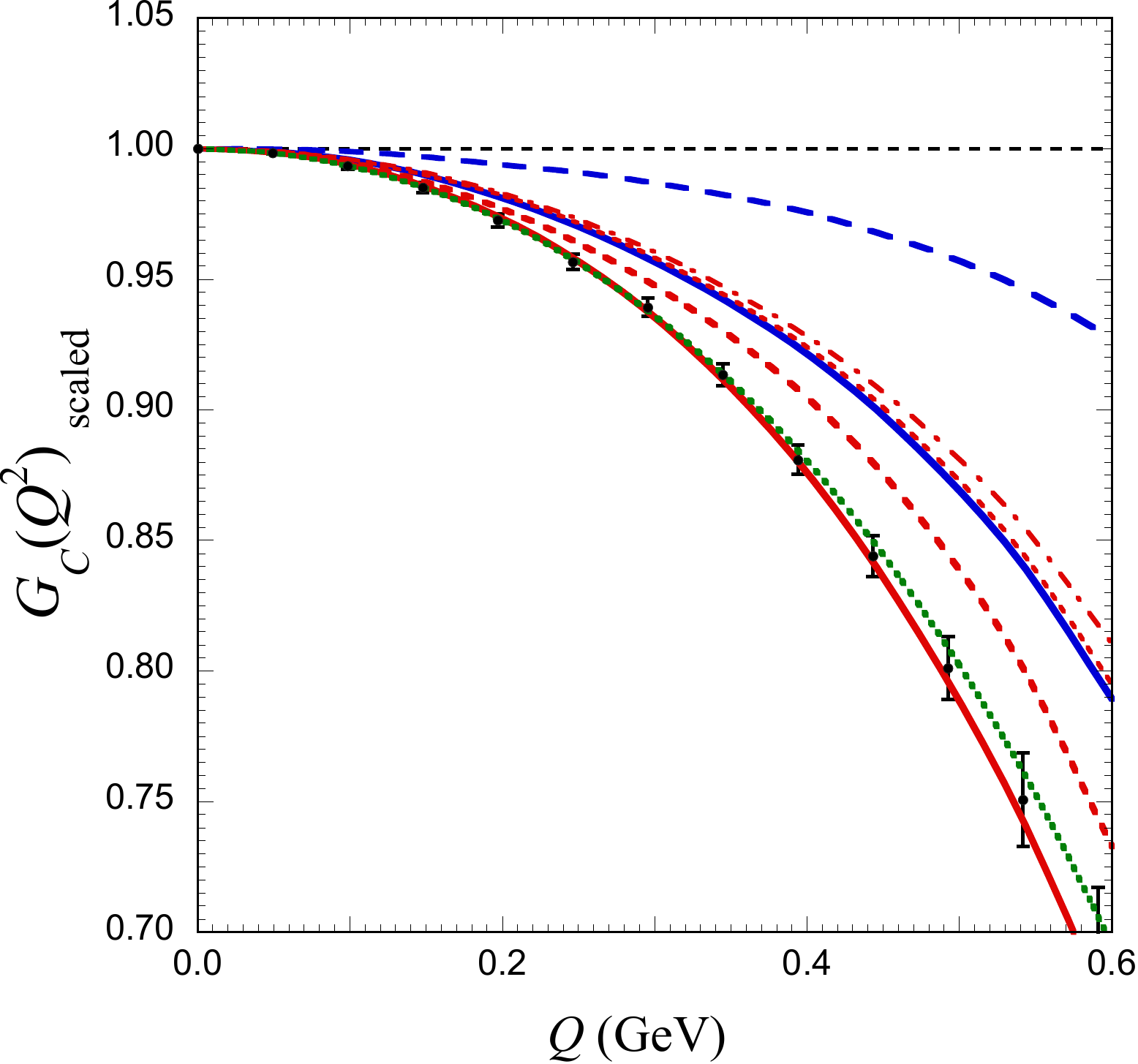} 
\hskip 0.5in  plus 0in minus 0in    
\includegraphics[width=3in]{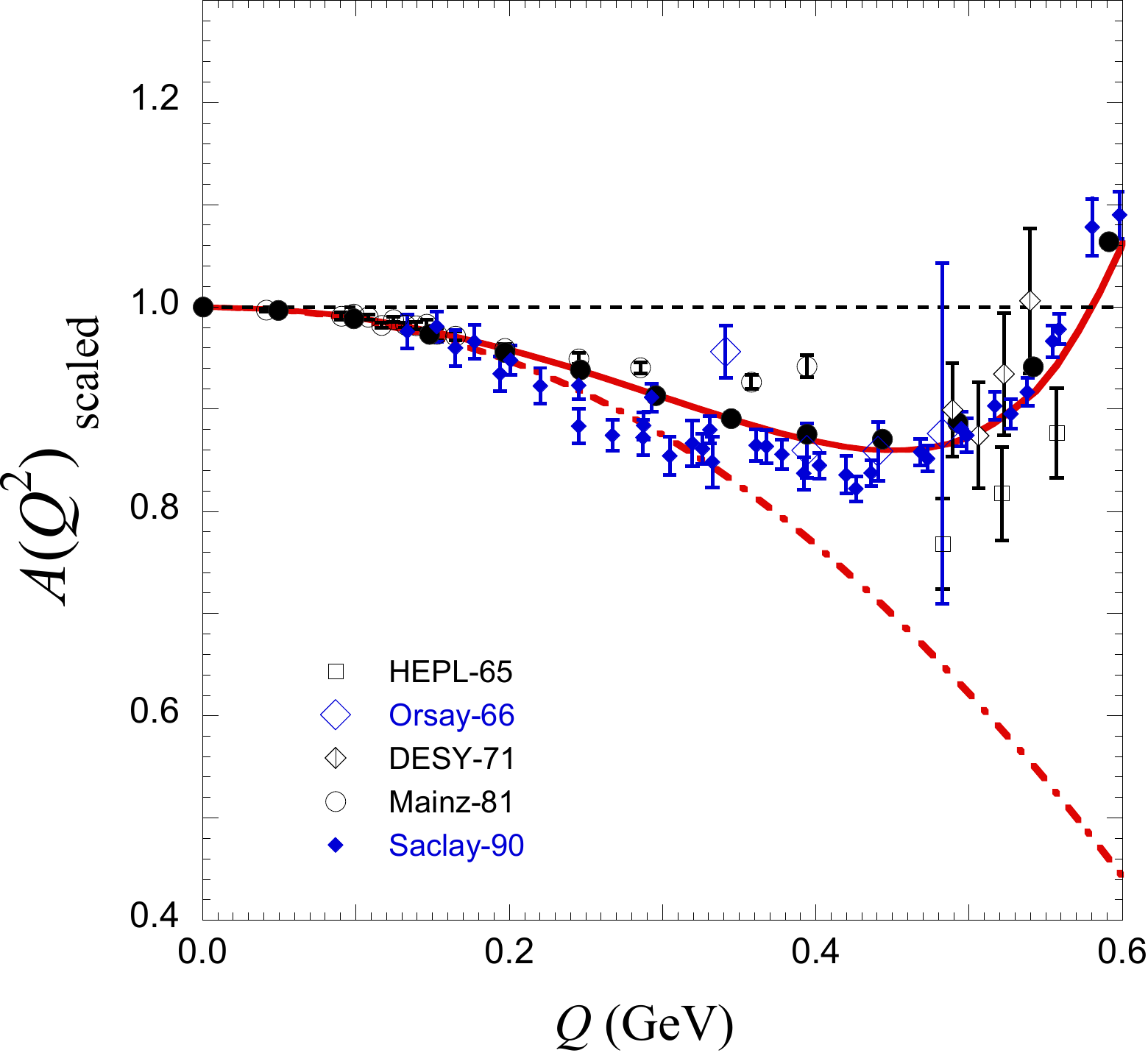} 
}
}
\caption{\footnotesize\baselineskip=10pt  Study of relativistic effects.  The left panel is scaled by $G_C^{\rm NR}$ given in Eq.~(\ref{eq:NRff}); the right panel is scaled by $(G_C^{\rm NR})^2$.  In both panels the small black dots are the Sick GA (with error bars) and the thick sold red lines are the predictions for model 2D.  The other curves are discussed in the text. 
}
\label{fig:RStudy}
\end{figure*} 

\section{Relativistic effects} \label{sec:Rel2}

Some in the electron scattering community still believe that relativistic effects are small in electron deuteron scattering and that it is possible to use deuteron wave functions calculated from the Schr\"odinger equation to study $ed$ elastic scattering.  Casper and I argued over 50 years ago \cite{Casper:1967zz}  that relativistic corrections were important when using deuteron scattering data to draw precise conclusions, and in this section I will review this issue in detail.


I focus only on the observables $G_C$ and $A$ at small $Q^2$, where it might be assumed that a nonrelativistic calculation would be reliable.  The nonrelativistic theory for $G_C$  gives
\bea
G_C^{\rm NR}&&(Q^2)= \int_0^\infty dr\left[u^2(r)+w^2(r)\right] j_0(\tau_0)
\nonumber\\
=&&\frac12\int_0^\infty k^2 dk \int_{-1}^1 dz \nonumber\\
&&\times\left[u(k_+)u(k_-)+P_2(\hat {\bf k}_+\cdot \hat{\bf k}_-)w(k_+)w(k_-)\right]\qquad \label{eq:NRff}
\eea 
where $\tau_0 = \frac12{rQ}$, $u$ and $w$ are the $S$ and $D$ state wave functions, and
\bea
&&{\bf k}_\pm={\bf k}\pm\frac14{\bf q}
\nonumber\\
&&k_\pm^2=k^2\pm \frac12 z\, k Q +\frac{Q^2}{16}
\nonumber\\
&&\hat {\bf k}_+\cdot \hat{\bf k}_-= \frac{16 k^2-Q^2}{16k_+k_-}\, . \label{eq:kpm}
\eea
[Beware that the ${\bf k}_\pm$ defined above differs significanty from the $\tilde{\bf k}_\pm$ defined in Eq.~(\ref{eq:ktildepm}).]    This momentum space nonrelativistic result emerges naturally from the nonrelativistic limit of the CIT.  This is a {\it very general\/} feature of this theory, and provides an excellent starting point for the study of relativistic effects.  To get the right limits, one must be very careful to use the correct nonrelativistic transformations: argument shift (\ref{eq:R2nr}) for the (A) diagram and (\ref{eq:R2nrB}) for the (B) diagram.  Both the (A) and (B) diagrams give exactly the same nonrelativistic limit, a limit where the RIA is accurate.

The size of various contributions, {\it scaled\/} by the nonrelativistic expression (\ref{eq:NRff}), is shown in the left panel of Fig.~\ref{fig:RStudy}.  The blue dashed line replaces the nonrelativistic  argument shifts that appear in (\ref{eq:NRff}), and were derived in (\ref{eq:R2nr}), with the fully relativistic ones (\ref{eq:R2}).  Note that this effect alone accounts for about a 4\% correction at $Q\simeq 0.4$ GeV, about eight times the size of the error in the Sick GA.  The blue solid line shows the result obtained from the full calculation of $G_C$ if only $u$ and $w$ wave functions are included.  At $Q\simeq 0.4$ this produces a discrepancy of almost 10\% with the nonrelativistic calculation.  All changes after this begin to go beyond relativistic kinematics.  Adding the $v_t$ and $v_s$ terms moves the result to the red dot dashed line, and adding the C contributions from the (B) diagram  moves the total to the red short dashed line, both small effects.  A bigger change occurs when we add in the A(2) diagram and all contributions from the $F_2$ nucleon form factor, bring the result to red longer dashed line.   Finally adding the contributions from the off-shell  form factors $F_3$ and $F_4$ brings us to the final result for model 2D, the heavy solid red line.  The green short dashed line is the function
\bea
G_C^{\rm fit}=1 - \frac{Q^2}{1.5}-\frac{Q^4}{2}\, ,
\eea
which gives a rough estimate of the size of all of the effects.  

The size of the relativistic argument shift alone is about 4 times smaller than the total shift, or about $Q^2/6$, comparable to the result $Q^2/8$ that Casper and I found over 50 years ago.  For comparison, the recoil effect of the deuteron itself is very much smaller
\bea
\frac1{D_0}=1-\frac{Q^2}{8m_d^2}\simeq 1-\frac{Q^2}{32}\, .
\eea
Because the kinematics and the relativistic shifts in the arguments of the wave functions (that add up to the solid blue line in Fig.~\ref{fig:RStudy}) can explain only about 1/2 of the total shift, it is clear that an accurate theoretical interpretation  of the data requires the use of a relativistic theory, even at the smallest values of $Q^2$.

The right panel of Fig.~\ref{fig:RStudy} shows theory and data for the structure function $A$, all scaled by $(G_C^{\rm NR})^2$.   The red dot-dashed line is $G_C^2$ of model 2D while the thick solid line is the full calculation of $A$ using all form factors from model 2D.  The panel shows that the other contributions to $A$ coming from $G_M^2$ and $G_Q^2$ begin to become important at $Q\gtrsim 0.2$ GeV.

\begin{table}
\caption{Separate contributions to the deuteron static moments from the diagrams shown in Fig.~\ref{Fig2}. 
} 
\label{tab:Ia}
\begin{ruledtabular}
\begin{tabular}{ll}
A & Total from diagram (A) with \\
&full $f_{00}, g_{00}$ given in Eq.~(\ref{eq:f0g0})\\[0.05in]
A$_0$ & Diagram (A) with $f_{00}=1, g_{00}=0$  \\[0.05in] 
A$-$A$_0$ & Total $h$ dependence from diagram (A)\\[0.05in]
A$_2$ & Diagram (A$^{(2)}$), calculated using \\
& Eq.~(\ref{eq:cstequation2}) with the interaction $V^{(2)}$\\[0.05in]
B, C & The two parts of  diagram (B) [the B and C terms in\\
&  the decomposition (\ref{eq:BandCparts})] with $\widehat \Gamma_{BS}$ calculated\\
&   using  Eq.~(\ref{eq:hatgamma2}) 
\\[0.05in] 
B$_0$, C$_0$ & The two parts of  diagram (B) with $k_0$ fixed  \\
&   at $E_k$ in $\widehat \Gamma_{BS}$,  but not in $h(p)$  \\[0.05in] 
B$_h$, C$_h$ & The dependence of $h(p)$ on $k_0-E_k$ in the   \\ 
& two  parts of diagram  (B)  
\\[0.05in]
B$_0-$B$_h$, & Removes the dependence of $h(p)$ on $k_0-E_k$ from \\
C$_0-$C$_h$& B$_0$ and C$_0$, leaving $k_0=E_k$ everywhere (on-shell) \\[0.1in]
%
\tableline \\
on-shell 
&  A$_0$ +  B$_0-$ B$_h$ +  C$_0 -$ C$_h$; \;$z_\ell^2$ terms, Refs.~II \cr
& 0.286 (1+$Q^\Delta_{\rm NR}+ Q_{\rm Rc}+Q_P+Q_\chi$), Ref.~III  \\[0.05in]
$h$ 
& A$-$A$_0$+ B$_h$ + C$_h$;\;  $a_\ell z_\ell^2$ terms, Ref.~II; \cr
&$0.286\, Q_{h'}$, Ref.~III\\[0.05in]
$V^{(2)}\qquad$ 
&  A$_2$;\;  $z_\ell z_\ell^{(2)}$ terms, Ref.~II;  0.286 $Q_{V_2}$, Ref.~III 
\\[0.05in] 
off-shell 
& B$-$B$_0$ + C $-$C$_0$;\;  $z_\ell \widehat z_\ell$ terms, Ref.~II; \cr
& 0.286 $(Q_{V_1}+Q_{\rm int})$, Ref.~III\cr
& (includes the $V^{(1)}$ current)\\[0.05in]
TOTAL & A+A$^{(2)}$+B+C\\[0.05in]
\end{tabular}
\end{ruledtabular}
\end{table}
%

\section{The static moments}  \label{sec:qis0}

The form factors at $Q^2=0$ give the charge, magnetic, and quadrupole moments in units reported in Eq.~(\ref{eq:A4}).  
Using the exact equations, there is no need to expand the analytic results around $Q^2=0$ as we did in Refs.~II and III.    However, comparison of the two different calculations uncovered some errors in Ref.~II, and I now find that the new value for the magnetic moment predicted by model WJC2 is in precise agreement with the measured result.  In addition the new, more accurate values of the quadrupole moment differ from the experimental values by over 1\%, with no significant  difference between the predictions of the two models, in disagreement with the conclusions of Ref.~III.  

Various contributions to the static moments are defined in Table  \ref{tab:Ia}.  Here, in order to provide details that may be of use to future investigators, I also report some contributions that I did not study in the previous references.  Tables \ref{tab:charge2} -- \ref{tab:quad2} compare the results obtained from the exact form factors with the results obtained from the approximate expansions reported in Refs.~II and III (and for the magnetic moment, in Appendix \ref{sec:errorsinII}). 

\subsection{Charge and magnetic moment}

In Ref.~II I conjectured that the errors in the expansions should be about 0.002.  As shown in Table \ref{tab:charge2}, the calculations of the charge agree to better than this, but the magnetic moment presents a more complicated picture.  I originally found such large disagreements with the expansions for the magnetic moment reported in Ref.~II that I redid them and found the corrected results given in Appendix \ref{sec:errorsinII}.  Table \ref{tab:mag2} shows that the new expansion disagrees with the exact results by about 0.002 for several terms but there are discrepancies as large as 0.007 (0.7\%) with others.  I believe that the major source of this discrepancy is the expansion of the nucleon kinetic energy
\bea
\frac{E_k}{m}\simeq 1+\frac{k^2}{2m^2} - \frac{k^4}{8m^4}+\cdots
\eea 
Since $k^4$ terms were dropped, the discrepancy could be as large as 0.007 if the terms conspire to make the coefficient of the $k^4$ term of the order of unity (and not 1/8) and the mean momentum of the nucleon is about 300 MeV.  In any case, the expansions are not as reliable as I expected.  The remarkable new result is that the magnetic moment for model  WJC2 is in very good agreement with experiment, differing by only 0.07\%.

\subsection{Quadrupole moment}

The comparison of the quadrupole moment with the expansions reported in Ref.~III does not fare much better.  Here I originally estimated the error to be about $0.2\%$ or a $\delta Q$ of 0.0006, and a comparison with Table \ref{tab:quad2} shows that this seems to be accurate for the small terms, but fails for the largest terms with an error of about 0.002, or about 1\% (similar to that found for the magnetic moment).  However. since all terms seem to have similar signs and magnitudes, there is no reason to expect an error as I did for the magnetic moment, and I did not recalculate the expansions given in Ref.~III.  The new conclusion here is that the two models have similar quadrupole moments, differing by about 1.5\% from the experimental result.

\begin{table}
\caption{The contributions to the deuteron charge (or normalization).  Since C$_h=0$ it is not shown.  }
\label{tab:charge2}
\begin{ruledtabular}
\begin{tabular}{lrrrr}
Quantity & \multicolumn{2}{c}{WJC1} & \multicolumn{2}{c}{WJC2} \\[0.05in]
 & 1B$\quad$ & Ref.~II & 2D$\quad$  & Ref.~II  \\[0.05in]
on-shell ($k_0=E_k$)  & 1.0547 & 1.055 &1.0231 &1.023 \\[0.05in]
$h$ dependence &  0.0245 & 0.025 & 0.0176  & 0.018 \\[0.05in]
$V^{(2)}$ current & $-$0.0228 &$-$0.023 & $-$0.0111 & $-$0.011\\[0.05in]
off-shell ($k_0\ne E_k$) & $-$0.0562 & $-$0.057&$-$0.0297 & $-$0.030\\[0.05in]
TOTAL & 1.0002 &1.000 &1.0000 & 1.000\\[0.05in]
\tableline\\
2 $\times$ A$_0$ &1.0547 &1.055&  1.0231 &   1.023\\[0.05in]
2 $\times$ (A$-$A$_0$) &0.0245 &0.025&  0.0176 &   0.018\\[0.05in]
2 $\times$ B &0.9693 &---&  0.9835 & ---\\[0.05in]
2 $\times$ C &$-$0.0025 &--- & $-$0.0021  & --- \\[0.05in]
2 $\times$ B$_0$ & 1.0816&---& 1.0428  & ---\\[0.05in]
2 $\times$ C$_0$ & $-$0.0025& ---&$-$0.0021   & --- \\[0.05in]
2 $\times$ B$_h$ &0.0245 &0.025& 0.0176  &   0.018\\[0.05in]
\end{tabular}
\end{ruledtabular}
\end{table}
%
 
\begin{table}
\caption{The contributions to the  deuteron magnetic moment, $\mu_d=
m G_M(0)/m_d$ (in nuclear magnetons)  The experimental value is 0.8574. }
\label{tab:mag2}
\begin{ruledtabular}
\begin{tabular}{lrrrr}
Quantity & \multicolumn{2}{c}{WJC1} & \multicolumn{2}{c}{WJC2} \\[0.05in]
 &1B$\quad$ & App \ref{sec:errorsinII}& 2D$\quad$   & App \ref{sec:errorsinII} \\[0.05in]
on-shell ($k_0=E_k$)   & 0.8985 & 0.8812 &0.8643 & 0.8630 \\[0.05in]
$h$ dependence & 0.0123 & 0.0145 & 0.0112 & 0.0092\\[0.05in]
$V^{(2)}$ current & $-$0.0156 & $-$0.0167 & 0.0004 & 0.0000 \\[0.05in]
off-shell ($k_0\ne E_k$)  & $-$0.0289 & $-$0.0170&  $-$0.0180 &$-$0.0129 \\[0.05in]
TOTAL & 0.8663 & 0.8620& 0.8580 & 0.8594\\[0.05in]
error & 0.0089 & 0.0046& 0.0006 & 0.0020 \\[0.05in]
error (\%) &1.04\% & 0.48\%& 0.07\% & 0.23\%\\[0.05in]
\tableline\\
2 $\times$ A$_0$ &0.9155 &&  0.8646 &   \\[0.05in]
2 $\times$ (A$-$A$_0$) & 0.0141&&  0.0158 &   \\[0.05in]
2 $\times$ B &0.7193 &&  0.7163 &   \\[0.05in]
2 $\times$ C & 0.1150& & $-$0.1183  &  \\[0.05in]
2 $\times$ B$_0$ & 0.7904&& 0.7553  &   \\[0.05in]
2 $\times$ C$_0$ & 0.1017& & 0.1154   &   \\[0.05in]
2 $\times$ B$_h$ &0.0145 && 0.0093  &   \\[0.05in]
2 $\times$ C$_h$ &$-$0.0039 && $-$0.0026  &   \\[0.05in]
\end{tabular}
\end{ruledtabular}
\end{table}
%

\begin{table}
\caption{The contributions to the deuteron quadrupole moment $Q_d=m_d^2 G_Q(0)$ (in fm$^{-2}$). The experimental value is 0.2859(6).  }
\label{tab:quad2}
\begin{ruledtabular}
\begin{tabular}{lrrrr}
Quantity & \multicolumn{2}{c}{WJC1} & \multicolumn{2}{c}{WJC2} \\[0.05in]
 & 1B$\quad$ & Ref.~III & 2D$\quad$   & Ref.~III  \\[0.05in]
on-shell ($k_0=E_k$) &0.2831 & 0.285& 0.2815 & 0.284 \\[0.05in]
$h$ dependence & 0.0011& 0.000& 0.0007 & 0.000  \\[0.05in]
$V^{(2)}$  current & $-$0.0009 & $-$0.001& $-$0.0002 & 0.000  \\[0.05in]
off-shell $k_0\ne E_k$  & $-$0.0014& $-$0.005& $-$ 0.0003 & 0.000 \\[0.05in]
TOTAL  & 0.2820 & 0.279 & 0.2817  & 0.284 \\[0.05in]
error  & $-$0.0039  & $-$0.007 & $-$0.0042  & $-$0.0019  \\[0.05in]
error (\%) & $-$ 1.38\%& $-$2.4\%  & $-$1.49\%& $-$0.7\%\\[0.05in]
\tableline\\
2 $\times$ A$_0$ &0.2825 &&  0.2815 &   \\[0.05in]
2 $\times$ (A$-$A$_0$) & 0.0014 &&  0.0008 &   \\[0.05in]
2 $\times$ B &0.2835 &&  0.2829 & \\[0.05in]
2 $\times$ C & $-$0.0017& & $-$0.0016  &  \\[0.05in]
2 $\times$ B$_0$ & 0.2863&& 0.2836  & \\[0.05in]
2 $\times$ C$_0$ &$-$0.0017 & &$-$0.0016   &  \\[0.05in]
2 $\times$ B$_h$ &0.0009 && 0.0006  &   \\[0.05in]
2 $\times$ C$_h$ &$-$0.0001 && $-$0.0000  &   \\[0.05in]
\end{tabular}
\end{ruledtabular}
\end{table}
%

\begin{table}
\begin{minipage}{3.5in}
\caption{Contributions to the deuteron radius for model 2D.  The experimental value of 2.130(10) fm is taken from Ref.~\cite{Sick:1998cvq}.  The last row of the table is model 2D with $dG_C/dQ^2=-19.36$ (GeV)$^{-2}$. }
\label{tab:rms}
\begin{ruledtabular}
\begin{tabular}{lcc}
Approximation  &$R^2_{\rm rms}$ (GeV)$^{-2}$ & $R_{\rm rms}$ fm\cr
NR  &116.1 &2.122 \cr
NR with (A) shift & 117.0 & 2.131 \cr
All $u, w$ &116.6 & 2.128 \cr
add $v_t, v_s$ & 116.7 & 2.128 \cr
add C terms &116.7 & 2.128\cr
add A$^{(2)}$ and $F_2$ & 116.3& 2.124\cr
add $F_3$ and $F_4$ &116.2 & 2.123 \cr
\end{tabular}
\end{ruledtabular}
\end{minipage}
\end{table}
%

\begin{figure}
\centerline{
\mbox{
\includegraphics[width=3in]{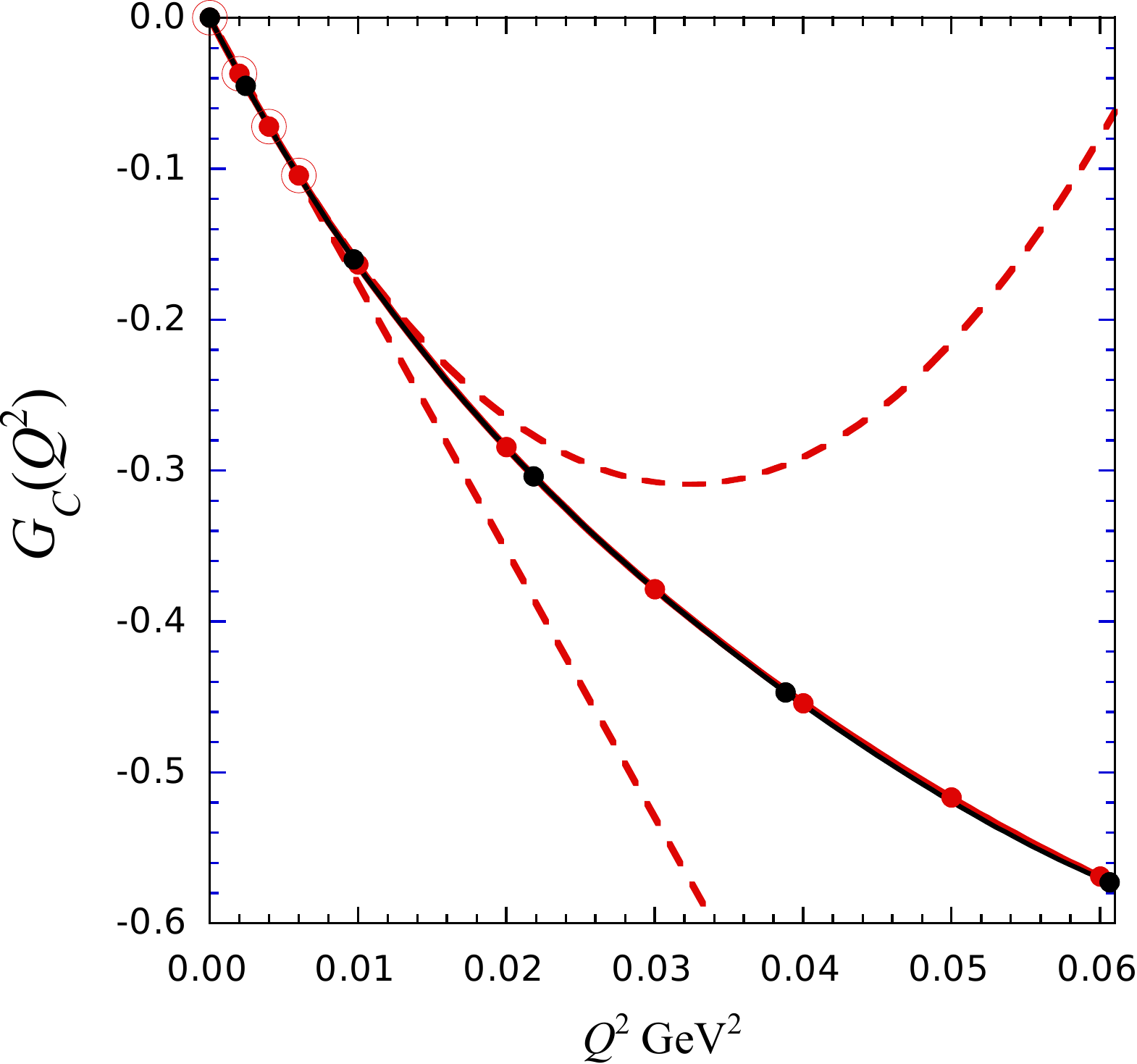} 
}
}
\caption{\footnotesize\baselineskip=10pt  Study of the dependence of the deuteron charge form factor on $Q^2$ at very small $Q^2$.  The two red dashed lines are linear and quadratic fits to the lowest four points represented by red dots surrounded by a red circle, all at $Q^2<0.1$.  The solid black lines are (indistinguishable) quadratic and cubic fits to all the red dots, including 6 beyond  beyond $Q^2=0.1$.   The black dots are the Sick GA.
}
\label{fig:RMS}
\end{figure} 

\subsection{Rms radius}

The rms radius of the deuteron is, by definition,
\bea
R^2_{\rm rms} = -6\frac{d}{dQ^2}G_C(Q^2)\, .
\eea
The values of $R^2_{\rm rms}$ (in GeV$^{-2}$) and $R_{\rm rms}$ (in fm) are shown in Table \ref{tab:rms}.  Note that the corrections from the relativistic effects discussed in Sec.~\ref{sec:Rel2} are very small.

Perhaps it is interesting to see how a linear fit to the $Q^2$ dependence of the form factor might affect how the radius would be extracted from experimental data.    Fig.~\ref{fig:RMS} shows four fits, with parameters listed in Table \ref{tab:rmsfits}, to a set of theoretical points calculated using model 2D.   The large variation in the derivative, $c_2$, shows how difficult it is to get the slope at $Q^2=0$ from the fits.  Using the 10 points seems to be less reliable than the four lowest points, and it is a surprise to me that the quadratic fit to the lowest points, which is completely unreliable at higher $Q^2$, gives a $c_2$ closest to the derivative.  

\begin{table}
\begin{minipage}{3.5in}
\caption{Fitting parameters for the four curves of the form $f(Q^2)=\sum_{n=1}^{n=3}c_n Q^{n}$ shown in Fig.~\ref{fig:RMS}.  Recall that direct calculation of the derivative gave $c_2=-19.36$.}
\label{tab:rmsfits}
\begin{ruledtabular}
\begin{tabular}{ccccc}
& 4 points & 4 points  &  10 points & 10 points \cr
$c_2$& $-17.667$ &$-19.202\;\;$ & $-15.701\;\;$& $-18.318\;\;$ \cr
$c_4$ & 0  &  293.13 & 106.05 & 229.76 \cr
$c_6$& 0 & 0 & 0 & $-1372.6\;\;$ \cr
\end{tabular}
\end{ruledtabular}
\end{minipage}
\end{table}
%

\section {conclusions and Discussion} \label{sec:5}

\subsection{Major new results}

This is the first time the deuteron form factors have been calculated using models WJC1 and WJC2, which give precision fits to the $np$ data base with $\chi^2$/datum $\approx 1$.  These models use a kernel with a dependence on the momentum of the off-shell particle 
and therefor require isoscalar interaction currents in order to conserve the two-body $np$ current.  At first it seems that the existence of these currents would make it impossible to make any unique predictions for the form factors, but I showed in Ref.~I that using principles of {\it simplicity\/} and {\it picture independence\/} it is possible to all but uniquely fix these currents in terms of the already determined parameters of the $np$ models.  These results fixed the currents at $Q^2=0$, and I show here that the exact calculations of  
the static moments of the deuteron, calculated  without adjustable parameters (assuming $F_4(0)=0$), give very good predictions.  [If $F_4(0)\ne0$ its effect on the magnetic moment is much larger than the quadrupole moment,  justifying the choice $F_4(0)=0$.]   

In addition, I believe that this is the first time anyone has obtained a {\it precision\/} fit to all of the deuteron elastic scattering data (where precision in this case also means $\chi^2$/datum $\approx1$).  I immediately qualify this remark: such a fit would be {\it impossible\/} without using the Global Analysis of Ingo Sick.  To obtain this Global Analysis, Sick reanalyzed all for the  data for the invariant functions $A(Q^2)$, $B(Q^2)$, and the polarization transfer function $T_{20}(Q^2)$.  My fit is actually to the Sick GA; as I have discussed briefly above, direct fits to the published data cannot give such a low $\chi^2$ because  the published data is not consistent to this level (recall Table \ref{tab:chi2A}).  These issues deserved to be reviewed by other scientists.

A third major new result is a {\it prediction\/} for the neutron charge form factor, $G_{En}(Q^2)$, in the region $Q^2\gtrsim 2$ (GeV)$^2$  where it has not been measured experimentally (see Fig.~\ref{fig:GEn}, and model CST1 in Table \ref{tab:GEnmodels}).  

The last new result I want to highlight is the determination of two new off-shell nucleon form factors $F_3(Q^2)$ and $F_4(Q^2)$, defined in Eqs.~(\ref{3.1}) and (\ref{eq:Fimua}).  These new form factors 
can contribute only when {\it both\/} the incoming and the outgoing nucleon is off-shell, and thus contribute only to the diagram  Fig.~\ref{Fig2}(A) where this is possible.  The form factor $F_3$, known for a long time, cannot be zero because current conservation {\it requires\/} $F_3(0)=1$.  Form factor $F_4$  (new to this paper and one of many that can appear in the most general expansion of the off-shell nucleon current), 
is purely transverse and hence cannot be constrained by current conservation in any way.  However, {\it balance\/} between the on-shell form factors $F_1(Q^2)$ and $F_2(Q^2)$ provides an {\it ab initio\/} argument for including $F_4$: since $F_3$ is required to complement $F_1$, it is not a stretch to argue that $F_4$ should be included to complement $F_2$, even though neither $F_2$ nor $F_4$ can be constrained by current conservation.  The data will determine these form factors; as it turns out $F_2$ can directly measured by electron nucleon scattering, while $F_4$ can only be measured by electron scattering from a composite nucleus, the deuteron being the simplest.   

In this paper model WJC2 uses the Sick GA at intermediate $Q^2$ to predict the  form factors $F_3$ and $F_4$.  The data is insensitive to precise values of $F_4$ at low $Q^2$ (I assumed $F_4(0)=0$, a value that would likely emerge from a comparison of the static moments, but not investigated here) and there is insufficient data at $Q^2 \gtrsim 2$ (GeV)$^2$ for a prediction, so I adjusted fits so that the large $Q^2$ behavior of these form factors would be small.  These introduce small uncertainties which I cannot estimate.  The reason for not using the model WJC1 to extract $F_3$ and $F_4$ was discussed in Sec.~\ref{sec:GEn}.  

Table \ref{tab:Body} gives numerical values for the 12 model 2D body form factors $D_{X,i}(Q^2)$ introduced in Eq.~(\ref{eq:expbody}).   The reader may use these to extract her own nucleon form factors from the data.  

\squeezetable
\begin{table*}
\caption{Body form factors for the model 2D, defined in Table \ref{tab:models}}
\hspace*{-0.2in}
\begin{tabular}{rrrrrrrrrrrrr}
Q \,\,& $D_{C1}\,\,$ & $D_{C2}\,\,$ & $D_{C3}\,\,$ & $D_{C4}\,\,\,\,$ & $D_{M1}\,\,$ & 
$D_{M2}\,\,$ & $D_{M3}\,\,$ & $D_{M4}\,\,\,\,$ & $D_{Q1}\,\,$ & $D_{Q2}\,\,$ & $D_{Q3}\,\,$ 
& $D_{Q4}\,\,$ \\[0.05in]
 0.001 &  1.003D+00 & -2.611D-07 & -3.334D-03 &  5.429D-10\;\; &  1.924D+00 &  1.793D+00 &  4.994D-03 & -2.247D-04\;\; &  2.548D+01 & -4.196D-01 & -2.725D-04 & -1.861D-04\;\;\cr
 0.101 &  8.576D-01 & -2.251D-03 & -3.308D-03 &  5.494D-06\;\; &  1.654D+00 &  1.527D+00 &  4.969D-03 & -2.402D-04\;\; &  2.189D+01 & -4.555D-01 & -1.918D-04 & -1.809D-04\;\;\cr
 0.201 &  5.988D-01 & -6.042D-03 & -3.232D-03 &  2.091D-05\;\; &  1.181D+00 &  1.064D+00 &  4.869D-03 & -2.844D-04\;\; &  1.540D+01 & -5.064D-01 & -1.728D-04 & -1.133D-04\;\;\cr
 0.301 &  3.847D-01 & -8.315D-03 & -3.109D-03 &  4.496D-05\;\; &  7.963D-01 &  6.922D-01 &  4.727D-03 & -3.554D-04\;\; &  1.017D+01 & -5.250D-01 & -1.888D-04 & -7.917D-05\;\;\cr
 0.401 &  2.354D-01 & -8.443D-03 & -2.945D-03 &  7.496D-05\;\; &  5.297D-01 &  4.401D-01 &  4.537D-03 & -4.483D-04\;\; &  6.642D+00 & -5.094D-01 & -1.921D-04 & -3.139D-05\;\;\cr
 0.501 &  1.356D-01 & -6.803D-03 & -2.745D-03 &  1.078D-04\;\; &  3.511D-01 &  2.759D-01 &  4.299D-03 & -5.575D-04\;\; &  4.369D+00 & -4.710D-01 & -1.559D-04 &  2.607D-05\;\;\cr
 0.601 &  6.999D-02 & -4.005D-03 & -2.517D-03 &  1.399D-04\;\; &  2.314D-01 &  1.696D-01 &  4.026D-03 & -6.759D-04\;\; &  2.905D+00 & -4.214D-01 & -1.329D-04 &  9.206D-05\;\;\cr
 0.701 &  2.753D-02 & -6.161D-04 & -2.271D-03 &  1.677D-04\;\; &  1.509D-01 &  1.010D-01 &  3.723D-03 & -7.969D-04\;\; &  1.954D+00 & -3.686D-01 & -9.700D-05 &  1.655D-04\;\;\cr
 0.801 &  4.582D-04 &  2.934D-03 & -2.014D-03 &  1.878D-04\;\; &  9.615D-02 &  5.674D-02 &  3.401D-03 & -9.133D-04\;\; &  1.330D+00 & -3.179D-01 & -5.520D-05 &  2.439D-04\;\;\cr
 0.901 & -1.602D-02 &  6.331D-03 & -1.755D-03 &  1.975D-04\;\; &  5.867D-02 &  2.803D-02 &  3.068D-03 & -1.018D-03\;\; &  9.156D-01 & -2.720D-01 & -2.272D-06 &  3.232D-04\;\;\cr
 1.001 & -2.542D-02 &  9.380D-03 & -1.503D-03 &  1.946D-04\;\; &  3.276D-02 &  9.419D-03 &  2.735D-03 & -1.105D-03\;\; &  6.341D-01 & -2.314D-01 &  5.797D-05 &  4.029D-04\;\;\cr
 1.101 & -2.979D-02 &  1.191D-02 & -1.264D-03 &  1.783D-04\;\; &  1.502D-02 & -2.510D-03 &  2.410D-03 & -1.171D-03\;\; &  4.398D-01 & -1.962D-01 &  1.235D-04 &  4.774D-04\;\;\cr
 1.201 & -3.084D-02 &  1.384D-02 & -1.043D-03 &  1.485D-04\;\; &  3.259D-03 & -9.659D-03 &  2.100D-03 & -1.213D-03\;\; &  3.051D-01 & -1.657D-01 &  1.914D-04 &  5.447D-04\;\;\cr
 1.301 & -2.998D-02 &  1.518D-02 & -8.456D-04 &  1.066D-04\;\; & -4.328D-03 & -1.347D-02 &  1.812D-03 & -1.229D-03\;\; &  2.110D-01 & -1.393D-01 &  2.594D-04 &  6.032D-04\;\;\cr
 1.401 & -2.780D-02 &  1.591D-02 & -6.730D-04 &  5.441D-05\;\; & -8.879D-03 & -1.509D-02 &  1.550D-03 & -1.220D-03\;\; &  1.457D-01 & -1.169D-01 &  3.250D-04 &  6.477D-04\;\;\cr
 1.501 & -2.479D-02 &  1.610D-02 & -5.266D-04 & -5.290D-06\;\; & -1.119D-02 & -1.529D-02 &  1.317D-03 & -1.189D-03\;\; &  1.004D-01 & -9.791D-02 &  3.850D-04 &  6.793D-04\;\;\cr
 1.601 & -2.191D-02 &  1.591D-02 & -4.047D-04 & -6.941D-05\;\; & -1.229D-02 & -1.462D-02 &  1.114D-03 & -1.136D-03\;\; &  6.860D-02 & -8.194D-02 &  4.353D-04 &  6.964D-04\;\;\cr
 1.701 & -1.899D-02 &  1.538D-02 & -3.066D-04 & -1.348D-04\;\; & -1.255D-02 & -1.351D-02 &  9.400D-04 & -1.068D-03\;\; &  4.652D-02 & -6.873D-02 &  4.755D-04 &  6.988D-04\;\;\cr
 1.801 & -1.622D-02 &  1.457D-02 & -2.300D-04 & -1.984D-04\;\; & -1.219D-02 & -1.215D-02 &  7.941D-04 & -9.879D-04\;\; &  3.138D-02 & -5.790D-02 &  5.046D-04 &  6.879D-04\;\;\cr
 1.901 & -1.371D-02 &  1.360D-02 & -1.723D-04 & -2.574D-04\;\; & -1.149D-02 & -1.070D-02 &  6.739D-04 & -8.993D-04\;\; &  2.088D-02 & -4.894D-02 &  5.229D-04 &  6.649D-04\;\;\cr
 2.001 & -1.154D-02 &  1.254D-02 & -1.296D-04 & -3.099D-04\;\; & -1.066D-02 & -9.310D-03 &  5.756D-04 & -8.059D-04\;\; &  1.375D-02 & -4.175D-02 &  5.296D-04 &  6.320D-04\;\;\cr
 2.101 & -9.675D-03 &  1.141D-02 & -9.989D-05 & -3.541D-04\;\; & -9.740D-03 & -7.972D-03 &  4.970D-04 & -7.120D-04\;\; &  8.848D-03 & -3.585D-02 &  5.265D-04 &  5.907D-04\;\;\cr
 2.201 & -8.117D-03 &  1.028D-02 & -7.934D-05 & -3.892D-04\;\; & -8.886D-03 & -6.797D-03 &  4.337D-04 & -6.199D-04\;\; &  5.557D-03 & -3.123D-02 &  5.139D-04 &  5.441D-04\;\;\cr
 2.301 & -6.862D-03 &  9.182D-03 & -6.584D-05 & -4.148D-04\;\; & -8.079D-03 & -5.735D-03 &  3.834D-04 & -5.318D-04\;\; &  3.260D-03 & -2.744D-02 &  4.938D-04 &  4.942D-04\;\;\cr
 2.401 & -5.746D-03 &  8.061D-03 & -5.774D-05 & -4.311D-04\;\; & -7.264D-03 & -4.790D-03 &  3.437D-04 & -4.494D-04\;\; &  1.817D-03 & -2.447D-02 &  4.681D-04 &  4.419D-04\;\;\cr
 2.501 & -4.754D-03 &  6.925D-03 & -5.210D-05 & -4.390D-04\;\; & -6.440D-03 & -3.932D-03 &  3.107D-04 & -3.733D-04\;\; &  8.937D-04 & -2.206D-02 &  4.370D-04 &  3.912D-04\;\;\cr
 2.601 & -3.896D-03 &  5.777D-03 & -4.913D-05 & -4.392D-04\;\; & -5.633D-03 & -3.152D-03 &  2.847D-04 & -3.060D-04\;\; &  2.644D-04 & -2.001D-02 &  4.041D-04 &  3.410D-04\;\;\cr
 2.701 & -3.160D-03 &  4.668D-03 & -4.683D-05 & -4.331D-04\;\; & -4.874D-03 & -2.440D-03 &  2.629D-04 & -2.460D-04\;\; & -1.439D-04 & -1.823D-02 &  3.694D-04 &  2.940D-04\;\;\cr
 2.801 & -2.499D-03 &  3.589D-03 & -4.483D-05 & -4.211D-04\;\; & -4.160D-03 & -1.814D-03 &  2.444D-04 & -1.929D-04\;\; & -4.162D-04 & -1.662D-02 &  3.338D-04 &  2.501D-04\;\;\cr
 2.901 & -1.957D-03 &  2.602D-03 & -4.285D-05 & -4.056D-04\;\; & -3.538D-03 & -1.287D-03 &  2.281D-04 & -1.482D-04\;\; & -5.767D-04 & -1.522D-02 &  2.996D-04 &  2.102D-04\;\;\cr
 3.001 & -1.456D-03 &  1.665D-03 & -4.031D-05 & -3.870D-04\;\; & -2.955D-03 & -8.248D-04 &  2.132D-04 & -1.097D-04\;\; & -6.466D-04 & -1.393D-02 &  2.660D-04 &  1.753D-04\;\;\cr
 3.101 & -1.087D-03 &  8.782D-04 & -3.741D-05 & -3.671D-04\;\; & -2.502D-03 & -4.903D-04 &  1.995D-04 & -7.772D-05\;\; & -6.775D-04 & -1.282D-02 &  2.346D-04 &  1.433D-04\;\;\cr
 3.201 & -8.082D-04 &  2.024D-04 & -3.415D-05 & -3.459D-04\;\; & -2.130D-03 & -2.430D-04 &  1.872D-04 & -5.123D-05\;\; & -6.945D-04 & -1.179D-02 &  2.051D-04 &  1.163D-04\;\;\cr
 3.301 & -5.949D-04 & -3.586D-04 & -3.030D-05 & -3.245D-04\;\; & -1.841D-03 & -6.112D-05 &  1.748D-04 & -2.929D-05\;\; & -6.672D-04 & -1.095D-02 &  1.781D-04 &  9.261D-05\;\;\cr
 3.401 & -4.416D-04 & -8.563D-04 & -2.668D-05 & -3.036D-04\;\; & -1.615D-03 &  5.790D-05 &  1.635D-04 & -1.184D-05\;\; & -6.497D-04 & -1.019D-02 &  1.541D-04 &  7.254D-05\;\;\cr
 3.501 & -3.380D-04 & -1.229D-03 & -2.288D-05 & -2.838D-04\;\; & -1.446D-03 &  1.078D-04 &  1.529D-04 &  1.821D-06\;\; & -6.073D-04 & -9.538D-03 &  1.323D-04 &  5.553D-05\;\;\cr
 3.601 & -2.548D-04 & -1.590D-03 & -1.884D-05 & -2.653D-04\;\; & -1.305D-03 &  1.403D-04 &  1.426D-04 &  1.276D-05\;\; & -5.735D-04 & -8.950D-03 &  1.127D-04 &  4.119D-05\;\;\cr
 3.701 & -3.003D-04 & -1.760D-03 & -1.476D-05 & -2.474D-04\;\; & -1.289D-03 &  7.839D-05 &  1.324D-04 &  2.140D-05\;\; & -5.790D-04 & -8.425D-03 &  9.546D-05 &  2.966D-05\;\;\cr
 3.801 & -2.796D-04 & -1.976D-03 & -1.119D-05 & -2.318D-04\;\; & -1.231D-03 &  5.794D-05 &  1.235D-04 &  2.756D-05\;\; & -5.466D-04 & -8.003D-03 &  8.061D-05 &  1.973D-05\;\;\cr
 3.901 & -2.138D-04 & -2.269D-03 & -7.747D-06 & -2.179D-04\;\; & -1.131D-03 &  6.653D-05 &  1.146D-04 &  3.219D-05\;\; & -5.126D-04 & -7.516D-03 &  6.789D-05 &  1.093D-05\;\;\cr
 4.001 & -1.926D-04 & -2.433D-03 & -4.371D-06 & -2.045D-04\;\; & -1.045D-03 &  3.904D-05 &  1.063D-04 &  3.530D-05\;\; & -4.771D-04 & -7.074D-03 &  5.643D-05 &  4.836D-06\;\;\cr

 \end{tabular}
\label{tab:Body} 
\end{table*}

\subsection{Assessment}

For this assessment I return to an issue I raised in Ref.~I:  can the CST make predictions?  Stated more forcefully: if I obtain a precision fit to the {\it three\/} independent sets of  deuteron data for $A$, $B$, and $T_{20}$ by adjusting another set of {\it three\/} independent functions $F_3$, $F_4$, and $G_{En}$, in what sense does this provide any understanding?  I will discuss this issue in 4 parts:

(i) First, the independent functions are multiplied by a body form factor, and hence are constrained by the values of the body form factor itself, which depends on the $np$ dynamics of the WJC models.  If the body form factors are small or have the ``wrong'' sequence of signs for $G_C, G_M, G_Q$, this will prevent the independent functions from giving a desirable fit to all three form factors.
 
(ii)  Next, the predictions for the static deuteron moments are absolute; they are free of any parameters (because $G_{En}(0)$ and $F_3(0)$ are known, and I constrain $F_4(0)=0$).  The  low $Q^2$ behavior of $A$, $B$, and $T_{20}$ (Figs.~\ref{fig:AallQ-scaled}, \ref{fig:BallQ-scaled}, and \ref{fig:T20}, respectively) all show a complete insensitivity to the independent functions for $Q\lesssim 0.5$ GeV.  This shows that the CST gives precise predictions for all low $Q^2$ observables, largely independent of the choice of the independent functions.

(iii) Determination  of the three independent functions using model WJC1 gives values of $G_{En}$ that disagree with the data for $G_{En}$ over the entire range of $Q^2$, as shown in Fig.~\ref{fig:GEn}.  In this sense model WJC1 fails, allowing me to conclude that the prediction obtained from model WJC2, which is consistent with the data for $G_{En}$ out to the highest $Q$ point measured ($Q\simeq 1.4$ GeV), is not an accident, but a real success (the body form factors for WJC2 have the correct properties).   An experimental confirmation of the prediction for $G_{En}$ at higher $Q$ would be a further success of model WJC2.

(iv)  Finally, note that no choice of $G_{En}$ can fit the GA for $A$ at the highest $Q^2$ points (recall the small circles in Fig.~\ref{fig:GEn}).  This is either an indication that model WJC2 fails at the highest $Q^2$, or might be an indication that the GA is inaccurate at the highest points, a possibility suggested by the largest JLabA measurement for $A$ at $Q^2\simeq 6$ GeV$^2$.  Further measurements at high $Q^2$ would clarify this.  

\subsection{Alternative interpretation} 

\begin{figure}
\centerline{
\mbox{
\includegraphics[width=3in]{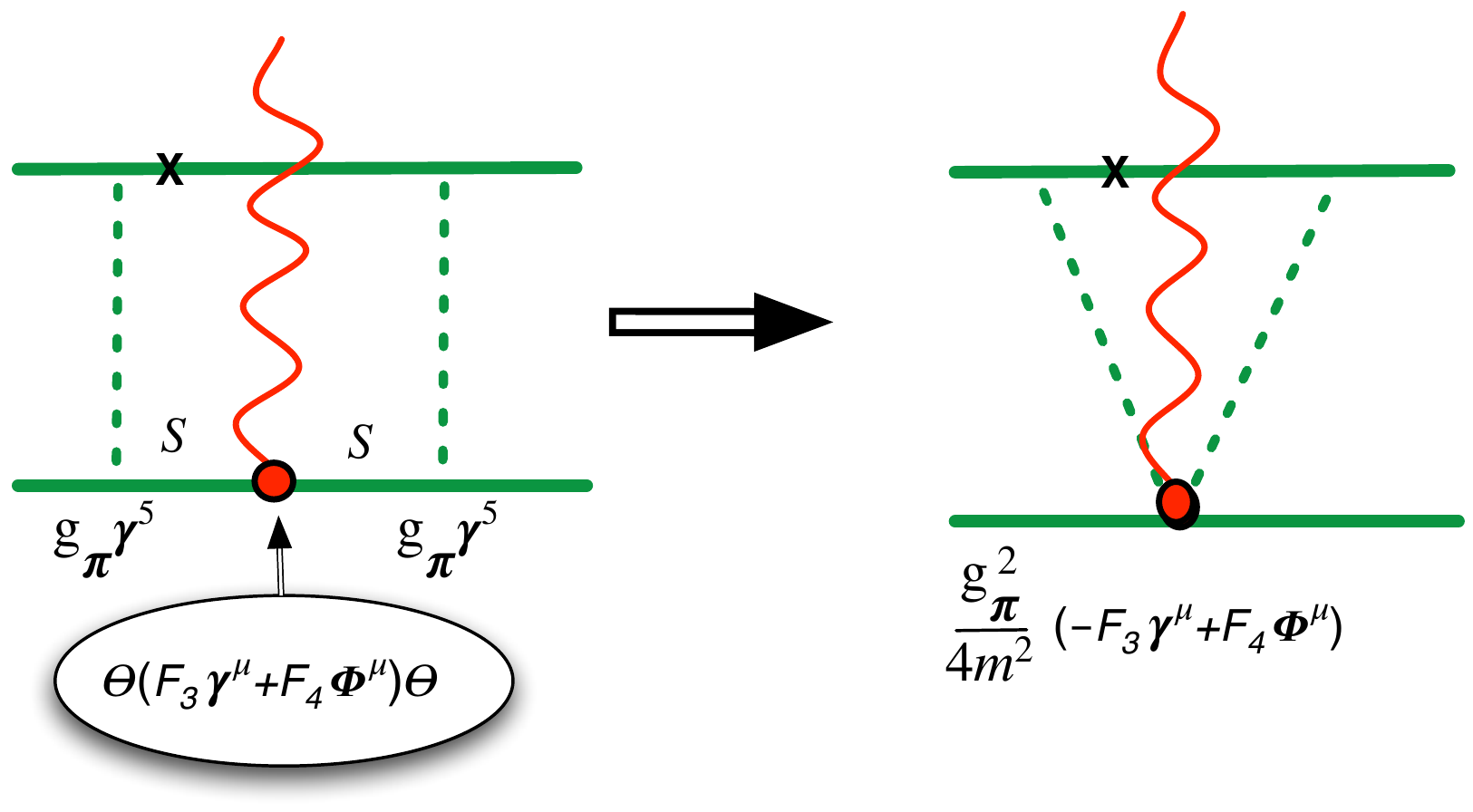} 
}
}
\caption{\footnotesize\baselineskip=10pt  Figure showing how the use of off-shell form factors (left panel) can generate IC diagrams (right panel).  In this case the removal of a one pion exchange interaction from the CST kernel with off-shell form factors is equivalent to another CST calculation with a two pion interaction current.  
}
\label{fig:Offshell}
\end{figure}

The central role played by the off-shell form factors $F_3$ and $F_4$ leads to the following question:  will the physics described by these form factors disappear in a formalism where the nucleons are always on-shell? 
The answer is ``no.''   The way the same physics is described in alternative formalisms is shown in Fig.~\ref{fig:Offshell}, where for shorthand I used
$\Phi^\mu = i \sigma^{\mu\nu}q_\nu/(2m)$.   The left panel shows, as an example, the case where the one pion exchange mechanism is the ``last'' interaction to be  factored out of the $NN$ iteration kernel, and the right panel shoes how the the projection operators $\Theta$ cancel the propagators $S$ leaving a two-pion exchange term with an effective interaction at the $2\pi NN$ vertex.  

This correspondence mirrors that shown in Fig.~8 of Ref.~I.  In that case the off-shell sigma coupling cancelled the nucleon propagators.  Here the details are very different, but the way in which off-shell projectors cancel propagators reducing the effective interaction of the off shell particle to a point interaction (modified by $F_3$ or $F_4$)  is the same.  It is another example of the theorem I proposed in Ref.~I:  a theory with {\it off-shell couplings\/} is equivalent  to  another theory with {\it no off-shell couplings plus an infinite number of very complex interaction currents\/}.

This comparison provides two further insights.  First, I showed in Ref.~I that the momentum dependent couplings in the kernel did not generate any two-pion exchange currents, while,  as the example in Fig.~\ref{fig:Offshell} shows, the off-shell form factors do.   Second, since the  comparison suggests the physical role for $F_4$ is to generate two-pion exchange currents  (as well as exchange currents involving other pairs of mesons) perhaps a more natural scale for the $\Phi^\mu$ factor multiplying $F_4$ is $1/m_\pi$ (instead of $1/m$ which is merely a carry over from the factors multiplying $F_2$).  If this were the case the $F_4$ form factor would be $m/m_\pi \sim  7$ times smaller that the curves shown in Figs.~\ref{fig:F3-F4} and \ref{fig:F4}.  The new $F_4$ would be more comparable in size to $F_3$.

\subsection{Outlook}

I remind the reader that model VODG provides a very good explanation of the data for $A, B$, and $T_{20}$.  
However, the revised model IIB which is the basis of the VODG calculation, does not  give a high precision fit to the $np$ data.  The newer high precision fits provided by models WJC1 and WJC2, with their momentum dependent couplings and accompanying exchange currents, required a completely new calculation.

The fits to the off-shell form factors and the prediction of a new high $Q^2$ behavior of $G_{En}$  completely fixes model WJC2, and allows for a precise prediction, without any free parameters, for the rescattering term in deuteron electrodisintegration at modest energy using the CST \cite{Adam:2002cn}.  In addition to being important in its own right, comparing this prediction to electrodisintegration data would be a decisive test of the CST.

Finally, extending the measurements of $A$ and particularly $B$ or $T_{20}$ to higher $Q^2$ would yield new information about the off-shell deuteron form factors, and perhaps (in the absence of direct measurements) the neutron charge form factor $G_{En}$.  This paper provides the predictions with which to compare experimental results.  Note in particular the CST prediction that $B$ will flatten out and reach a secondary maximum [recall Fig.~(\ref{fig:BallQ})].  The large  size of $B$ in this region may make measurements less difficult than previously anticipated.


\acknowledgements

This work was partially supported by Jefferson Science
Associates, LLC, under U.S. DOE Contract No. DE-AC05-
06OR23177.  I especially want to thank Ingo Sick for sharing his Global Analysis (GA), without which it would have been more difficult to complete this work.  It is also a pleasure to thank my colleagues for their support and collaboration over the years.  In particular, I thank J.~W.~Van Orden, Alfred Stadler, and Carl Carlson for their significant contributions to the development of the CST as applied to $NN$ scattering, and to previous calculations of the deuteron form factors.

\appendix

\section{Short review of the theory} \label{app:theory}

This Appendix reviews some details of the calculations of the deuteron form factors discussed in several previous papers, but also includes some new analysis useful for a detailed understanding of this paper. 


\subsection{Form factors and helicity amplitudes} \label{sec:1B}

The most general form of the covariant deuteron electromagnetic vector current illustrated in Figs~\ref{Fig1} and \ref{Fig2} can be expressed in terms of three deuteron form factors 
\bal
\left<P_+\,\lambda\right|J^\mu&\left|P_-\,\lambda'\right>\nonumber\\
&=-2D^\mu\bigg\{G_1 \,\xi^*_{\lambda}\cdot\xi'_{\lambda'}-G_3\frac{(\xi^*_{\lambda}\cdot q)(\xi'_{\lambda'}\cdot q)}{2m_d^2}\bigg\} \nonumber\\
&-G_M\Big[\xi'^\mu_{\lambda'}(\xi^*_{\lambda}\cdot q)-\xi^{*\mu}_{\lambda}(\xi'_{\lambda'}\cdot q)\Big]\, ,
\label{eq:G123}
\end{align}
where the form factors $G_1$, $G_3$, and $G_M=G_2$ are all functions of the square of the  momentum transfer $q=P_+-P_-$, with $Q^2=-q^2$, $D^\mu=\frac12(P_++P_-)^\mu$, 
and $\xi'_{\lambda'}$ ($\xi_{\lambda}$) are the four-vector polarizations of the incoming (outgoing) deuterons with helicities $\lambda'$ ($\lambda$).  The polarization vectors satisfy the well known constraints
\bal
&P_+\cdot \xi_\lambda = P_-\cdot\xi'_{\lambda'}=0 \nonumber\\
&\xi^*_\lambda\cdot \xi_\rho=-\delta_{\lambda\rho} 
\nonumber\\
&\xi'^*_{\lambda'}\cdot \xi'_{\rho'}=-\delta_{\lambda'\rho'}\, .
\end{align}
This notation agrees with that used in  Ref.~\cite{Gilman:2001yh}, except that now $\lambda$ denotes the helicity of the {\it outgoing\/} deuteron and  $\lambda'$ the helicity of the {\it incoming\/} deuteron.

The form factors $G_1$ and $G_3$ are usually replaced by the charge and quadrupole form factors, defined by 
\bea
&&G_C=G_1+\frac{2}{3}\eta G_Q \nonumber\\
&&G_Q=G_1+(1+\eta)G_3-G_M\, , \label{eqLgcgq}
\eea
with $\eta$ defined in Eq.~(\ref{eq:eta}).  At $Q^2=0$, the three form factors $G_C$, $G_Q$, and $G_M$ give the charge, quadrupole moment, and magnetic moment of the deuteron
\bea
\begin{array}{ll}
G_C(0)=\;1\;=G_1(0) & ({\rm units\;of}\;e)\cr
G_M(0)=\mu_d=G_2(0) &  ({\rm units\;of}\;e/2m_d)\cr
G_Q(0)=Q_d=G_3(0)+1-\mu_d  &  ({\rm units\;of}\;e/m^2_d)\, .\qquad\quad
\end{array}
\label{eq:A4}
\eea

Contracting the vector current (\ref{eq:G123}) with the photon helicity vectors
\bea
\epsilon_0^\mu&=&\{0,0,0,1\}
\nonumber\\
\epsilon_\pm^\mu&=&\{0,\mp1,-i,0\}/\sqrt{2}
\eea
gives the helicity amplitudes, denoted by
\bea
G^{\lambda_\gamma}_{\lambda\lambda'}\equiv
\left<P_+\,\lambda\right|J_\mu\left|P_-\,\lambda'\right>\epsilon_{\lambda_\gamma}^\mu \, . \label{eq:current1}
\eea
%
The properties of the helicity amplitudes are discussed in Sec.~III of Ref.~II, where  it was
shown that only three  of the possible 27 amplitudes are independent, so the form factors can be expressed  in terms of the three combinations 
\bea
&& {\cal J}_1\equiv G^0_{00}=2D_0\left(G_C+\frac{4}{3}\eta\, G_Q\right)\nonumber\\
&& {\cal J}_2\equiv G^0_{+-}=2D_0\left(G_C-\frac{2}{3}\eta \,G_Q\right)\nonumber\\
&&{\cal J}_3\equiv \frac12(G^+_{+0} +G^-_{0-})=Q\frac{D_0}{M_d}\,G_M\, , \label{eq:dffmatrix}
\eea
where the symmetrised sum in the definition of ${\cal J}_3$ is used for convenience.   To calculate the deuteron form factors, it therefore sufficient to calculate the ${\cal J}_n$ (with $n=1,2,3$).  

The experimental observables $A$, $B$, and $\widetilde T_{20}$ were defined in terms of the form factors in Eqs.~(\ref{AandB}) and (\ref{eq:T20}). 
%

\subsection{Mathematical form of the current}
\label{sec:math}

The helicity amplitudes of the current, ${\cal J}_n(q)$, are the sum of the three types of contributions shown in Fig.~\ref{Fig2}
\bea
{\cal J}_n(q)={\cal J}^A_n(q)+{\cal J}^{(2)}_n(q)+{\cal J}^B_n(q)\, . \label{eq:traces}
\eea 
The ${\cal J}^A_n$ and ${\cal J}^{(2)}_n$ contributions were combined  in Eq.~(3.28) of Ref.~II;   here I find it convenient to write them as two separate terms.  Including the (B) diagrams from Eq.~(3.36) of Ref.~II, all three contributions can be written in a compact form: 
\begin{widetext}
\begin{subequations}
\bea
{\cal J}_n^A(q)&&=e_0 \int_k\Big\{ f_0(p_+,p_-) \sum_{i=1}^2\Big[ F_i(Q^2){\cal A}_{n,i}(\Psi_+\Psi_-)\Big] +\frac{g_0(p_+,p_-)}{4m^2}\sum_{i=3}^4\Big[F_{i}(Q^2){\cal A}_{n,i}(\Gamma_+\Gamma_-) \Big]\Big\}  \label{eq:currentA}
\\
{\cal J}^{(2)}_n(q)&&=-e_0\int_k \sum_{i=1}^2F_i(Q^2) \Big[ \frac{h_+}{h_-}{\cal A}_{n,i}(\Psi_+\Psi^{(2)}_-) +  \frac{h_-}{h_+}{\cal A}_{n,i}(\Psi^{(2)}_+\Psi_-)\Big]
\\
{\cal J}_n^B(q)&&=e_0 \int_k\Bigg\{\left[\frac{mE_k}{k_zQ}\right] \sum_{i=1}^2 F_i(Q^2)\left(\frac{{\cal B}_{n,i}(k_0)}{k_0}\Big|_{-}- \frac{{\cal B}_{n,i}(k_0)}{k_0}\Big|_{+}\right) -\frac1{m} {\cal O}_n\sum_{i=1}^2 F_i(Q^2){\cal C}_{n,i}(\Gamma\,\widehat\Gamma_{\rm off})\Bigg\} \label{eq:currentB}
\eea
\end{subequations}
\end{widetext}
where the integral is
\bea
\int_k=\int\frac{d^3k}{(2\pi)^3}\frac{m}{E_k}\, , \label{eq:volint}
\eea
the operator ${\cal O}_n X(q)=X(q)+\epsilon_{n3} X(-q)$, with the phase $\epsilon_{n3}=(1-2\delta_{n3})$, and  $|_{\pm} \to |_{k_0=E_\pm}$, where $E_\pm$ was defined in Eq.~(\ref{eq:epm}).   The coefficient of the  $g_0$ term in Eq.~(\ref{eq:currentA}) differs from that reported in Ref.~II; in includes a sum over {\it two\/} off-shell nucleon form factors, 
$F_3$ and $F_4$,  defined in Eq.~(\ref{3.1}).   
The quantities ${\cal A}, {\cal B},$ and ${\cal C}$ are traces over products of pairs of covariant wave functions (or vertex functions), summarized in Table \ref{tab:ingredients}, one for the initial and one for the final deuteron, and are multiplied by one of the four from factors describing the interaction of the virtual photon with the off-shell nucleon.  The detailed formulae for these traces are given in Ref.~II:  Eqs.~(B1) and (B2) for ${\cal A}$,  Eqs.~(B6) and (B7) for ${\cal B}$, and Eqs.~(B9) and (B10) for ${\cal C}$.    I found corrections to these formulae that are reported in Appendix \ref{sec:errorsinII}.  

The  three types of wave functions or vertex functions that enter into the traces (\ref{eq:currentA}) -- (\ref{eq:currentB}) are  $\Psi, \,\Psi^{(2)}$, and $\widehat \Gamma_{\rm BS}$.  
The equation for the bound state wave function 
with particle 1 on shell  
is
\bea
S^{-1}(p)&&{\Psi}(\hat k,P)
=-\int_{k'} \overline V(\hat k,\hat k';P){\Psi}(\hat k',P)\, , \qquad \label{eq:cstequation}
\eea
where $\overline{V}$ is the symmetrized one boson exchange (OBE) kernel (introduced in Ref.~\cite{Gross:2008ps} and discussed in detail in Ref.~I) and the volume integral was defined in (\ref{eq:volint}). 
The wave function $\Psi^{(2)}(\hat k,P)$ and the subtracted vertex function $\widehat \Gamma_{\rm BS}(\widetilde k,P)$ (where $\widetilde k=\{k_0,{\bf k}\}$ can be off-shell) are obtained from an iteration of the basic equation (\ref{eq:cstequation}) using the kernels $\overline V^{(2)}$ and $\overline V-\overline V^{(1)}$
%
\begin{subequations}
\begin{align}
&S^{-1}(p){\Psi}^{(2)}(\hat k,P)
=-\int_{k'} \overline V^{(2)}(\hat k,\hat k';P){\Psi}(\hat k',P)  \label{eq:cstequation2} \\
&\widehat \Gamma_{\rm BS}(\widetilde k,P)=-\int_{k'} [\overline V-\overline V^{(1)}](\widetilde k,\hat k';P){\Psi}(\hat k',P)\label{eq:hatgamma2}\, ,
\end{align}
\end{subequations}
%
where $ \overline V^{(1)}$ and $ \overline V^{(2)}$ are kernels constructed from the momentum dependence of the meson-$NN$ vertiex couplings to particle 1 and 2 as described in Ref. I.  

The off-shell subtracted vertex function $\widehat \Gamma_{\rm BS}$ is composed of two parts with a different matrix structure.  These were previously defined in Eq.~(\ref{eq:BandCparts}). The B part of the vertex function appears in the ${\cal B}$ traces and the C part in the ${\cal C}$ traces.    (The reader is warned not to confuse the B term in Eq.~(\ref{eq:BandCparts}) with the total contribution to the (B) diagrams.)   Note that each of the ${\cal B}$ trace terms is singular when $Q\to 0$, and only through the cancellation of the two terms at $k_0=E_\pm$ is this singularity removed.  
This cancellation is required by the physical behavior of this contribution, as discussed in detail in Sec.~IIF of Ref.~I.
The C term vanishes when particle 1 is on-shell, and is interesting because it is a measure of contributions from off-shell terms that do not contribute to the on-shell two-body CST equation used to fix the parameters of the kernel. 


\subsection{Relativistic effects due to shifts in the arguments of the wave functions} \label{sec:trans}

The wave functions and vertex functions (referred to collectively as wave functions in the following discussion) that enter into the relativistic formulae have arguments shifted by the relativistic kinematics.  It is of considerable interest in itself to study the size of these affects, and this is the focus of this subsection.

\subsubsection{Arguments for the A diagrams}

 As discussed in Sec.~IIC of Ref.~II, when one particle is on-shell, the wave functions depend on only one variable, which I have chosen to be  ${\bf k}^2$ (the square of the three momentum of the on-shell particle 1).  
 When boosted to the rest frame, this variable is denoted by $R^2$, which is then either the momentum of particle 1 or the relative momentum of both particles (identical in the rest frame). 
 The quantity $R$ is a function of ${\bf k}^2$, $k_z$ (the component of ${\bf k}$ in the direction of ${\bf q}$), and $Q^2$.  

For the A diagrams, with the momenta labeled as in Fig.~\ref{Fig1}(A), the exact expression for this argument is [
using $R_A^2$ for rest frame values from diagram (A)]
\begin{subequations}
\bea
(R_A^\pm)^2&=&\frac{(P_\pm\cdot  \hat k)^2}{m_d^2}-m^2 \label{eq:R2first}
\\
&=&{\bf k}^2 
\mp k_z\,Q \frac{D_0 E_k}{m_d^2} +\eta\left(E_k^2+k_z^2\right) \label{eq:R2}
\\
&\to& \left({\bf k}\mp \frac14{\bf q}\right)^2\qquad m, m_d\to\infty
\qquad
\label{eq:R2nr}
\eea
\end{subequations}
where $(R_A^-)^2$ [$(R_A^+)^2$] is the rest frame value of $R_A^2$ obtained from a moving incoming (outgoing) deuteron in the Breit frame.  

The last expression, Eq.~(\ref{eq:R2nr}), is the value of the rest frame momentum $(R_A^\pm)^2$ in the infinite mass (nonrelativistic) limit, and shows that, nonrelativistically,  these momenta must be interpreted as the {\it relative\/} momenta, ${\bm \rho}=\frac12({\bf k}_1-{\bf k}_2)$, because  before and after the collision with the photon, the assignment of momenta that correctly describes this process is
\bea 
{\rm before} \qquad&&\begin{cases} {\bf k}_1={\bf k} &  \cr
{\bf k}_2=-{\bf k}-\frac12 {\bf q} & \cr
{\bm \rho} = {\bf k}+\frac14{\bf q} & \end{cases}
\nonumber\\
{\rm after} \qquad && \begin{cases} {\bf k}_1={\bf k} &  \cr
{\bf k}_2=-{\bf k}+\frac12 {\bf q} & \cr
{\bm \rho} = {\bf k}-\frac14{\bf q} & \end{cases} \label{eq:nrmomA}
\eea
%
Note the reassuring fact that Eq.~(\ref{eq:R2first}) 
gives the same result if  $\hat k$ is replaced by the relative momentum in the moving frame
\bea
(R_A^{\pm'})^2&=&\frac{(P_\pm\cdot  (\hat k-\frac12 P_\pm))^2}{m_d^2}-(\hat k-\frac12 P_\pm)^2
\nonumber\\
&=&\frac{(P_\pm\cdot  \hat k)^2}{m_d^2}-(P_\pm\cdot \hat k)+\frac{m_d^2}{4}
\nonumber\\
&&-m^2+(P_\pm\cdot \hat k)-\frac{m_d^2}{4}
\nonumber\\
&=&(R_A^\pm)^2\, .
\eea
The lesson from this discussion is that the effective rest frame momentum, $(R_A^\pm)^2$, is the same whether or not one starts in the moving frame from the four-momentum of particle 1, or the relative four-momentum of the two particles; this  must be true, of course, since the two are indistinguishable in the rest system.

\begin{figure}
\centerline{
\mbox{
\includegraphics[width=3.2in]{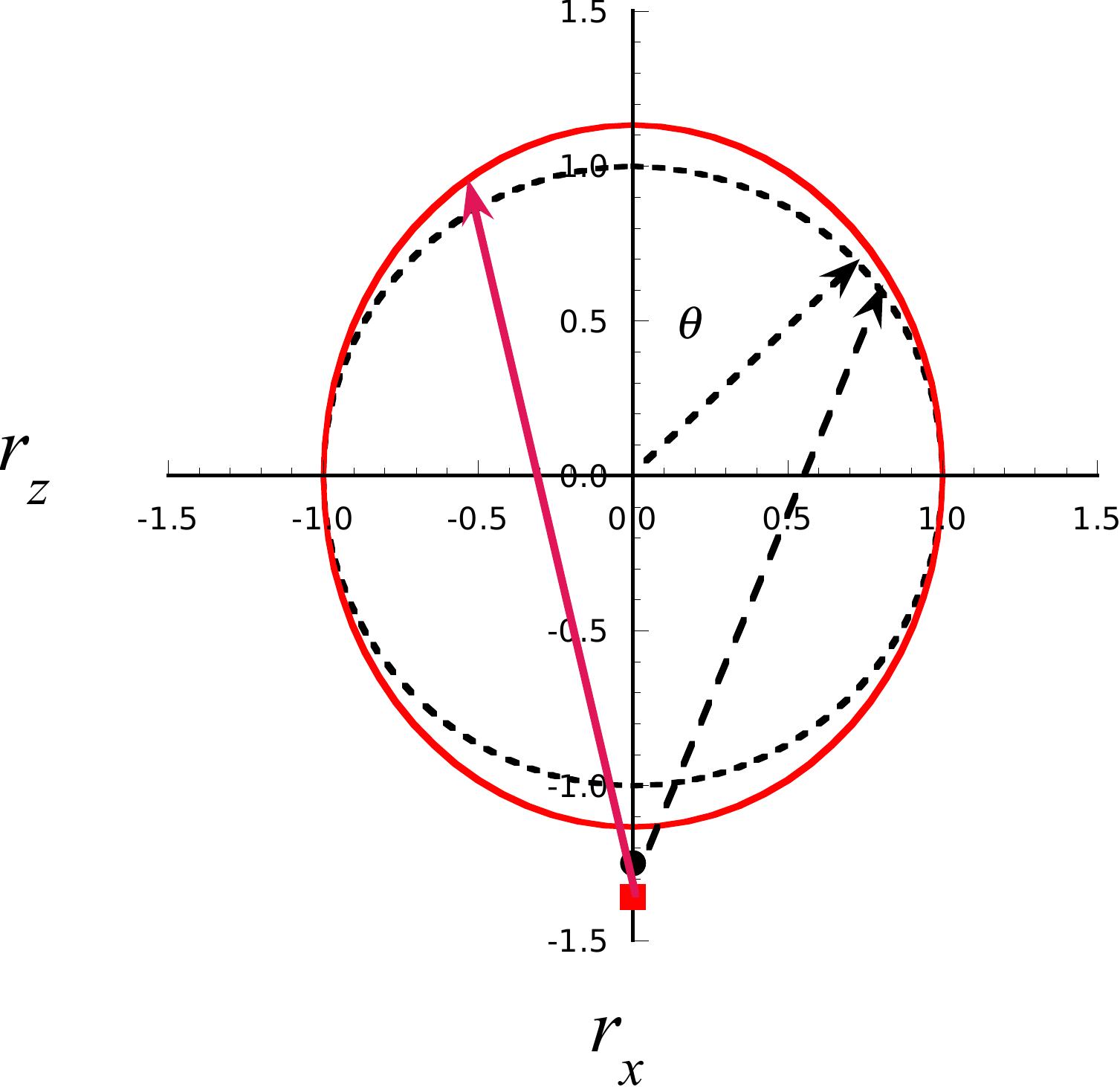}
}
}

\caption{\footnotesize\baselineskip=10pt  
Locus of the points $r_z$ 
and $r_x=r_\perp$, Eq.~(\ref{eq:ra}), 
as a function of $\theta$.  The two closed curves and the solid round and square reference points near $r_z\sim -1$ are discussed in the text.  I chose $Q=2$ GeV and $k=400$ MeV when the curves or reference points depended on $Q$ or $k$.    
}
\label{fig:Rplus}
\end{figure} 

To better understand the results (\ref{eq:R2}) and (\ref{eq:R2nr}), it is useful to obtain the longitudinal and transverse components of ${\bf R}^\pm_A$ by directly  transforming the components of $\hat k=\{E_k,k_\perp,k_z\}$ from the moving system to the rest system, using the relations
\bea
(R^\pm_A)_\perp&=&k_\perp=k\sin\theta
\nonumber\\
(R^\pm_A)_z&=&\frac{D_0}{m_d} k\cos\theta\mp \frac{Q}{2m_d}E_k  \label{eq:RRa}
\eea
where it is easily shown that $(R_A^\pm)^2=(R^\pm_A)_\perp^ 2+(R^\pm_A)_z^2$.  In  Fig.~\ref{fig:Rplus} I show the related components
\bea
r_\perp&=&\frac{(R^-_A)_\perp}{k}=\sin\theta
\nonumber\\
r_z&=& \frac1{k}\Big[(R^-_A)_z-\frac{QE_k}{2m_d}\Big]=\sqrt{1+\eta}\, \cos\theta \label{eq:ra}
\eea
plotted in the $r_x=r_\perp, r_z$ plane.    

Two cases are shown in the figure.  The first is the nonrelativistic limit of the boost (with $\eta=0$). It is described by the dashed black circle which can be described as the locus of points swept out by the unit vector ${\bf r}$ (represented by the dashed arrow fixed to the origin) rotating through polar angle $\theta$ with the $\hat z$ axis.  However, the  {\it same\/} circle is also the locus of points swept out by ${\bf R^-_A}/k$, represented by the longer black dashed vector, with one end fixed at the reference point (represented by the round black dot at $r_z^0=-Q/(4k)=-1.25$ in the figure)  and the other end following the locus of points swept out by dashed circle.   This vector, which is the  rest frame momentum in the nonrelativistic limit, divide by $k$, must change length in order to track the dashed circle. 

The second case shown is the solid red ellipse, which shows the behavior of the relativistic  A-type rest-frame momentum.  Because $\eta\ne0$  ($\eta\simeq 0.02835$ for the parameters chosen)  the curve changes from a circle to an ellipse and the reference point (represented by the solid red square) shifts to $r_z^0=-QE_k/(2km_d)\simeq -1.359$.  The vector that sweeps out the relativistic momentum ${\bf R^-_A}/k$ is represented by the red arrow that connects the solid red square reference point to the ellipse.  One can see clearly how the relativistic transformation changes the effective rest-frame momentum; the shift in the reference point is due to the role that the particle energy plays in the transformation, and change from a circle to an ellipse is due to the dilation factor in the transformation.  

Fig.~\ref{fig:Rplus} shows the relativistic ellipse {\it expanding\/} in the $\hat z$ direction, rather than contracting.  This is because we are keeping $k$ constant in the {\it moving frame\/}.   To obtain this condition, we must start from an expanded ellipse in the rest-frame, so that when it is contracted by the  transformation to the moving system it will be compressed back into a circle.  Hence the transformation behaves as expected after all.  This expansion explains qualitatively the behavior of the relativistic argument shift shown by the dashed blue line shown in Fig.~\ref{fig:RStudy}.

\subsubsection{Arguments for the B diagrams} \label{app:Bargs}

For the B diagrams, when both particles can be off-shell, the wave functions can depend on an additional variable, which was previously chosen to be the energy of particle 1 in the moving frame, $k_0$, which  transformed to $R_0$ in the rest frame.  The momenta are labeled in Fig.~\ref{Fig2}(B) and following the discussion in Sec.~\ref{sec:off-shell}, I make the substitution 
\bea
\widetilde k_0^\pm=x_{10}^\pm \tilde E_\pm    \label{eq:x10pm}
\eea 
into the momenta given in Ref.~II (with the change in notation $\tilde R\to R_B$)
\bea
[R_B^\pm(\tilde{\bf k},x_{10} ^\pm)]^2&=&\frac{\left(P_\pm\cdot  \widetilde k_\pm \right)^2}{m_d^2}-\widetilde k_\pm^2 = \tilde k_\perp^2 +\tilde k_{\pm z}^2
\nonumber\\
&& \mp x_{10}^\pm \, \tilde k_{\pm z} Q\frac{D_0 \tilde E_\pm}{m_d^2}
+\eta [(x_{10}^\pm \, \tilde E_\pm)^2+\tilde k_{\pm z}^2]
\nonumber\\
&\to& \tilde k_\perp^2 +\tilde k_{\pm z}^2 \mp  \frac12 x_{10}^\pm \tilde k_{\pm z} Q + x_{10}^{\pm 2} \frac{Q^2}{16}
\nonumber\\
&=& \Big[{\bf\tilde{k}}\pm\frac14 (2- x_{10}^\pm){\bf q}\Big]^2 \nonumber
\qquad  m, m_d\to\infty 
\eea
\bea
R^{\pm}_0(\tilde{\bf k},x_{10}^\pm )&=&\frac{P_\pm\cdot  \widetilde k_\pm }{m_d}=\frac1{2m_d}[2x_{10}^\pm  D_0\tilde E_\pm\mp \tilde k_{\pm z}Q]
\nonumber\\
&\to& x_{10}^\pm m+\frac{x_{10}^\pm}{2m}\Big[\tilde k_\perp^2+\tilde k_{\pm z}^2 +\frac18 Q^2\Big] \mp\frac{\tilde k_{\pm z}Q}{4m}
\nonumber\\
&=&x_{10}^\pm m+\frac{1}{2m}\Big[x_{10}^\pm \tilde{k}^2\pm \frac12 \tilde k_z Q(2 x_{10}^\pm-1)
\nonumber\\
&&+\frac1{16}(5 x_{10}^\pm-4) Q^2\Big] \quad m, m_d\to\infty
\, , \qquad \quad\label{eq:129}
\eea
where [recalling that the momenta in Fig.~\ref{Fig2}(B) are labeled with a tilda to distinguish them from momenta in Fig.~\ref{Fig2}(A)]
\bea
&&\widetilde k_\pm=\{x_{10}^\pm  \tilde E_\pm, \tilde  {\bf k}\pm \frac12{\bf q}\}
\nonumber\\
&&\tilde k_{\pm z}=\tilde k_z\pm\frac12 Q  \label{eq:ktilde}\, .
\eea
%
I prefer using the variable $x_{10}$ instead of the unconstrained  energy $k_0$ because 
when particle 1 is on-shell, $x_{10}=1$, independent of momenta.  
Hence
\bea
[R_B^\pm(\tilde {\bf k},1)]^2&=&\tilde k_\perp^2 +\tilde k_{\pm z}^2 \mp \tilde k_{\pm z} Q\frac{D_0 \tilde E_\pm}{m_d^2}
+\eta [\tilde E_\pm^2+\tilde k_{\pm z}^2]\quad
\nonumber\\
&\to& \Big({\bf \tilde k}\pm\frac14 {\bf q}\Big) \qquad  m, m_d\to\infty
\nonumber\\
R^{\pm}_0(\tilde {\bf k},1)&=&\frac1{2m_d}[2D_0 \tilde E_\pm \mp \tilde k_{\pm z} Q]
\, , \qquad \nonumber\\
&\to& m+\frac1{2m}\Big({\bf\tilde k}\pm\frac14{\bf q}\Big)^2 \quad  m, m_d\to\infty \, .
\label{eq:R2nrB}
\eea
 It is easy to see that $R^{\pm}_0$ is constrained by the  mass shell condition
\bea
R^{\pm}_0(\tilde {\bf k},1)=\sqrt{m^2+[R_B^\pm(\tilde {\bf k},1)]^2}\, , \label{eq:constraint}
\eea
as required by relativity.   The condition $x_{10}^\pm=1$ is now a simple, momentum independent way to specify that particle 1 is on-shell.

In the calculation of the (B) diagram, only the values of $x_{10}^\pm$ given in Eq.~(\ref{eq:x10actual}) are needed. To order $1/m^2$ these are
\bea
x_{10}^\pm=\zeta^\pm&\simeq& 1+\frac{\tilde k_\mp^2}{2m^2}-\frac{\tilde k_\pm^2}{2m^2}
\nonumber\\
&=& 1 \mp\frac{\tilde kz Q}{m^2}\, .
\eea
When this is substituted into (\ref{eq:129}), the result to order $1/m^2$ is
\bea
R^{\pm}_0(\tilde{\bf k},\zeta^\pm )&\simeq&m\Big(1\mp\frac{\tilde kz Q}{m^2}\Big) +\frac1{2m}\Big[\tilde k^2 \pm \frac12 \tilde k_z Q +\frac{Q^2}{16}\Big]
\nonumber\\
&=&m +\frac1{2m}\Big[\tilde k^2 \mp \frac12 \tilde k_z Q +\frac{Q^2}{16}\Big] 
\nonumber\\
&\simeq& \sqrt{m^2 +\Big(\tilde{\bf k}\mp \frac14 {\bf q}\Big)^2} \, ,
\eea
showing that the mass shell condition holds to order $1/m^2$.

While $x_{10}^\pm$ is the appropriate quantity describing the off-shell behavior in the moving frame, the quantity that describes this in the  rest frame of each state, denoted by $X^\pm_{10}(\tilde {\bf k},x_{10}^\pm)$, is defined by the relations
\bea
R^{\pm}_0(\tilde {\bf k},x_{10}^\pm)=X_{10}^\pm (\tilde {\bf k},x_{10}^\pm) \sqrt{m^2+[R^{\pm}_B(\tilde{\bf k},x_{10} ^\pm)]^2}\, .\qquad \label{eq:x10pma}
\eea
This can be derived directly by transforming $\tilde k_0^\pm$ and $\tilde E_\pm$ in Eq.~(\ref{eq:x10pm}) to the {\it rest frame\/} and requiring the transformation of $x_{10}^\pm\to X_{10}^{\pm}$ to maintain the equation.   Note that setting $x_{10}^\pm=1$ in (\ref{eq:x10pma}) and using the relation (\ref{eq:constraint}) gives the result $X_{10}^{\pm} (\tilde {\bf k},1) = 1$.

It is satisfying to observe that nonrelativistic limits of the momenta (\ref{eq:129}) [or (\ref{eq:R2nrB})] correctly describe the process in which the photon is absorption on particle 1 instead of particle 2:
\bea 
{\rm before} \qquad&&\begin{cases} {\bf k}_1={\bf k}-\frac12 {\bf q} &  \cr
{\bf k}_2=-{\bf k} & \cr
{\bm \rho} = {\bf k}-\frac14{\bf q} & \end{cases}
\nonumber\\
{\rm after} \qquad && \begin{cases} {\bf k}_1={\bf k} +\frac12 {\bf q} &  \cr
{\bf k}_2=-{\bf k} & \cr
{\bm \rho} = {\bf k}+\frac14{\bf q}\, . & \end{cases} \label{eq:nrmomB}
\eea
This is a consequence of the fact that the nonrelativistic limit of $[R_B^\pm(\tilde {\bf k},1)]^2$ is {\it not equal\/} to  $(R_A^\pm)^2$. 

As I did in the previous subsection, it is useful to derive (\ref{eq:129}) directly from the Lorentz boost.  Starting from the off-shell four-momentum of particle 1 in the moving frame, $\tilde k_\pm=\{x_{10}^\pm\tilde E_\pm, \tilde {\bf k}\pm\frac12Q\}$ (where $\tilde k_\pm$ pairs with momenta $P_\pm$),  and transforming to the rest frame using $R^{\pm}_B=\Lambda_\mp \tilde k_\pm$, gives
\bea
(R^{\pm}_B)_\perp&=&\tilde k_\perp = \tilde k\sin\theta
\nonumber\\
(R^{\pm}_B)_z
&=&\frac{D_0}{m_d} \tilde k_{\pm z}
\mp x_{10}^\pm\frac{Q\tilde E_\pm}{2m_d}  
\nonumber\\
R^{\pm}_0&=&\mp\frac{Q \tilde k_{\pm z}}{2m_d} 
+x_{10}^\pm\frac{D_0 \tilde E_\pm}{m_d}  \, . \qquad\label{eq:RRb}
\eea
This result for $R^{\pm}_0$ agrees immediately with (\ref{eq:129}), and it is also easy to show that $[R^{\pm}_B]^2=(R^{\pm}_B)_\perp^2+(R^{\pm}_B)_z^2$ as expected.  

\begin{figure}
\centerline{
\mbox{
\includegraphics[width=3.2in]{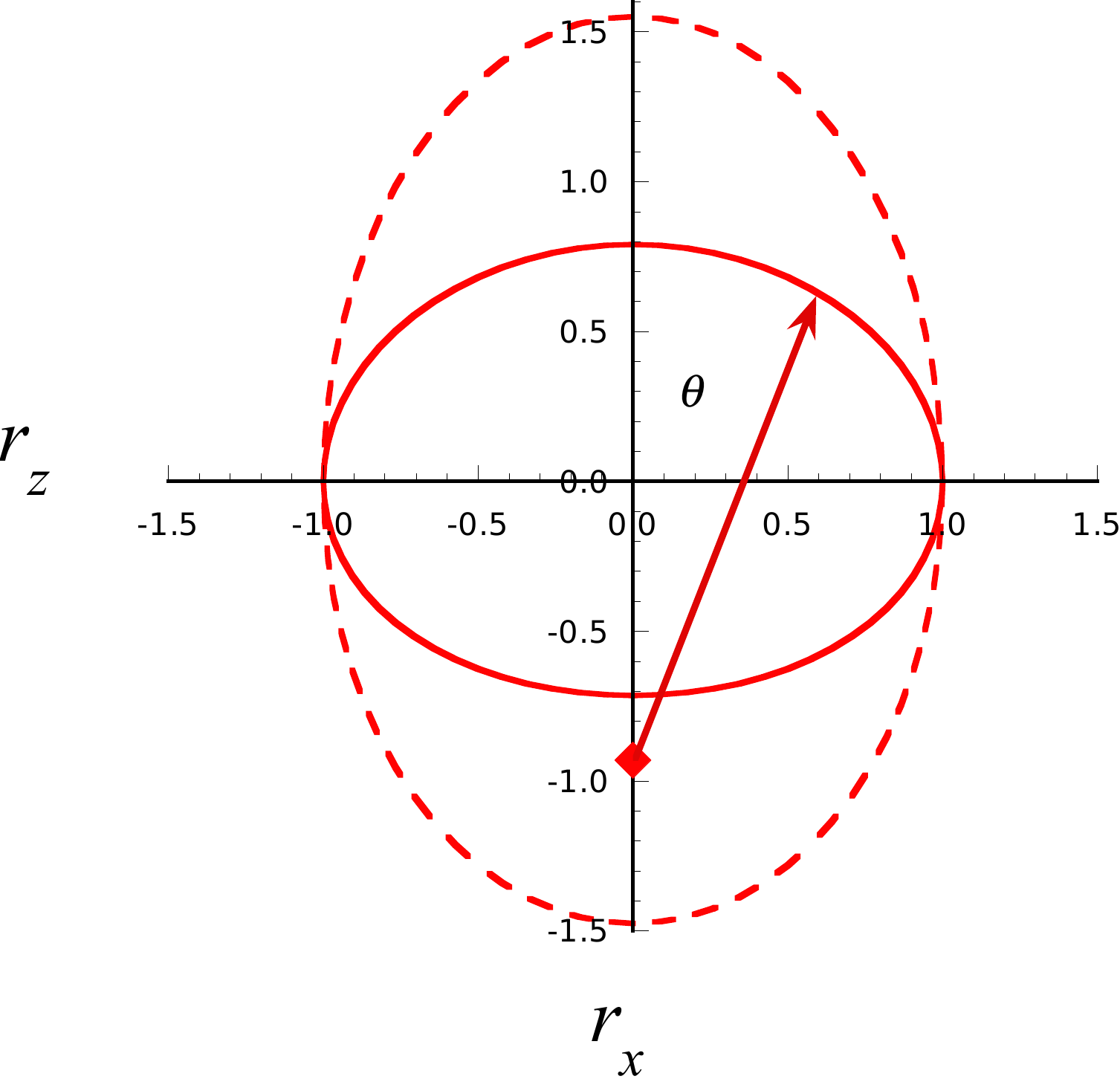}
}
}

\caption{\footnotesize\baselineskip=10pt 
Locus of the points $\{r_{z}^b, r_\perp^{b}\}$ of Eq.~(\ref{eq:rboff}) (the solid line ellipse) compared to $\{r_z, r_\perp\}$ of Eq.~(\ref{eq:ra}) (the dashed-line ellipse, identical to the one shown in Fig.~\ref{fig:Rplus}) as a function of $\theta$.  The diamond reference point near $r_z\sim -1$ is the point from which the magnitude of the transformed momentum is measured.  I choose $Q=2$ GeV and $k=400$ MeV for this example.   
}
\label{fig:Roff}
\end{figure} 

To represent the behavior of the spacial components when particle 1 is on shell, Fig.~\ref{fig:Roff} shows the behavior of the two components that enter into the (B$_+$) diagram when $x_{10}=1$
\bea
r_\perp^{b}&=&\frac1{\widetilde k} [R^+_B(1)]_\perp=\sin\theta
\nonumber\\
r_z^{b}&=& \frac1{\widetilde k}\Big[[R^+_B(1)]_z -\frac{Q}{2m_d}(D_0-E_0)\Big]
\nonumber\\
&=&\sqrt{1+\eta}\, \cos\theta -\frac{Q}{2km_d}(E_+-E_0) \, .\label{eq:rboff}
\eea
In defining $r^b_z$, I introduced a new subtraction term depending on the energy $E_0$,
\bea
E_0\equiv\sqrt{m^2+k^2+\frac{Q^2}{4}}=E_+(\cos\theta=0)\, ,
\eea
chosen to be independent of $\theta$ and of the correct size to center the elliptical locus of the points $\{r_\perp^{b}, r_z^{b}\}$ at the origin

The solid red ellipse shown in Fig.~\ref{fig:Roff} is the locus of points swept out by the vector ${\bf R^+_B}/\widetilde k$, which reaches from the new reference point (the subtraction term) shown as a solid red diamond located on the $\hat z$ axis at $-0.930$ (for the same values $Q=2$ GeV and $k=400$ MeV used in Fig.~\ref{fig:Rplus}) to the smaller of the red ellipses, and has a different length than the one for diagram (A).  For comparison, the ellipse for the transformation of diagram (A) shown in   Fig.~\ref{fig:Rplus} is the red dashed ellipse in Fig.~\ref{fig:Roff}. The Lorentz transformation of the (A) and (B) diagrams {\it have very different behaviors\/},  which can be traced to the different behavior of the energies of particle 1 in the two cases.  

I return now to the issue of how far off-shell particle 1 is forced by the kinematics in the \ref{Fig2}(B) diagrams.  This was already addressed nonrelativistically in Sec.~\ref{sec:off-shell} above.  As I showed there, the relevant values of $x_{10}^\pm$, denoted $\zeta^\pm$,  were given in Eq.~(\ref{eq:x10actual}).  The correct quantity is therefore
\bea
X_{10}^{\pm} (\tilde {\bf k},\zeta^\pm)&=&\frac{R^{\pm}_0(\tilde {\bf k},\zeta^\pm)}{\sqrt{m^2+[R^{\pm}_B(\tilde{\bf k},\zeta ^\pm)]^2}}
\nonumber\\
&=&
\frac{2D_0\tilde E_\mp\mp \tilde k_{\pm z} Q}{\sqrt{(2D_0\tilde E_\mp\mp \tilde k_{\pm z} Q)^2\pm 8 m_d^2 \tilde k_z Q}}\qquad
\eea
The maximum of $X^{\pm}_{10}$ occurs when $\tilde k_z=\mp  k$ (I drop the tilde at this point), giving
\bea
X^{\pm{\rm max}}_{10}=\frac{2D_0 E_+(Q)+ k_- Q}{\sqrt{(2D_0 E_+(Q) + k_- Q)^2 - 8 m_d^2 k\, Q}}
\label{eq:x10exact}
\eea
where  
\bea
E_\pm(Q)=\sqrt{m^2+\left(k\pm\frac12 Q\right)^2}\, .
\eea
As for the non-boosted case, the minimum is found by changing $k\to-k$ (or $\theta=0$ to $\pi$).  Also, note that 
\bea
\lim_{m_d\to\infty}X^{\pm{\rm max}}_{10}=\frac{E_+(Q)}{E_-(Q)}
\eea
which, once $k_{\rm max}$ has been found, agrees with the result (\ref{eq:x10a}).

For each $Q$, this limiting function is has a maximum at a particular value of $k$ (which is best found numerically) so the maximum and minimum for all $k$ can be shown as a function of $Q$ only.  This limiting value was already shown in  Fig.~\ref{fig:X0}. The figure shows how far off-shell the particle can be forced, 
even at modest values of $Q$.

\subsection{Calculation of the off-shell invariant functions} \label{app:H}

The wave and vertex functions can be expanded into scalar functions.   For recent discussion of the decomposition of $\Psi$, see Appendix A of Ref. II  and Sec.~III of Ref.~\cite{Gross:2010qm}.  The wave function $\Psi$  can be expressed in terms of four scalar functions:  the two familiar nonrelativistic S and D-state wave functions, $u$ (S-state), $w$ (P-state), and the two small P-state components of relativistic origin,$v_t$ (spin triplet P-state), and $v_t$ (spin singlet P-state), or alternatively in terms of the helicity amplitudes $z_\ell^{\rho_1\rho_2}$ (with $\rho_1=+$; see the discussion below).  The momentum dependence of these wave functions was shown in Figs. 6 and 7 of Ref.~\cite{Gross:2010qm}.

In this subsection, I discuss a few technical details that I found helpful in carrying out the numerical calculations.  

First, instead of using the amplitudes $z_\ell^{\rho_1\rho_2}$, where
\bea
z_0^{++}&=&\frac1{\sqrt{6}}(u+\sqrt{2}w)
\nonumber\\
z_1^{++}&=&\frac1{\sqrt{6}}(\sqrt{2}u-w)
\nonumber\\
z_0^{+-}&=&-\frac1{\sqrt{2}}v_s
\nonumber\\
z_1^{+-}&=&-\frac1{\sqrt{2}}v_t \, ,\label{eq:uz}
\eea
I use the realted amplitudes
\bea
y^{\rho_1\rho_2}_\ell(k,k_0)&=&\delta_{\rho_2} z^{\rho_1\rho_2}_\ell(k,k_0)\, ,
\label{eq:yvertex}
\eea
where $\rho=\pm$ and $\delta_\rho$ is related to the inverse of the nucleon propagator $G^\rho$ for positive and negative energy nucleon states.  These propagators are 
\bea
G^\rho
&=&\frac\rho{E_k+\rho(k_0-m_d)}
\nonumber\\
&=&\frac\rho{E_k(1+\rho\, x_{10})-\rho \,m_d)}\equiv \frac\rho{\;\,\delta_\rho}
\, .\label{eq:propk0}
\eea

The reason for using the $y$'s instead of the $z$'s is that the propagator appears naturally when the $z$'s are extended off-shell.  Since  $G^+$ is singular at 
 $\delta_+=0$, or at 
\bea
x_{10} = \frac{m_d-E_k}{E_k}\simeq 1- \frac{k^2}{m^2} \, ,
\eea
the $z$'s are singular at these points, and it is hard to compute them numerically around these singularities.  The problem becomes critical because these singularities are very close to $x_{10}=1$ at small $k^2$.  These singularities are cancelled in the amplitudes $y$, which  are very smooth near $x_{10}=1$ and provide a much better input for numerical solutions.

The eight invariant functions that define the Dirac-space form of the vertex function can be expressed in terms of the helicity amplitudes  $y^{\rho_1\rho_2}_\ell$.  The results in terms of the $z$'s was given in Eq.~(A27) of \cite{Gross:2014wqa}.  When expressed in terms of the $y$'s  the relations  become 
\bea
\frac{F}{{\cal C}_0}&=&(E_k+k_0)\Big[\delta_-y^{++}_1-\frac{m}{k} \delta_+y^{+-}_1\Big]
\nonumber\\
&&-(E_k-k_0)\Big[\delta_+y^{--}_1+\frac{m}{k}\delta_-y^{-+}_1\Big]
\nonumber\\
\frac{k^2 G}{m {\cal C}_1}&=&(E_k+k_0)\Big[\delta_- \Big(E_k y^{++}_0-\frac{m y^{++}_1}{\sqrt{2}}\Big)
-\frac{k\delta_+y^{+-}_1}{\sqrt{2}}\Big]
\nonumber\\
&-&(E_k-k_0)\Big[\delta_+ \Big(E_k y^{--}_0-\frac{m y_1^{--}}{\sqrt{2}}\Big)+\frac{k \delta_- y^{-+}_1}{\sqrt{2}}\Big]
\nonumber\\
\frac{k\,H}{mE_k{\cal C}_0}&=&-(E_k+k_0)\,y^{+-}_1-(E_k-k_0)\,y^{-+}_1
\nonumber\\
%
\frac{k^2 I}{m^2{\cal C}_1}&=&(E_k+k_0)\Big[ m \,y^{++}_0-\frac{E_k}{\sqrt{2}}y^{++}_1+k \,y^{+-}_0\Big]
\nonumber\\&&
-(E_k-k_0)\Big[ m \,y^{--}_0-\frac{E_k}{\sqrt{2}}y^{--}_1-\,k \,y^{-+}_0\Big]
\nonumber\\
\frac{k K_1}{mE_k{\cal C}_0}&=&-\delta_+y^{+-}_1-  \delta_-y^{-+}_1
\nonumber\\
%
\frac{k^2 K_2}{m^2{\cal C}_1}&=&m \,\delta_-y^{++}_0-E_k\frac{\delta_-y^{++}_1}{\sqrt{2}}-k \delta_+y^{+-}_0
\nonumber\\&& -m\,\delta_+y^{--}_0+E_k\frac{\delta_+y^{--}_1}{\sqrt{2}}-k \delta_-y^{-+}_0
\nonumber\\
%
\frac{k K_3}{m^2{\cal C}_0}&=&- k\,y^{++}_1-m\,y^{-+}_1+k\,y^{--}_1- m\,y^{+-}_1
\nonumber\\
\frac{k^2 K_4}{m^3{\cal C}_0}&=&\sqrt{2}\,E_k y^{++}_0-m y^{++}_1 +k y^{-+}_1
\nonumber\\&&
- \sqrt{2} \, E_k y^{--}_0+m y^{--}_1 +k y^{+-}_1 \, , \label{eq:AtoK}
\eea
where 
\bea
{\cal C}_0=\frac{\sqrt{3}\,{\cal K}}{2E_k m_d}\qquad{\cal C}_1=\sqrt{2}\,{\cal C}_0
\eea
with ${\cal K}=\pi\sqrt{2m_d}$.  

It turns out that only the four amplitudes $y^{+\rho_2}_\ell$ need to be considered; the amplitudes $y^{- \rho_2}_\ell$ will never contribute to the final result. The argument is in two steps.    First, when particle 1 is on-shell, $k_0=E_k$, and the four invariant functions $F, G, H, I$ do not depend on the amplitudes $y^{-\rho_2}_\ell$.   Next, when both particles are off-shell, only subtracted amplitudes $\hat H,\cdots \hat K_4$ contribute, and I have found that the {\it subtracted\/} amplitudes $\hat y_\ell^{-\rho_2}$ are numerically so small as to be nearly zero, and can be discarded from the calculation.   In this case the subtracted $\hat K_i$ are not zero, but depend only on the amplitudes $\hat y_\ell^{+\rho_2}$.  I have not looked for a proof of the relation $\hat y_\ell^{-\rho_2}=0$, which I believe to be true.

All of the invariants are regular as $k\to0$, yet the expressions for all but $F$ show a possible singularity at $k=0$.   To avoid this there must be relations between the $y$'s near $k=0$.  
To examine this, drop all of the $y_\ell^{-\rho_2}$ terms, and substitute $k_0=x_{10} E_k$ and examine the $k\to0$ limits, dropping all terms proportional to $k^2$ or higher, since they are finite.   Taking $m_d\to 2m$, and using the expansions
\bea
\delta_+&\simeq&m(x_{10}-1)
\nonumber\\
\delta_-&\simeq& m(3-x_{10})
\eea
gives
\bea
\lim_{k\to0}G&=& \frac{G_0}{k^2}(1+x_{10})\Big\{(3-x_{10})\big[y_0^{++}-\frac{y_1^{++}}{\sqrt{2}}\big]
\nonumber\\
&&-(x_{10}-1)\frac{k}{m}\frac{y_1^{+-}}{\sqrt{2}}\Big\}
\nonumber\\
\lim_{k\to0}H&=&\frac{H_0}{k}(1+x_{10})\,y_1^{+-}
\nonumber\\
\lim_{k\to0}I&=& \frac{I_0}{k^2}(1+x_{10})\Big(y_0^{++}-\frac{y_1^{++}}{\sqrt{2}}+\frac{k}{m}y_0^{+-}\Big)
\nonumber\\
\lim_{k\to0}K_1&=&\frac{K_{10}}{k}(x_{10}-1)\,y_1^{+-}
\nonumber\\
\lim_{k\to0}K_2&=& \frac{K_{20}}{k^2}\Big\{(3-x_{10})\big[y_0^{++}-\frac{y_1^{++}}{\sqrt{2}}\big]
\nonumber\\&&
-(x_{10}-1)\frac{k}{m}y_0^{+-}\Big\}
\nonumber\\
\lim_{k\to0}K_3&=& \frac{K_{30}}{k}\Big\{\frac{k}{m}y_1^{++}+y_1^{+-}\Big\}
\nonumber\\
\lim_{k\to0}K_4&=& \frac{K_{40}}{k^2}\Big\{\sqrt{2}\,y_0^{++}-y_1^{++}
+\frac{k}{m}y_1^{+-}\Big\}\, .
\eea 
Requiring that these seven invariants be regular at $k=0$ gives conditions on the four vertex functions.  Near $k=0$, and independent of $x_{10}$, we require
\bea
&&\lim_{k\to0}(y_1^{++}-\sqrt{2}y_0^{++}) \to a_1 k^2
\nonumber\\
&&\lim_{k\to0}y_1^{+-} \to a_2 k
\nonumber\\
&&\lim_{k\to0}y_0^{+-} \to a_3 k
\eea
Note that these limits are satisfied by the usual behavior of the momentum space wave functions.  
Using the definitions [taken from  (\ref{eq:uz}) multiplied by $\delta_{\rho_2}$] gives
\bea
&&y_1^{++}-\sqrt{2}y_0^{++}=-\sqrt{\frac32} w_v= -\sqrt{\frac32} \delta_+ w
\nonumber\\
&&y_{0}^{+-}=-\frac1{\sqrt{2}}\delta_- v_s
\nonumber\\
&&y_{1}^{+-}=-\frac1{\sqrt{2}}\delta_- v_t
\eea
which shows that the standard $k^\ell$ behavior of the $P$ and $D$ state wave functions will satisfy the necessary conditions.

\section{Nonrelativistic form factor}\label{app:NRff}

\subsection{Wave functions momentum space}

To prepare for the discussion of the nonrelativistic form factor, I write the nonrelativistic wave functions in the form
\bea
 Z_{\ell m}({\bf r})&=&\frac{i^\ell \,z_\ell(r)}{r} Y_{\ell m}(\hat{\bf r})
 \nonumber\\
  Z_{\ell m}({\bf k}) &=&z_\ell(k) Y_{\ell m}(\hat{\bf k})
\eea
where $z_\ell$ a generic name for the radial wave functions, $u (\ell=0)$ or $w (\ell=2)$, and $Y_{\ell m}$ is the spherical harmonic with relations
\bea
&&\int d\Omega_{\hat{\bf r}} \,Y_{\ell m}(\hat{\bf r}) Y^*_{\ell' m'}(\hat{\bf r})=\delta_{mm'}\delta_{\ell \ell'}
\nonumber\\
&&\frac{4\pi}{2\ell+1}\sum_{m=-\ell}^{\ell} Y_{\ell m}(\hat{\bf r})Y^*_{\ell m}(\hat{\bf k})=P_\ell (\hat{\bf k}\cdot\hat{\bf r})
\nonumber\\
&&\frac{4\pi}{2\ell+1}\sum_{m=-\ell}^{\ell} Y_{\ell m}(\hat{\bf r})Y^*_{\ell m}(\hat{\bf r})=1
\eea
 Note that the wave functions in coordinate space are {\it reduced\/} and the wave functions in coordinate and momentum spaces are only distinguished is by their arguments ($r$ for coordinate space and $k$ for momentum space).  For a  discussion of the phase $i^\ell$, see Eq.~(3.36) and the last paragraph of Sec.~III C in Ref.~\cite{Gross:2010qm}.   This phase, which comes from  the familiar plane wave expansion, is need to keep the $z_\ell$'s real.  The standard Fourier transform links the two spaces
\bea
Z_{\ell m}({\bf r})&&=\frac{1}{(2\pi)^{\frac32}}\int d^3 k \exp(i{\bf k}\cdot {\bf r})\,Z_{\ell m}({\bf k})
\nonumber\\
Z_{\ell m}({\bf k})&&=\frac{1}{(2\pi)^{\frac32}}\int d^3 k \exp(-i{\bf k}\cdot {\bf r})\,Z_{\ell m}({\bf r})\, .
\label{eq:FT}
\eea
Using the familiar plane wave expansion  
\bea
\exp(i {\bf k}\cdot {\bf r})&&=\sum_{\ell=0}^\infty (2\ell+1)i^\ell j_\ell(kr)P_\ell (\hat{\bf k}\cdot\hat{\bf r})
\nonumber\\
&&=4\pi\sum_{\ell=0}^\infty i^\ell j_\ell(kr)\sum_{m=-\ell}^\ell Y_{\ell m}(\hat{\bf k}) Y^*_{\ell m}(\hat{\bf r})
\, , \qquad\label{eq:planewave}
\eea
where  $j_\ell(kr)$ is the spherical Bessel function,  I reduce the Eqs.~(\ref{eq:FT}) to
\begin{widetext}
\bea
&&\frac{1}{(2\pi)^{\frac32}}\int d^3 k \exp(i{\bf k}\cdot {\bf r})\,z_\ell(k) Y_{\ell m}(\hat{\bf k}) 
= Z_{\ell m}({\bf r})
= \frac{i^\ell z_\ell(r)}{r}Y_{\ell m}(\hat{\bf r})= i^\ell\,\sqrt{\frac2{\pi}} \int_0^\infty r^2\,dr j_\ell(kr) z_\ell(k) Y_{\ell m}(\hat{\bf r})
\nonumber\\
&&\frac{i^\ell}{(2\pi)^{\frac32}}\int d^3 r \exp(-i{\bf k}\cdot {\bf r})\frac{z_\ell(r)}{r} Y_{\ell m}(\hat{\bf r}) 
= Z_{\ell m}({\bf r})
= z_\ell(k)Y_{\ell m}(\hat{\bf k})= \sqrt{\frac2{\pi}} \int_0^\infty r^2\,dr j_\ell(kr) \frac{z_\ell(r)}{r} Y_{\ell m}(\hat{\bf k})\qquad
\eea
%
Dropping the common factors gives the relations [see Eq.~(A32) of Ref.~II]
\bea
\frac{z_\ell(r)}{r}&=&\sqrt{\frac{2}{\pi}}\int_0^\infty k^2 dk \,j_\ell(kr) z_\ell(k)\, , \label{eq:zell}
\nonumber\\
z_\ell(k)&=&\sqrt{\frac{2}{\pi}}\int_0^\infty r^2 dr \,j_\ell(kr) \frac{z_\ell(r)}{r}\, .
\eea

\subsection{Derivation of the charge form factor in momentum space}  

The nonrelativistic charge form factor is
%
\bea
G_C^{\rm NR}=\int_0^\infty dr \big[u^2(r)+w^2(r)\big] j_0(\tau_0) \label{eq:GCNR}
\eea
where $\tau_0=\frac12 Qr$.  For convenience I give the derivation of the momentum space expression, which is not widely known.

Each factor can be most easily be transformed to momentum space by first restoring the missing angular integrals: 
%
\bea
\int_0^\infty&& r^2 dr  j_0(\tau_0)\frac{z_\ell^2(r)}{r^2}=\frac1{2\pi}\int_0^\infty r^2 dr \int d\Omega_{\hat{\bf r}}\; \exp\Big[\frac{ i{\bf q}\cdot {\bf r}}{2}\Big]\frac{z_\ell^2(r)}{r^2}=
\frac{2}{2\ell+1}\int d^3 r\, \exp\Big[\frac{ i{\bf q}\cdot {\bf r}}{2}\Big] \frac{z_\ell^2(r)}{r^2}\sum_{m=-\ell}^\ell Y_{\ell m}(\hat{\bf r}) Y^*_{\ell m}(\hat{\bf r})
\nonumber\\&&
=\frac{2}{2\ell+1}\int d^3r\int d^3r'\;\delta({\bf r}-{\bf r'}) \, \exp\Big[\frac{ i{\bf q}\cdot {\bf r}}{2}\Big]  \frac{z_\ell(r)}{r}\frac{z_\ell(r')}{r'}\sum_{m=-\ell}^\ell Y_{\ell m}(\hat{\bf r})  Y^*_{\ell m}(\hat{\bf r}')  \qquad
\nonumber\\&&
=\frac{2}{(2\ell+1)(2\pi)^3}\sum_{m=-\ell}^\ell\int d^3k\int d^3r\;\exp  \Big[i({\bf k}+\frac12{\bf q})\cdot {\bf r}\Big]  \frac{z_\ell(r)}{r} Y^*_{\ell m}(\hat{\bf r})\int d^3r'\;\exp\Big[-i {\bf k}\cdot{\bf r'}\Big]\frac{z_\ell(r')}{r'} Y_{\ell m}(\hat{\bf r}')  \qquad
\nonumber
\eea
\bea
&&=\frac{2}{(2\ell+1)(2\pi)^3}\sum_{m=-\ell}^\ell\int d^3k\int d^3 k'\;\delta({\bf k'}-\frac12{\bf q}-{\bf k})\int d^3r\;\exp  \Big[i {\bf k'}\cdot {\bf r}\Big]  \frac{z_\ell(r)}{r} Y^*_{\ell m}(\hat{\bf r})
\nonumber\\
&&\qquad\qquad\qquad\qquad\times \int d^3r'\;\exp\Big[-i {\bf k}\cdot{\bf r'}\Big]\frac{z_\ell(r')}{r'} Y_{\ell m}(\hat{\bf r}')  \qquad
\nonumber\\&&
=\frac{2}{2\ell+1}\sum_{m=-\ell}^\ell\int d^3k\int d^3 k'\;\delta({\bf k'}-\frac12{\bf q}-{\bf k}) z_\ell(k') Y_{\ell m}^*(\hat{\bf k'}) z_\ell(k)Y_{\ell m}(\hat{\bf k}) 
\nonumber\\
&&=\frac{2}{2\ell+1}\sum_{m=-\ell}^\ell\int d^3k\; z_\ell(k_{\frac12}) z_\ell(k) Y_{\ell m}^*(\hat{\bf k}_{\frac12})Y_{\ell m}(\hat{\bf k})=\frac1{2\pi}\int d^3k P_\ell({\bm k_+}\cdot {\bm k_-})  z_\ell(k_+)z_\ell(k_-)
\eea
where ${\bf k}_{\frac12}={\bf k}+\frac12{\bf q}$ and ${\bf k}_\pm$ was already defined in Eq.~(\ref{eq:kpm}).  This final result was already given in Eq.~(\ref{eq:NRff}).

\end{widetext}

\section{Extraction of $F_3$ and $F_4$ from data for $G_M$ and $T_{20}$} \label{app:F3F4nextract}

Here I present details of how the unknown off-shell from factors $F_3$ and $F_4$ are determined from a simultaneous fit to the Sick GA ``data" for $G_M$ and $T_{20}$.  To this end recall the expansion (\ref{eq:expbody}).  Dropping explicit mention of the $Q^2$ arguments, this expansion is  rewritten in a form that isolates the $F_3$ and $F_4$ contributions
\bea
G_X=G_{X,0}+\sum_{i=3}^4F_i D_{X,i} \, , \label{eq:GX34}
\eea
where $G_X$ is the value of form factor at each Sick GA point $G_X=G_X(Q^2_i)$.
A similar expansion for the parameter  $y$ of Eq.~(\ref{eq:yT20}) that fixes $T_{20}$ can be written
\bea
&&3y G_C=2\eta\, G_Q\rightarrow
\sum_{i=3}^4F_ia_{i}= a_0\label{eq:y34}
\eea
where
\bea
a_i&=&3 y D_{C,i}-2\eta\,D_{Q,i}\qquad i=\{3,4\}
\nonumber\\
a_0&=&2\eta\,G_{Q,0}-3yG_{C,0} \, .
\eea
Solving Eqs.~(\ref{eq:GX34}) (with $X\to M$) and (\ref{eq:y34}) (for $i=\{3,4\}$ and $j=\{4,3\}\ne i$) gives 
 \bea
 F_i=\frac{1}{D_{ij}}\left[a_0 D_{M,j}+a_j(G_{M,0}-G_M)\right]\, . \label{eq:solutionF3F4}
 \eea
 were
 \bea
 D_{ij}=a_iD_{M,j}-a_jD_{M,i}=-D_{ji}\, .
 \eea
 Note that when $y\to\pm\infty$, $F_i$ becomes
 \bea
 F_i\to \frac{G_{C,0}D_{M,j}+D_{C,j}(G_M-G_{M,0})}{D_{M,i}D_{C,j}-D_{C,i}D_{M,j}}\, ,
 \eea
 independent of the sign of $y$, insuring that the $F_i$ are continuous at the point where $T_{20}=-1/\sqrt{2}$.
 
  Assuming there are no errors other than the error $\delta G_M$  in $G_M$ and $\delta y$ in the $y$ parameter, the errors in $F_3$ and $F_4$ can be obtained by expanding Eqs.~(\ref{eq:GX34}) and (\ref{eq:y34}) to first order, giving
 \bea
 \sum_{i=3}^4&&\delta F_i\,D_{M,i}=\delta G_M
 \nonumber\\
 \sum_{i=3}^4&&\delta F_i \,a_i=-3\delta y \left(G_{C,0}+F_3 D_{C,3}+F_4D_{C,4}\right)
 \nonumber\\ &&\phantom{\delta F_i \,a_i}\equiv -3\delta y \,b_0
 \eea
 with the solution
 \bea
 \delta F_i&=&-\frac{1}{D_{ij}}(a_j\delta G_M+3\delta y D_{M,j}b_0)
 \nonumber\\
 &\to& \left| \frac{1}{D_{ij}}\right|\,\Big(\left| a_j\delta G_M\right|+\left|3\delta y K_{jM}b_0\right|\Big]
 \eea
where the second expression ensures that each error is treated as a positive contribution.

\section{Extraction of $G_{En}$ from data for $A$} \label{app:GEnextract}  

While it is straightforward to extract the predicted values of $G_{En}$ from the data for $A$, it is still useful to outline here the way in which this was done.  I begin by isolating the $G_{En}$ contribution  from the expansion (\ref{eq:expbody}).  Dropping the $Q^2$ arguments, the new expansion is
\bea
G_X&=& G_{E} D_{X,E}+G_{M} D_{X,M} + \sum_{i=3}^4 F_{i} D_{X,i}
\nonumber\\
&\equiv& G_{E} J_{1X}+J_{0X}\, ,\qquad \label{eq:exGE}
\eea
where $J_{1X}=D_{X,E}$, $J_{0X}$ is defined by the expression, and, as before, all nucleon form factors contributing to the deuteron are {\it isoscalar\/}, so that here $G_E=G_{Es}=G_{Ep}+G_{En}$.  The first two on-shell nucleon form factors, $F_{1}$ and $F_{2}$  are related to the nucleon electric and magnetic form factors in the usual way
\bea
G_{E}(Q^2)&=&F_{1}(Q^2)-\tau F_{2}(Q^2)
\nonumber\\
G_{M}(Q^2)&=&F_{1}(Q^2)+ F_{2}(Q^2)\, , 
\eea
where $\tau=Q^2/(4m^2)$.  Hence
\bea
D_{X,1}(Q^2)&=& D_{X,E}(Q^2)+D_{X,M}(Q^2)
\nonumber\\
D_{X,2}(Q^2)&=& D_{X,M}(Q^2)-\tau D_{X,E}(Q^2)\, ,
\eea
or, in terms of the calculated body form factors,
\bea
D_{X,E}&=&\frac{D_{X,1}- D_{X,2}}{1+\tau}
\nonumber\\
D_{X,M}&=&\frac{\tau D_{X,1}+D_{X,2}}{1+\tau}\, . \label{eq:F1F2Ge}
\eea
This defines all of the coefficients in the expansion (\ref{eq:exGE}).

\begin{table*}
\begin{minipage}{6.5in}
\caption{These vector products used in the definitions of the ${\cal C}_{n,i}$ traces, originally defined in Table VI in Ref.~II, are redefined as a consequence of the transformation (\ref{eq:CC3}).  
} 
\label{tab:Acoeffs}
\begin{ruledtabular}
\begin{tabular}{lcccc}
$a$'s & $n=1\;(J_{00}^0)$ & $n=2\; (J_{+-}^0)$ & $n=3_+\; (J_{+0}^+)$
& $n=3_-\; (J_{0-}^-)$\\[0.02in]
\tableline\\[-0.08in]
$a_+$&  $(E_kQ-2k_{Cz}D_0)/(2m_d)$& $\frac1{\sqrt{2}}(k_x-ik_y)$&  $\frac1{\sqrt{2}}(k_x-ik_y)$ &
$(E_kQ-2k_{Cz}D_0)/(2m_d)$ \cr
& $\to (E_+Q-(2k_z+Q)D_0)/(2m_d)$& & & $\to (E_+Q-(2k_z+Q)D_0)/(2m_d)$\\[0.05in]
$a_-$&$-(E_kQ+2k_{Cz}D_0)/(2m_d)$& $\frac1{\sqrt{2}}(k_x+ik_y)$ & $-(E_kQ+2k_{Cz}D_0)/(2m_d)$& $\frac1{\sqrt{2}}(k_x+ik_y)$\cr
& $\to -(E_+Q+(2k_z+Q)D_0)/(2m_d)$& &  $\to-(E_+Q+(2k_z+Q)D_0)/(2m_d)$ & \\[0.05in]
$a_0$&$E_k\to E_+$ & $E_k\to E_+$  & $\frac1{\sqrt{2}}(k_x+ik_y)$ & $-\frac1{\sqrt{2}}(k_x-ik_y)$\\[0.05in]
\end{tabular}
\end{ruledtabular}
\end{minipage}
\end{table*}
%

The quadratic dependence go $A$ on $G_{E}$ can now be expressed in a compact form 
\bea
A=G_{E}^2 C_2+G_{E} C_1+C_0\, . \label{eq:GN}
\eea
The coefficients $C_i$ (all functions of $Q^2$) are 
\bea
C_2&=& J_{1C}^2+\frac89\eta^2 J_{1Q}^2+\frac23\eta J_{1M}^2
\nonumber\\
C_1&=&2J_{1C}J_{0C}+\frac{16}9\eta^2 J_{1Q}J_{0Q}+\frac43\eta J_{1M}J_{0M}\qquad
\nonumber\\
C_0&=& J_{0C}^2+\frac89\eta^2 J_{0Q}^2+\frac23\eta J_{0M}^2\, .
\eea
The solution to (\ref{eq:GN}) is
\bea
G_{E}=\frac1{2C_2}\Big(\sqrt{4C_2(A-C_0)+C_1^2}-C_1\Big)
\eea
where the sign of the square root was chosen to give a positive $G_{E}$ when $G_{M}\to0$.

The error in $G_{E}$ comes from both the error in $A$ and the errors in $F_3$ and $F_4$.  Since $F_3$ and $F_4$ contribute only to $J_{0X}$, its contribution to the error is contained in the factors
\bea
\delta J_{0X}=\sum_{i=3}^4\delta F_i {D}_{X,i}
\eea
which contribute the following errors to the $C_i$
\bea
\delta C_0&=&2 \Big[J_{0C} \delta J_{0C}+\frac89\eta^2J_{0Q} \delta J_{0Q}+\frac23\eta J_{0M}\delta J_{0M} \Big]\qquad
\nonumber\\
\delta C_1&=&2\Big[J_{1C} \delta J_{0C}+\frac89\eta^2J_{1Q} \delta J_{0Q}+\frac23\eta J_{1M} \delta J_{0M}\Big] .
\eea
combining these errors with the Experimental error in $A$ gives the following estimate for the error in $G_{E}$
\bea
\delta G_{E}=\frac{\delta A-\delta C_0-\delta C_1 G_{E}}{\sqrt{4C_2(A-C_0)+C_1^2}}\, .
\label{eq:errorGEs}
\eea

With these results in hand, we find the solution for $G_{En}$ from the solution for  $G_{E}$  by subtracting $G_{Ep}$, which is also assumed to have no error.  Hence $\delta G_{En} = \delta G_{E}$ as given in (\ref{eq:errorGEs}).

\section{Redefinitions of the ${\cal C}$ traces} \label{app:Ctraces}

In Refs. II and III the arguments for the ${\cal B}$ and ${\cal C}$ traces were chosen differently.  This is inconvenient for the numerical calculations performed in this paper, and can be easily avoided by some redefinitions.  Specializing to the case when the outgoing  particle 1 is on-shell, momenta used for the ${\cal B}$ were
\bea
\tilde k^B_+&=&\big\{E_+,{\bf k}+\frac12{\bf q}\big\}
\nonumber\\
\tilde k^B_-&=&\big\{E_+,{\bf k}-\frac12{\bf q}\big\}
\nonumber\\
\tilde p^B_-&=&\big\{D_0-E_+,-{\bf k}\big\}
\eea 
while for the ${\cal C}$ traces I previously used
\bea
k^C_+&=&\big\{E_k,{\bf k}_C\big\}
\nonumber\\
k^C_-&=&\big\{E_k,{\bf k}_C-{\bf q}\big\}
\nonumber\\
p^B_+&=&\big\{D_0-E_k,-{\bf k}_C+\frac12{\bf q}\big\}
\eea 
where here, to avoid confusion, I labeled the ${\bf k}$ momenta used for the ${\cal C}$ diagrams by ${\bf k}_C$, and therefore in these expressions $E_k\equiv E_{k_C}$.  
%
%
%
%
The ${\cal C}$ momenta can be transformed into the ${\cal B}$ momenta by the simple transformation 
\bea
{\bf k}_C\to {\bf k}+\frac12{\bf q}\, . \label{eq:CC3}
\eea
In this Appendix I show the effect of this transformation on the formulae for the ${\cal C}$ traces published in Ref.~II.

First, consider the argument shifts for the ${\cal C}$ traces.  In Ref.~II the arguments of the $K_i$ are shifted to
\bea
&&\hat R_-^2
=\frac{1}{m_d^2}\big[D_0E_k+\frac12(k_{Cz}-Q)\,Q\big]^2
-(m^2+2k_{Cz}Q-Q^2) 
\nonumber\\&&\qquad
\to \frac{1}{m_d^2}\big[D_0E_+ +\frac12(k_{z}-\frac12Q)\,Q\big]^2
-(m^2+2k_{z}Q)
%
\nonumber\\
&&\hat R_0^-=\frac1{2m_d}\big[2D_0E_k+(k_{Cz}-Q)\,Q\big]
\nonumber\\&&\qquad
\to \frac1{2m_d}\big[2D_0E_+ +(k_{z}-\frac12 Q)\,Q\big]
%
%
\eea
while the argument of the outgoing generic on-shell $Z_+$ is
\bea
R_+^2&=&\frac1{m_d^2}\Big[D_0E_k-\frac12k_{Cz}Q\Big]^2 - m^2
\nonumber\\
&\to& \frac1{m_d^2}\Big[D_0E_+-\frac12(k_{z}+\frac12 Q)Q\Big]^2 - m^2\, .
%
\eea
If the outgoing state in the ${\cal B}$ traces has particle 1 on shell, so that $k_0=E_+$,  these expressions are identical to the argument shifts given in Eq.~(\ref{eq:129}), showing that the transformation (\ref{eq:CC3}) transforms the shifts for ${\cal C}$ into those for ${\cal B}$.

The expressions for the ${\cal C}$ traces depend on the coefficients defined in Table IX of Ref.~II (some of which are defined in Tables VI and VII of that reference).  The only coefficients that depend on ${\bf k}$ are the $a$'s, with transformations summarized in Table \ref{tab:Acoeffs}, and the coefficients $c_0'$, $c_q$, and the particle 1 momentum squared, $p_+^2$, all of which transform to
\bea
c_0'&=&D_0E_k\to D_0E_+
\nonumber\\
c_q&=&-Qk_{Cz}\to -Q(k_z+\frac12 Q)
\nonumber\\
p_+^2&=&(P_+- k)^2= m_d^2+m^2-2D_0E_k+Qk_{Cz}
\nonumber\\
&\to& m_d^2+m^2 -2D_0E_+ + Q(k_z+\frac12Q) \, .
\eea
Finally, the volume integral transforms to
\bea
\int_k\equiv \int\frac{d^3k_C}{(2\pi)^3}\frac{m}{E_k}\to \int\frac{d^3k}{(2\pi)^3}\frac{m}{E_+}
= \int_k\frac{E_k}{E_+}
\eea
where in the final expression, I return to the definition $E_k=\sqrt{m^2+{\bf k}^2}$ used everywhere.

With these substitutions, the same four-vector $k_+=\{E_+,{\bf k}+\frac12{\bf q}\}$ is used for both the ${\cal C}$ and ${\cal B}_+$ traces.

\section{Corrected treatment of the angular integrals when $x_{10}\ne 1$} \label{app:x10corr}

Evaluation of the he angular integrals was discussed in detail in Appendix B of Ref.~\cite{Gross:2010qm}, but the discussion there was not accurate for cases when $x_{10}\ne 1$.  In that paper we introduced the variable $x_0$ to scale for the off-shell energy dependence to the {\it relative\/} energy, where 
\bea
p_0=x_0(E_p-\frac12 W)
\eea
(c.f. Eq.~(A16) of Ref.~\cite{Gross:2010qm}).  However, since the relative energy $p_0$ can be large when $p\to0$, the quantity $x_0$ defined in this way can also become quite large, making numerical calculations using this quantity difficult to carry out accurately.  In this paper I have chosen to scale the off-shell energy of particle 1, using the relation
\bea
p_{10}&=&x_{10}E_p
\nonumber\\
p_{20}&=&W-x_{10}E_p\, .
\eea
It follows immediately that the relative energy, expressed in tin terms of $x_{10}$ is
\bea
p_0=\frac12(p_{10}-p_{20})= x_{10}E_p-\frac12 W
\eea 
so that
\bea
x_0=\frac{2x_{10}E_p-W}{2E_p-W}\, .
\eea
This correspondence can be used quite successfully in many places, but for the discussion of the angular integrals it is best to work directly with $x_{10}$.


As  an example, consider how the treatment of the direct terms must be modified when the {\it both\/} nucleons are off-shell. Now the the momentum transfer depends on $x_{10}$
\bea
q^2(x_{10})&=&(x_{10}E_p-E_{p'})^2-p^2-p'^2+2pp'z
\nonumber\\
&=&2pp'(z_0-z)
\eea
where I assume that the initial state (with momentum $p'$) has particle 1 on-shell.  This momentum transfer is zero at the critical cosine
\bea
z_0=\frac{p^2+p'^2-(x_{10}E_p-E_{p'})^2}{2pp'}\, .
\eea
The angular integrals are strongly peaked at $z=z_0$.  
 When $x_{10}=1$, $z_0\ge1$ and approaches 1 only when $p\to p'$.  This singularity can be handled by the methods used in Ref.~\cite{Gross:2010qm}.  However,  for $x_{10}\ne 1$, $z_0$ can be less than 1 and the angular integrals can peak inside of the region of integration.   This requires a mapping of the type used for the exchange terms, described in Appendix B3, Eq.~(B9) of Ref.~\cite{Gross:2010qm}.

 \section{Errata in Ref.~II} \label{sec:errorsinII}



There are errors in the magnetic moment results reported in Ref.~II.  
As I did in Ref.~II, here I present the difference between the expansion of the relativistic  calculation and $\mu_s=0.880 = 1+\kappa_s$ to obtain the ``corrections'' to the magnetic moment coming from the relativistic calculation.  
Multiplying the  normalization condition by 0.880, written in the form  (approximating  $E_k-M_d\to -m$ and $M_d\to 2m$ in the  $a(p^2)$ terms)
\begin{widetext}
\bea
0.880 = (1+\kappa_s)&&\int_0^\infty k^2 dk\Big\{
u^2+w^2+v_t^2+v_s^2 +4a(p^2)m\Big[\delta_k(u^2+w^2)-2m(v_t^2+v_s^2)\Big]
\nonumber\\
&&-u[\delta_+\hat u]_{k_0}-w[\delta_+\hat w]_{k_0}+v_t [\delta_-\hat v_t]_{k_0}+v_s [\delta_-\hat v_s]_{k_0}-uu^{(2)}-ww^{(2)}-v_tv_t^{(2)}-v_sv_s^{(2)}\Big\}\, .
\eea
and subtracting this from the predictions of the (A) + (A$^{(2)}$) + (B) diagrams (multiplied by 1/2) gives the following corrections to the magnetic moment %
\bea
\Delta \mu_d =\sum_{X=A,B}\int_0^\infty k^2 dk\frac12\Big\{\delta\mu_{\rm NR}^X+\delta\mu_{R_c}^X+\delta\mu_{h'}^X+\delta\mu_{V_2}^A+\delta\mu_{V_1}^B+\delta\mu_{\rm int}^A+\delta\mu_P^X\Big\} \label{eq:deltamus}
\eea
where the expression reflects the fact that the only nonzero contributions to $\delta\mu_{V_1} (\delta\mu_{V_2})$ come from the (B) [(A)$^{(2)}$)] diagrams and $\delta\mu_{\rm int}^B$, while not zero, is of lower order and can be dropped.  The non-zero contributions  are therefore
\bea
\delta\mu_{\rm NR}^A&=&\delta\mu_{\rm NR}^B=-\frac34 (1+2\kappa_s)w^2
\nonumber\\
\delta\mu_{R_c}^A&=&-\left[\frac{E_k-m}{6E_k}\right]\Big[2(1+\kappa_s)u^2+\sqrt{2}(1-2\kappa_s)uw-(2-\kappa_s)w^2\Big]
\nonumber\\
\delta\mu_{R_c}^B&=& -\kappa_s \left[\frac{E_k-m}{6E_k}\right](2u^2+w^2-2\sqrt{2}uw)
\nonumber\\
\delta\mu_{h'}^A&=& -a(p^2) m \Big\{(1+2\kappa_s)(3\delta_k w^2-4m\,v_s^2) -4 \kappa_s m v_t^2 +6\sqrt{2}m\,v_tv_s\Big\}
\nonumber\\
\delta\mu_{h'}^B&=& -a(p^2) m \Big\{(1+2\kappa_s)\Big[3\delta_k w^2-2m( v_t^2+2v_s^2)\Big] - 8m\sqrt{2}(1+\kappa_s)v_t v_s\Big\} \nonumber\\
\delta\mu_{V_2}^A&=&\frac32 (1+2\kappa_s)w w^{(2)} +
\frac12(5+6\kappa_s) v_t v_t^{(2)}+ (3+2\kappa_s)v_s v_s^{(2)} + \sqrt{2}\kappa_s(v_t v_s^{(2)}+v_s v_t^{(2)})
\nonumber\\
&&-\frac{m}{\sqrt{6}}\bigg\{u^{(2)}(v_t'-\sqrt{2}v_s')+w^{(2)}(\sqrt{2}v_t'+v_s')
-v_t^{(2)}(u'+\sqrt{2}w')+v_s^{(2)}(\sqrt{2}u'-w')
\nonumber\\
&&+\frac1k \Big[2u^{(2)}( v_t-\sqrt{2}v_s)-w^{(2)}(\sqrt{2}v_t+v_s)-  3w(\sqrt{2}v_t^{(2)}+ v_s^{(2)})\Big]\bigg\} 
\nonumber\\
\delta\mu_{V_1}^B&=&(1+\kappa_s)\Big[3w[\delta_+w]_{k_0}- v_t [\delta_-v_t]_{k_0}-2v_s [\delta_-v_s]_{k_0}
-\sqrt{2}(v_t [\delta_-v_s]_{k_0}+v_s [\delta_-v_t]_{k_0})\Big]
\nonumber\\
\delta\mu_{\rm int}^A&=&- \frac{m}{\sqrt{6}}\Big[u'(v_t-\sqrt{2}v_s)+w'(\sqrt{2}v_t+v_s)+\frac3k w(\sqrt{2}v_t+ v_s)\Big]
%
%
\nonumber\\
\delta\mu_P^A&=&-\frac14 (5+6\kappa_s)v_t^2-\frac12(3+2\kappa_s)v_s^2-\sqrt{2}\kappa_s v_tv_s
\nonumber\\
\delta\mu_P^B&=& \frac14 \Big[-(3+2\kappa_s)(v_t^2+2v_s^2)+2\sqrt{2}(1-2\kappa_s)v_t v_s\Big]
\eea
Combining these terms allows us to compare then with Eq.~(5.6) of Ref.~II. The sums (divided by 2) are
\bea
\mu_{R_c}&=&\int_0^\infty k^2dk\left[\frac{E_k-m}{E_k}\right]\Big\{-\frac13\mu_s(u^2-\sqrt{2}uw+\frac12 w^2)+\frac16u^2-\frac{5}{6\sqrt{2}}uw +\frac13w^2\Big\}
\nonumber\\
\mu_{h'}&=&\int_0^\infty k^2dk \,a(p^2) m\left\{ 2\mu_s \Big[-3\delta_k w^2 +2m(v_t^2+\sqrt{2}v_tv_s+2v_s^2)\Big] +3\delta_k w^2-3m(v_t^2+\sqrt{2}v_tv_s+\frac43 v_s^2) \right\}
\nonumber\\
\mu_{V_2}&=&\int_0^\infty \frac{k^2dk}{2} \left\{(2 \mu_s-1) \frac32 w w^{(2)} +\mu_s(3v_tv_t^{(2)}+2v_sv_s^{(2)})+(\mu_s-1)\sqrt{2}(v_tv_s^{(2)}+v_sv_t^{(2)})-\frac12v_tv_t^{(2)}+v_sv_s^{(2)}  -m'^{(2)}\right\}
\nonumber\\
\mu_{V_1}&=&\int_0^\infty k^2dk\; \mu_s\left\{\frac32 w[\delta_+w]_{k_0}- \frac12 v_t [\delta_-v_t]_{k_0}-v_s [\delta_-v_s]_{k_0}
-\frac1{\sqrt{2}}(v_t [\delta_-v_s]_{k_0}+v_s [\delta_-v_t]_{k_0})\right\}
\nonumber\\
\mu_{\rm int}&=&-\frac{m}{2\sqrt{6}}\int_0^\infty k^2dk\; \left\{u'(v_t-\sqrt{2}v_s)- w(\sqrt{2}v_t+ v_s)'+\frac1{k}w(\sqrt{2}v_t+v_s)\right\}
\nonumber\\
\mu_{P}&=&\int_0^\infty k^2dk\; \left\{-\mu_s\Big(\frac98v_t^2+\sqrt{2}v_tv_s +\frac54 v_s^2\Big)  +\frac14 v_t(v_t+5\sqrt{2}v_s)\right\}
\eea
\end{widetext}
Note that $\mu_{\rm NR}$, $\mu_{V_2}$ and $\mu_{\rm int}$ agree with Ref.~II, but the others do not.  The $\mu_\chi$ terms have been found to be negligible and were not recalculated.

In the process of computing the form factors, the following errata we discovered in the equations reported in Ref.~II:

\begin{enumerate}
\item Eq.~(C3) should read  $\widetilde{Z}_\pm=\widetilde{Z}-(E_k-k_0)\widetilde{Z}_{k_0}$
\item Eq.~(B1): 
\begin{itemize}
\item the coefficient of the $D_+D_-$ term should be divided by an additional factor of $2$
\item  the coefficient of the $C_+C_-$ term should be divided by an additional factor of $2m^2$
\end{itemize}
\item Eq.~(B2): 
\begin{itemize}
\item 
in the  coefficient of $C_+A_-$, the term $-2b_0a_+b_+$ should be replaced by $-2b_0a_-b_+$
\item a closing parentheses, ), is missing from the  coefficient of the $C_+D_-$ term; it belongs just before $-z_+$, so that the coefficient of $b_0$ includes the $b_+$ and $a_+$ terms  but not the $z_+$ term
\item a similar closing parentheses, ), is missing from the  coefficient of the $D_+C_-$ term; it belongs just before $-z_-$
\item in the  coefficient of $C_+C_-$, the coefficient of $(4m^2+m_d^2)$ is $(b_+z_-+b_-z_+)$, and {\it NOT\/} $(b_+z_-+b_-c_+)$
\end{itemize}

\item Eq.~(B.7): 

\begin{itemize}
\item the $H I$ terms are divided by $m^4$ ({\it not} $m^2$)
\item in the coefficient of the $\widetilde F_+ \widetilde F_-$ term replace $(2X_4-X_5)$ by $(X_4-X_5)$ and in the coefficient of the $\widetilde H_+ \widetilde H_-$ term replace $X_1X_5$ by  $X_1(X_5+X_4)$
\end{itemize}

\item Eq.~(B10):

\begin{itemize}
\item the $G K_2$ terms should be divided by $m^2$
\item divide the entire trace by an extra factor of $2m$ (so the coefficient in front is $\zeta_B/(4m^2)$, {\it not\/} $\zeta_B/(2m)$
\end{itemize}
\end{enumerate}

These errors arose when the original Mathematica formula were transcribed into text.  Fortunately, all of the results of Refs.~II and III are unaffected by these errors because they were derived directly from the correct Mathematica formula.

\end{document}